\documentclass[a4paper,11pt]{article}
\pdfoutput=1 
\usepackage{jheppub} 
\usepackage[T1]{fontenc} 
\usepackage{amsmath,amsfonts,amssymb,amsthm}
\usepackage{psfrag}
\usepackage{enumerate}
\usepackage{mathrsfs}
\usepackage{graphicx}
\usepackage{wrapfig}
\usepackage{xcolor}
\usepackage{afterpage}

\newcommand{\be}{\begin{equation}}
\newcommand{\ee}{\end{equation}}
\newcommand{\beq}{\begin{eqnarray}}
\newcommand{\eeq}{\end{eqnarray}}

\def\({\left (}
\def\){\right )}

\def\r2{\sqrt{2}}

\newcommand{\bbibitem}[1]{\bibitem{#1}\marginpar{#1}}

\newcommand{\myeq}[1]{\begin{equation} #1 \end{equation}}

\newtheoremstyle{problemstyle}  
        {\topsep}                                               
        {\topsep}                                               
        {}                               
        {}                                                  
        {\bfseries}                 
        {:}         
        {\newline}                                          
        {\thmname{#1}\thmnumber{ #2}\thmnote{ (#3)}%
        \addcontentsline{toc}{subsection}{#1~\theproblem:~#3}}

\theoremstyle{problemstyle}
\newtheorem{problem}{Exercise}

\newtheoremstyle{solutionstyle}  
        {\topsep}                                               
        {\topsep}                                               
        {}                               
        {}                                                  
        {\bfseries}                 
        {:}         
        {\newline}                                          
        {}

\theoremstyle{solutionstyle}
\newtheorem{solution}{Solution}

\makeatletter

\newcommand\listexercisename{List of Exercises}
\newcommand\listofexercises{%
  \section*{\listexercisename}\@starttoc{problem}}
\makeatother

\def\noLabels{\let\Label=\label}
\def\nobbibitem{\let\bbibitem=\bibitem}

\usepackage[normalem]{ulem}
\usepackage{etoolbox}
\usepackage{cancel}

\newtoggle{editing}
	\togglefalse{editing}

\iftoggle{editing}{%
    
	\newcommand{\BCcom}[1]{{\textbf{\textcolor{magenta}{#1 -- BC}}}}
	\newcommand{\BCout}[1]{{\textbf{\textcolor{magenta}{\sout{#1}}}}}

	\newcommand{\JSout}[1]{{\textbf{\textcolor{blue}{\sout{#1}}}}}
	\newcommand{\JSeqout}[1]{{\mathbf{\textcolor{blue}{\cancel{#1}}}}}
	\newcommand{\JSc}[1]{{\textbf{\textcolor{blue}{[#1 -- JS]}}}}

	\newcommand{\LLc}[1]{{\textbf{\textcolor{green}{#1 -- LL}}}}

	\newcommand{\SRMout}[1]{{\textbf{\textcolor{purple}{\sout{#1}}}}}
	\newcommand{\SRMc}[1]{{\textbf{\textcolor{purple}{[#1 -- Sam]}}}}
	
	\def\Label#1{\label{#1}}
}{%
	
	\newcommand{\BCcom}[1]{}
	\newcommand{\BCout}[1]{}

	\newcommand{\JSout}[1]{}
	\newcommand{\JSeqout}[1]{}
	\newcommand{\JSc}[1]{}

	\newcommand{\LLc}[1]{}

	\newcommand{\SRMout}[1]{}
	\newcommand{\SRMc}[1]{}
	
	\def\Label#1{\label{#1}}

}

\subheader{SU-ITP-15/07}
\title{Integral Geometry and Holography}

\author[a]{Bart{\l}omiej Czech}
\author[a]{Lampros Lamprou}
\author[a]{Samuel McCandlish}
\author[b]{James Sully}

\affiliation[a]{Stanford Institute for Theoretical Physics, Department of Physics, Stanford University\\
Stanford, CA 94305, USA}
\affiliation[b]{Theory Group, SLAC National Accelerator Laboratory\\ Menlo Park,
CA 94025, USA}

\emailAdd{czech@stanford.edu}
\emailAdd{llamprou@stanford.edu}
\emailAdd{smccandlish@stanford.edu}
\emailAdd{jsully@slac.stanford.edu}

\abstract{
We present a mathematical framework which underlies the connection between information theory and the bulk spacetime in the AdS$_3$/CFT$_2$ correspondence. A key concept is \emph{kinematic space}: an auxiliary Lorentzian geometry whose metric is defined in terms of conditional mutual informations and which organizes the entanglement pattern of a CFT state. When the field theory has a holographic dual obeying the Ryu-Takayanagi proposal, kinematic space has a direct geometric meaning: it is the space of bulk geodesics studied in integral geometry. Lengths of bulk curves are computed by kinematic volumes, giving a precise entropic interpretation of the length of any bulk curve. We explain how basic geometric concepts -- points, distances and angles -- are reflected in kinematic space, allowing one to reconstruct a large class of spatial bulk geometries from boundary entanglement entropies. In this way, kinematic space translates between information theoretic and geometric descriptions of a CFT state. As an example, we discuss in detail the static slice of AdS$_3$ whose kinematic space is two-dimensional de Sitter space. 
}

\begin{document}
\maketitle
\flushbottom

\section{Introduction} 
\Label{intro}

The last decade has taught us to be mindful of entanglement. One way to see the importance of entanglement in local quantum field theories is to contemplate the area law \cite{area1, area2}: the physically relevant states obeying the area law are a small island in the morass of the full Hilbert space, most of which comprises states with too much or too little entanglement. Traditional formalisms of quantum field theory ignore this fact; they treat entangled and unentangled states on an equal footing. One might imagine, however, that it is possible to formulate quantum field theory in a novel way, wherein quantum states are described and organized according to the amount and structure of their entanglement. If successful, such a reformulation would radically alter our understanding of what quantum field theory is.

A promising place to start looking for such a formulation of quantum field theory is holographic duality \cite{maldacena97}. 
This is because the bulk gravitational dual encodes the boundary field theory entanglement in an explicit and convenient way in terms of areas of minimal surfaces \cite{rt1, rt2, hrt}. As a result, central tenets of gravitational physics -- Einstein's equations \cite{einsteineq1, einsteineq2, einsteineq3}, energy conditions \cite{tomography, markslast}, even connectedness of spacetime \cite{marksessay, mark2, bianchimyers, hole, erepr, holeography} -- have been reduced to statements about entanglement in the boundary theory. This development should be compared with another route toward an entanglement-based understanding of field theory: tensor networks. Swingle suggested  \cite{briansessay, brianspaper} that the MERA tensor network \cite{mera, merathermal} may be viewed as a discretization of the spatial slice of the holographically dual spacetime, because the latter -- like MERA -- also organizes the entanglement of the field theory state according to scale.\footnote{The analogy between holography and tensor networks was further advanced in \cite{tadashisnetwork, hartmanmaldacena, xiaoliang, shocks, errorcorrectingnetwork}; see also \cite{critique} for a critique of associating MERA directly with the spatial slice of AdS$_3$.} It is clear that holography has much to teach us about entanglement in field theory, but the lessons have not yet become apparent. Conversely, by understanding the structure of entanglement we may develop a powerful new framework for studying quantum gravity.

In this paper we introduce a new geometric language, which clarifies the holographic connection between geometry and entropy. A key trick -- which resonates with our philosophy of treating entanglement as a central quantity -- is to describe the bulk geometry entirely through its set of minimal surfaces.\footnote{See \cite{porratirabadan} for a pre-Ryu-Takayanagi effort in the same direction.} A natural way to do this is to consider a sort of integral transform: we generate an auxiliary dual geometry whose points describe integrated quantities in the real space. This approach is the central motif of \emph{integral geometry} \cite{santalo}.

With an eye to applications to the AdS$_3$/CFT$_2$ correspondence, we will examine the space of geodesics---\emph{kinematic space}---on 2-dimensional manifolds with boundary.\footnote{The boundary may be a conformal boundary.} 
The manifold in question is the spatial slice of a holographic, time-reflection symmetric, asymptotically AdS$_3$ spacetime.
While the mathematical literature on integral geometry has focused on symmetric spaces, we show that the formalism is easily extended to more general settings. Of particular interest are those manifolds whose tangent space is completely covered by geodesics anchored on the boundary. As we show, in this case the data in kinematic space suffices to determine the real space geometry.

In the AdS$_3$/CFT$_2$ correspondence the kinematic space ushers in two ostensibly distinct, powerful simplifications. On the geometric side, it translates lengths of bulk curves into boundary quantities, which have an entropic interpretation in the field theory. This property of kinematic space was first discovered in  \cite{holeography} under the guise of the differential entropy formula. But for an information theorist, the kinematic space organizes the data about subsets of the field theory Hilbert space and the pattern of their entanglement. Its causal structure tells us which intervals contain which other intervals (Sec.~\ref{causalstr}) while its volume form (eq.~\ref{gencrofton}) captures the conditional mutual informations of neighboring intervals (Sec.~\ref{cmi}). The miracle of holography is that these two {\it a priori} distinct descriptions of the field theory state are unified in a single object: kinematic space. In fact, the benefits of kinematic space have already found a compelling application: in an upcoming paper \cite{upcoming} we explain that the MERA tensor network encodes the conditional mutual informations and containment relations among intervals in exactly the same way as kinematic space does. Thus, MERA is best seen as a discretization of kinematic space rather than of the spatial slice of the holographic geometry.
\begin{figure}[t]
\centering
\includegraphics[width=.9\textwidth]{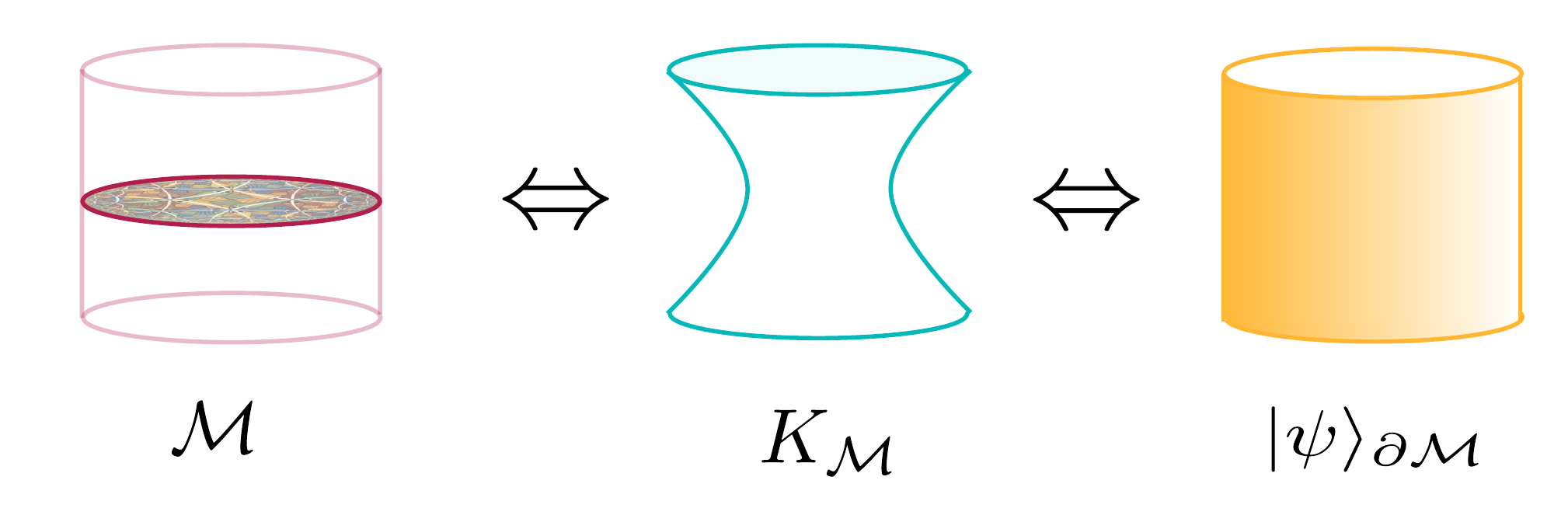}
\caption{Kinematic space, $K_{\mathcal{M}}$, acts as an intermediary that translates between the boundary language of information theory describing the state $\psi$ and the bulk language of geometry for the manifold $\mathcal{M}$.}
\Label{fig:schematic}
\end{figure}

The present text is intended to introduce the (appropriately generalized) formalism of integral geometry to the community of holographers, field theorists and information theorists. The generalization extends integral geometry to spaces with less symmetry, away from pure AdS$_3$. We also explain how to use the formalism to reconstruct the bulk from boundary data (Appendix~\ref{point-curve-derivation}). We advocate that kinematic space should be understood as a translator that converts the boundary language of information theory into the bulk language of geometry (see Figure \ref{fig:schematic}).

For an initial orientation, we start with some classic results in integral geometry in two spatial dimensions -- first in flat space, then on the hyperbolic plane. These will give the reader a flavor of what to expect next.
The central theme of the introductory expositions is describing objects on the (Euclidean and hyperbolic) plane in terms of straight lines (geodesics) that intersect them. This is relevant for AdS$_3$/CFT$_2$, because in that case the Ryu-Takayanagi proposal equates boundary entanglement entropies with lengths of geodesics. 

\subsection{Integral geometry on the Euclidean plane}
\Label{flatplane}

\begin{figure}[t]
\centering
\includegraphics[width=.3\textwidth]{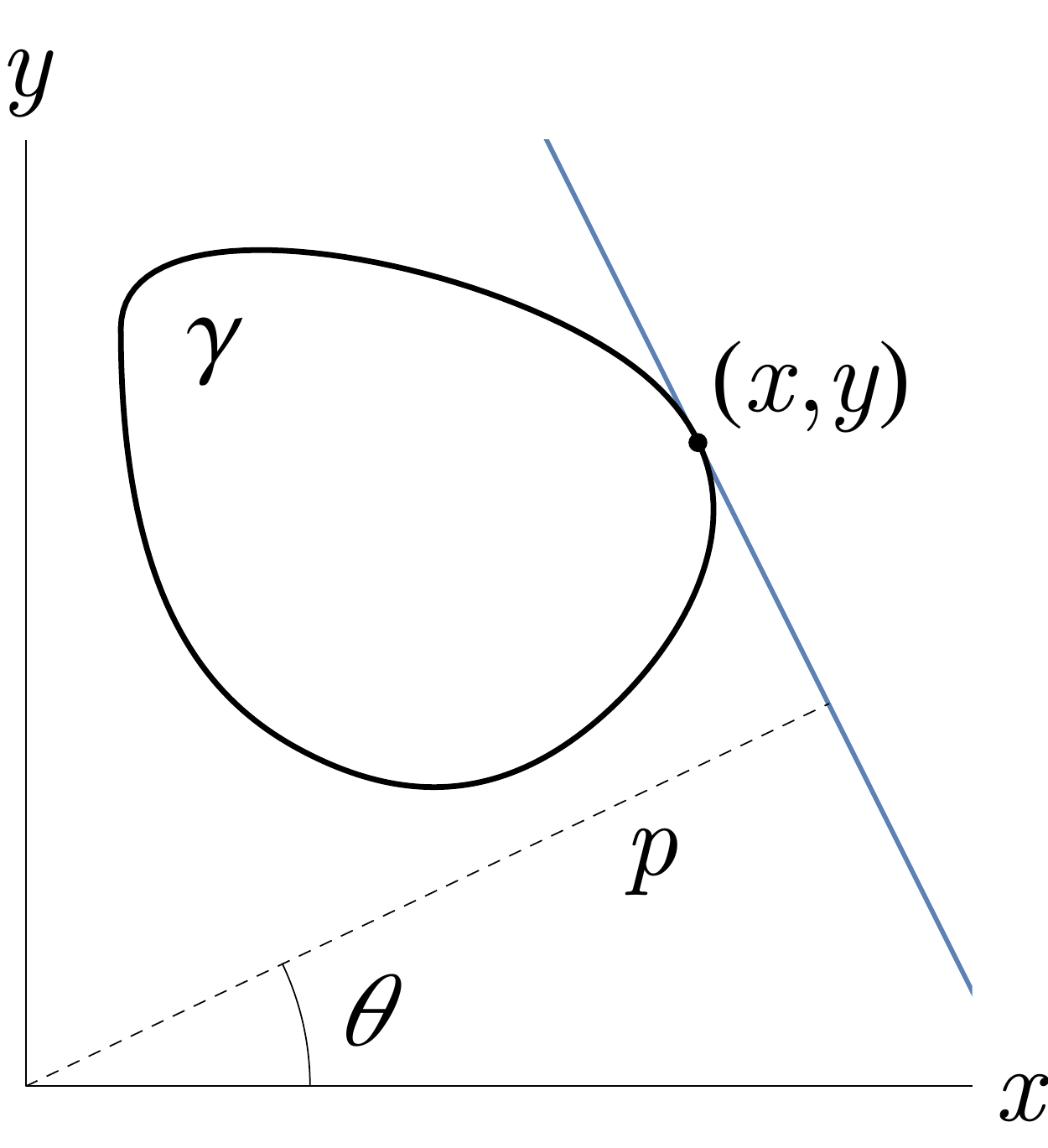}
\caption{The notation of Sec.~\ref{flatplane}.}
\Label{croftonfig}
\end{figure}

Our presentation is adapted from \cite{santalo}. 
Consider a closed, smooth, convex curve $\gamma$ on the Euclidean plane. We wish to relate the circumference of $\gamma$ to its tangent straight lines; see Figure~\ref{croftonfig}. 
Any straight line can be written in the form
\begin{equation}
x \cos \theta + y \sin\theta - p = 0\,,\Label{straight}
\end{equation}
where $\theta$ is the polar angle on the plane. In this representation, $p$ is the distance of the straight line from the origin. 
Note that if we continuously change $\theta \to \theta + \pi$ and $p \to -p$, the straight line maps to itself but its asymptotic endpoints are exchanged. 
Thus, $(\theta, p)$ parameterize the set of \emph{oriented} straight lines.

As we follow the curve $\gamma$, the $\theta$- and $p$-values of the straight line tangent to $\gamma$ change. 
We represent their dependence as a function $p(\theta)$, which for a convex curve is single-valued. 
In the appendices, we collect useful basic results as exercises, the first of which, Exercise~\ref{deriveflatcrofton}, shows that the circumference of $\gamma$ can be expressed in terms of $p(\theta)$ as:
\begin{equation}
\textrm{circumference of $\gamma$} = \int_0^{2\pi} d\theta\, p(\theta). 
\Label{flatcrofton}
\end{equation}
Eq.~(\ref{flatcrofton}) holds for closed, convex curves; the smoothness assumption is unnecessary.

Instead of working with straight lines tangent to $\gamma$, it is possible to re-express eq.~(\ref{flatcrofton}) in terms of straight lines that intersect it (Exercise~\ref{derivefinalflat}). 
The result is the celebrated Crofton formula:
\begin{equation}
\textrm{length of $\gamma$} = \frac{1}{4} \int_0^{2\pi} d\theta \int_{-\infty}^{\infty} dp \, n_\gamma(\theta, p)\,,
\Label{flatfinalcrofton}
\end{equation}
where $n_\gamma(\theta,p)$ is the intersection number with $\gamma$. 
The weighting by $n_\gamma$ makes the Crofton formula manifestly additive under concatenation of curves, which allows us to lift the assumptions of Exercise~\ref{derivefinalflat}: $\gamma$ need not be convex or closed, which is why we replaced the word `circumference' with `length.'    
The Crofton formula (\ref{flatfinalcrofton}) holds for all finite length curves on the Euclidean plane.  
The space $\theta\in[0,2\pi],\ p\in[-\infty,\infty]$ of oriented geodesics is known as the \emph{kinematic space}, $K$, of the Euclidean plane.

\begin{figure}[t]
\centering
\includegraphics[height=.3\textwidth]{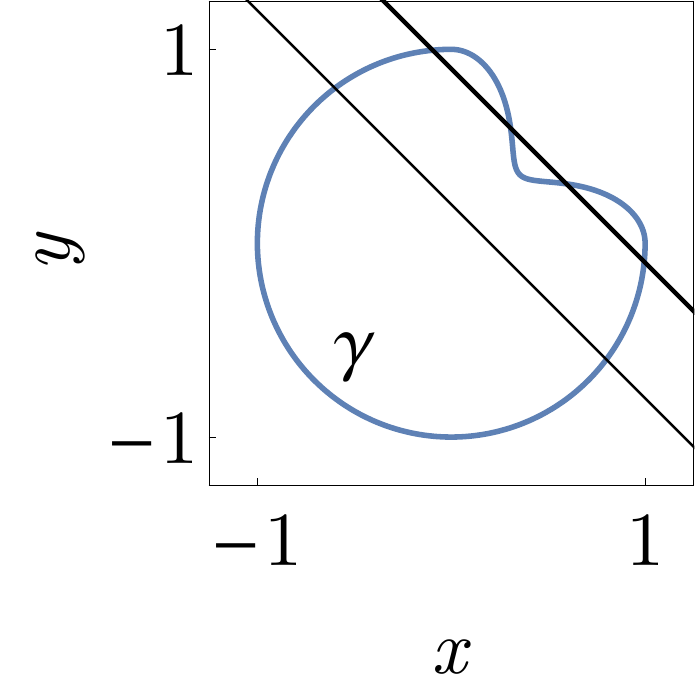}
\hspace{0.1\textwidth}
\includegraphics[height=.3\textwidth]{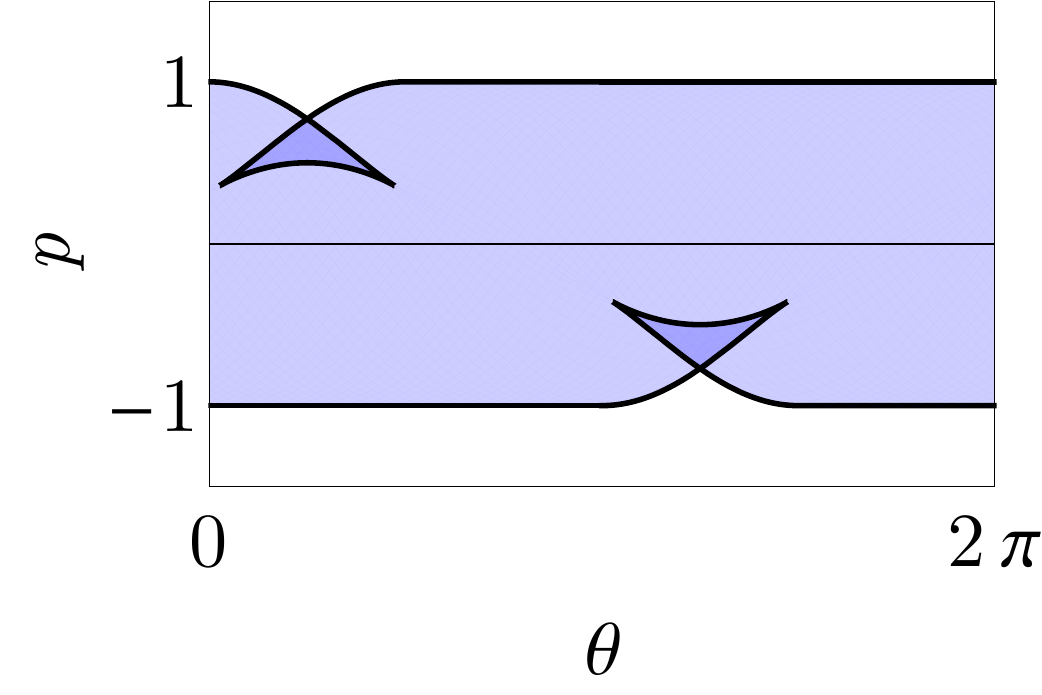}
\caption{Left: A nonconvex curve $\gamma$ in flat space.  Some lines intersect the curve twice, while others intersect four times.
Right: The region of kinematic space showing the set of curves intersecting $\gamma$.  The black curves correspond to geodesics tangent to $\gamma$, and they form the boundary of the region of intersecting curves.  The light-shaded region corresponds to geodesics intersecting twice, while the dark-shaded region corresponds to geodesics intersecting four times.  This intersection number enters the Crofton formula (\ref{flatfinalcrofton}).}
\Label{nonconvex-flat}
\end{figure}

In summary, we obtain an equation of the form
\begin{equation}
\textrm{length of $\gamma$} = \frac{1}{4} \int_K \omega(\theta, p)\, n_\gamma(\theta,p) \, .
\Label{flatcroftonformal}
\end{equation}
The integral on the right hand side is done over the kinematic space $K$, the space of all oriented straight lines on the Euclidean plane. It is well defined, because the measure of integration (the Crofton form),
\begin{equation}
\omega = dp \wedge d\theta \, ,
\Label{flatcroftonform}
\end{equation}
is invariant under the isometries of the Euclidean plane. 
The integrand is the intersection number of the curve $\gamma$ and the straight line. 
In effect, our formula translates the length of a curve into a volume in the space of straight lines. When the curve is not convex, the volumes contribute with different multiplicities; see Fig. \ref{nonconvex-flat}.

\subsection{A preview of integral geometry on the hyperbolic plane}
\Label{introhyp}

The flat space results mentioned above have nice analogues on the hyperbolic plane $\mathbb{H}^2$. In coordinates
\begin{equation}
ds^2 = d\rho^2 + \sinh^2 \rho\, d\tilde\theta^2
\Label{h2metric}
\end{equation}
the equation of a geodesic is:
\begin{equation}
\tanh\rho \, \cos(\tilde\theta - \theta) = \cos \alpha.
\Label{geodesic}
\end{equation}
The parameters that label the oriented geodesic are $\theta \in [0, 2\pi]$ and $\alpha \in [0, \pi]$; they replace the flat space parameters $\theta$ and $p$. 
Pairs of geodesics differing only in orientation are now related by
\begin{equation}
\theta\, \leftrightarrow\, \theta + \pi
\qquad {\rm and} \qquad
\alpha \, \leftrightarrow \, \pi - \alpha.
\end{equation}
The parameterization is illustrated in Fig. \ref{alphatheta}.

\begin{figure}[t]
\centering
\includegraphics[height=.3\textwidth]{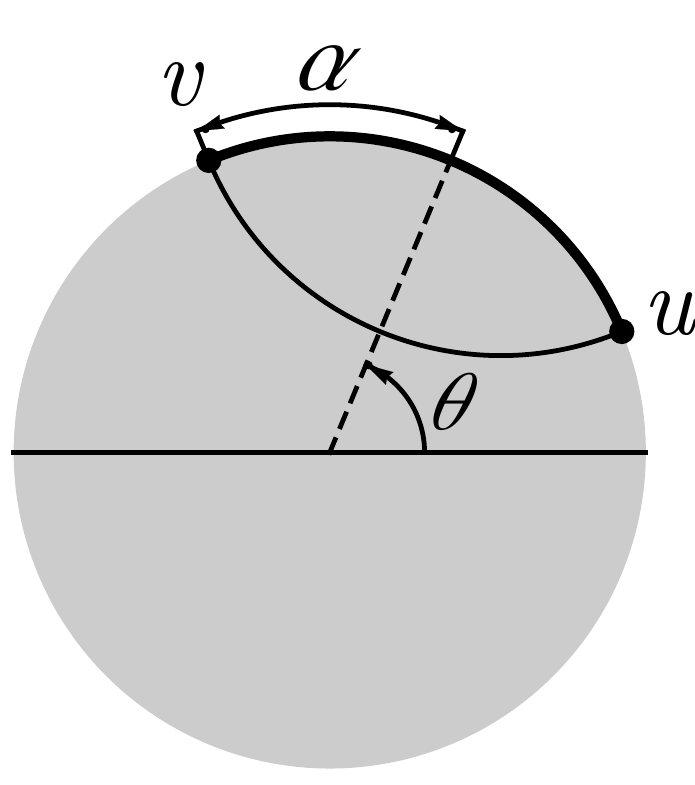}
\caption{The parameterization of kinematic space for the hyperbolic plane.  We denote the opening angle of a geodesic by $\alpha$ and the angular coordinate of the center of the geodesic by $\theta$. The endpoints are labeled $u$ and $v$.}
\Label{alphatheta}
\end{figure}

Eq.~(\ref{flatcroftonformal}) carries over to the hyperbolic plane essentially unchanged \cite{santalo, solanes}:
\begin{equation}
\textrm{length of $\gamma$} = \frac{1}{4} \int_K \omega(\theta, \alpha)\, n_\gamma(\theta,\alpha) \,.
\Label{hypfinalcrofton}
\end{equation}
The non-trivial part is to find the measure $\omega(\theta, \alpha)$ on the space of geodesics. One can make a good guess by considering the flat space limit of hyperbolic space (Exercise~\ref{makeguess}). To verify the guess, one checks that $\omega(\theta,\alpha)$ is invariant under the isometries of the hyperbolic plane (Exercise~\ref{proveguess}). The final answer is:
\begin{equation}
\omega(\theta, \alpha) = - \frac{1}{\sin^2\alpha}\, d\alpha \wedge d\theta \,.
\Label{atcroftonform}
\end{equation}
For the purposes of this paper, it is more useful to parameterize the space of geodesics in terms of the $\tilde\theta$-coordinates of their endpoints on the cutoff surface:
\begin{equation}
u = \theta - \alpha \qquad {\rm and} \qquad v = \theta + \alpha \, .
\Label{defuv}
\end{equation}
In these coordinates, the Crofton form becomes:
\begin{equation}
\omega(u,v) = \frac{1}{2 \sin^2\left(\frac{v-u}{2}\right)}\, du \wedge dv\,.
\Label{uvcroftonform}
\end{equation}

The key facts are eqs.~(\ref{uvcroftonform}) and eq.~(\ref{hypfinalcrofton}). These formulas encode all there is to know about the geometry of the hyperbolic plane.
They gain a profound physical significance in the AdS$_3$/CFT$_2$ correspondence, where the Ryu-Takayanagi proposal \cite{rt1, rt2} identifies bulk geodesic lengths with boundary entanglement entropies. We will see below that the ``density of geodesics'' (\ref{uvcroftonform}) is directly related to the strong subadditivity of entanglement entropy in the dual boundary theory. The Crofton formula (\ref{hypfinalcrofton}) then relates lengths of bulk curves to the strong subadditivity of the infrared degrees of freedom. In the remainder of this paper we explain these and other intriguing connections between integral geometry, holography and quantum information theory.

\subsection{What we do next}
\Label{plan}

In the previous sections we introduced the reader to the language of integral geometry. We used classic results about symmetric 2-dimensional spaces for illustration. We now wish to generalize the machinery of integral geometry to describe the bulk geometry in AdS$_3$/CFT$_2$ holography. 

In Sec.~\ref{croftongeneral} we define the kinematic space for an arbitrary background, departing from the highly symmetric settings of the hyperbolic plane. Remarkably, a choice of the Crofton form that is naturally suggested by holography (eq.~\ref{uvcroftonform}) has a beautiful interpretation in information theory: it captures the conditional mutual informations of neighboring intervals in the dual CFT.
That, in turn, motivates equipping the kinematic space with a metric structure, which is necessarily Lorentzian. This device will turn out to be useful for making contact with tensor network approaches to holography, which is the subject of our upcoming paper. 
The connection between information theory and the kinematic space will allow us to translate a number of geometric objects into the language of entropy.

This translation is the subject of Sec.~\ref{vackin}. 
We discuss how points, distances, curves, and angles are represented in kinematic space. 
In particular, we will see that curves in the bulk correspond to collections of intervals on the cutoff surface. 
The length of the curve is an aggregate measure of how far these intervals are from saturating the strong subadditivity of entanglement entropy. 
Eq.~(\ref{kgammageod}) is an example of this statement.
We explain how to use our formalism to reconstruct the bulk from boundary data in Appendix~\ref{point-curve-derivation}.

Sec.~\ref{exads3} illustrates the formalism in the most familiar settings -- the static slice of pure AdS$_3$. In this special instance our results can be obtained through a different route that exploits the maximal symmetry of the hyperbolic plane. In Appendices~\ref{geometric-concepts} and \ref{exercises} we present a broader class of facts and examples, some arranged in the form of exercises with solutions. We hope that this method of presentation will benefit the reader who intends to put integral geometry to practical use.
This arXiv submission also includes a Mathematica visualization tool.

We conclude with a discussion. 

\section{The Crofton form and conditional mutual information}
\Label{croftongeneral} 

The discussion in Sec.~\ref{introhyp} relied on the symmetry of the hyperbolic plane. In this section we extend this formalism beyond the vacuum while expressing the results of integral geometry in the language of holography.

We claim that for every static holographic spacetime in the AdS$_3$/CFT$_2$ correspondence it is possible to choose a Crofton form $\omega(\theta, \alpha)$ that makes eq.~(\ref{hypfinalcrofton}) correct. In other words, in a static situation we always know how to select the correct measure on the space of geodesics to obtain the schematic equation
\begin{equation}
\textrm{length of $\gamma$} = \int_\textrm{intersect $\gamma$} \mathcal{D}{\rm (geodesics)}\,,
\Label{eqschematic}
\end{equation}
which applies to every curve $\gamma$ in the bulk. Once more, for non-convex curves $\gamma$ the geodesics on the right hand side count with multiplicity.

Because our motivation is holographic, it is wise to restore units of length. We denote the asymptotic curvature of the spacetime $L$. Its magnitude in terms of the Planck length is proportional to the central charge of the dual CFT \cite{brownhenneaux}:
\begin{equation}
c = 3L / 2G \Label{brhenn}
\end{equation}
In these units, we are looking for the Crofton form $\omega(u,v)$ that satisfies
\begin{equation}
\frac{\textrm{length of $\gamma$}}{4G} = \frac{1}{4} \int_K \omega(u, v)\, n_\gamma(u,v)
\Label{ourclaim}
\end{equation}
for every curve $\gamma$ on the static slice of the holographic spacetime. The coordinates $u$ and $v$ are related to $\alpha$ and $\theta$ by eq.~(\ref{defuv}).

\subsection{The Crofton form}
\Label{seccrofton}

\begin{figure}[t]
\centering
\includegraphics[height=.4\textwidth]{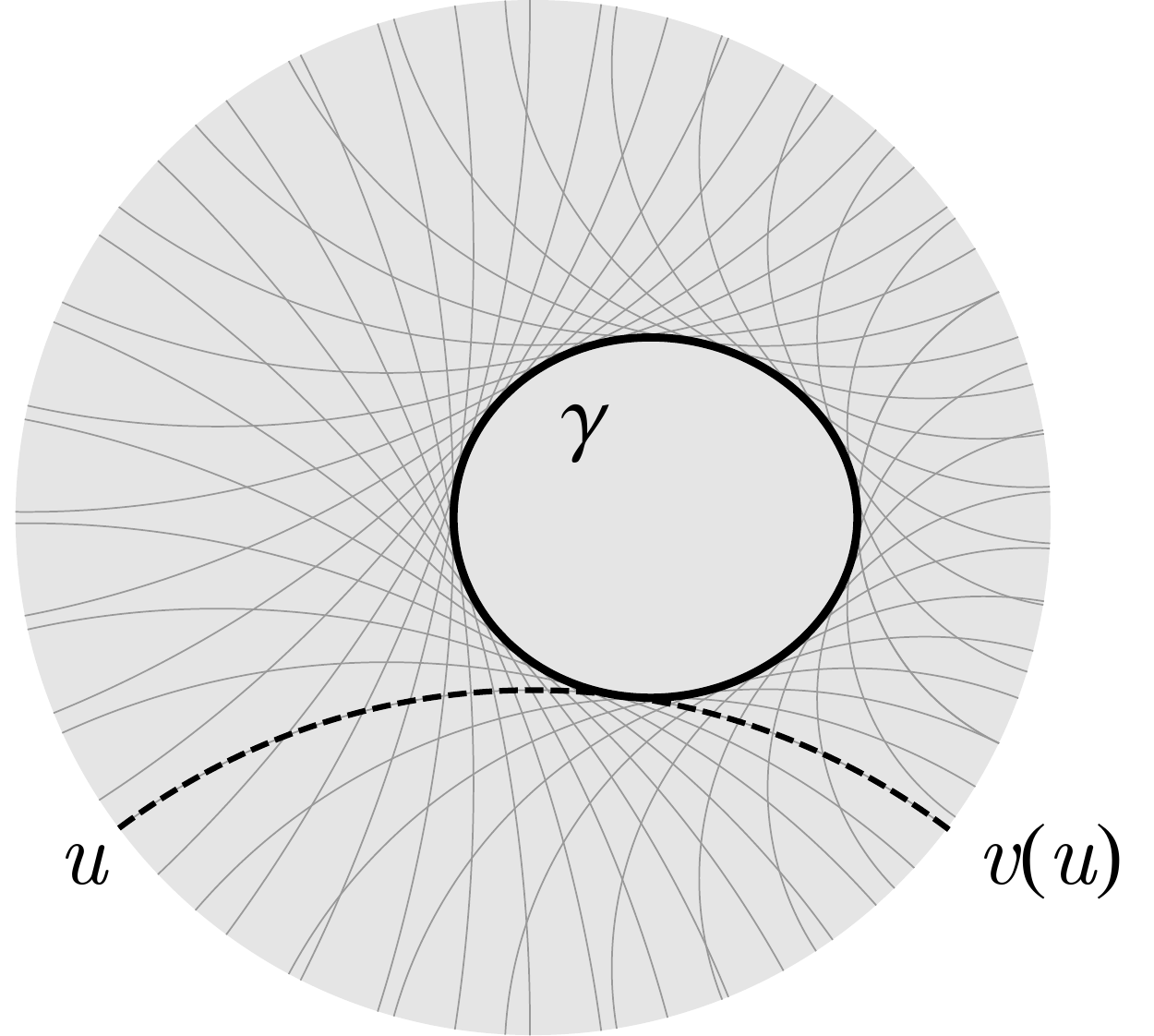}
\caption{We characterize a convex curve $\gamma$ with a function $v(u)$ chosen so that the geodesic from $u$ to $v(u)$ is tangent to $\gamma$. When $\gamma$ is not differentiable, the geodesic meets $\gamma$ at an isolated point.}
\Label{dualcurve}
\end{figure}

The correct choice of measure in the space of geodesics will be:
\begin{equation}
\omega(u,v) = \frac{\partial^2 S(u,v)}{\partial u\, \partial v} \, du \wedge dv\, .
\Label{gencrofton}
\end{equation}
Here $S(u,v)$ is the length of the geodesic connecting the boundary points $\tilde\theta = u$ and $\tilde\theta = v$ on the cutoff surface. When the Ryu-Takayanagi formula applies, $S(u,v)$ is the entanglement entropy of the interval $(u,v)$.

We will verify this is the correct measure by substituting a closed convex test curve $\gamma$ into eq.~(\ref{hypfinalcrofton}). To do so, we need a workable parameterization of $\gamma$. In analogy to the function $p(\theta)$ in Sec.~\ref{flatplane}, we describe the curve using a function $v(u)$. When the curve $\gamma$ is differentiable, $(u, v(u))$ are endpoints of a geodesic tangent to $\gamma$; for non-differentiable $\gamma$ we require that the geodesic connecting $u$ and $v(u)$ intersect $\gamma$ at an isolated point. See Fig.~\ref{dualcurve} for an illustration.

With this parameterization of $\gamma$, we can perform integral (\ref{ourclaim}): \begin{equation}
\frac{1}{4}  \int_K \omega(u, v)\, n_\gamma(u,v) 
= \frac{1}{4} \cdot 2 \int_0^{2\pi} du \int_{v(u)}^{u+\pi} dv \,\,\frac{\partial^2 S(u,v)}{\partial u\, \partial v} \cdot 2 
= - \int_0^{2\pi} du\, \frac{\partial S(u,v)}{\partial u} \big|_{v = v(u)}
\end{equation}
The explicit limits of integration account for each physical geodesic only once, without regard to orientation; this is corrected by the first factor of 2. The second factor of 2 is the intersection number of the geodesic and curve $\gamma$, which follows from assuming that $\gamma$ is convex. 

The right hand side is the well-known differential entropy formula \cite{holeography} for the length of a closed convex curve:
\begin{equation}
\frac{\textrm{circumference of $\gamma$}}{4G} = - \int_0^{2\pi} du\, \frac{\partial S(u,v)}{\partial u} \big|_{v = v(u)}
\Label{diffent}
\end{equation}
As was first proved in Ref.~\cite{robproof}, this equation applies in any spacetime satisfying the Ryu-Takayanagi proposal.\footnote{For completeness, in Appendix \ref{diffent-review} we present a compact general proof of differential entropy in the spirit of \cite{wien}.}

This establishes eq.~(\ref{ourclaim}) for closed, convex curves $\gamma$. The exercises in the appendices illustrate that its validity extends to nonconvex and open curves. In effect, we have shown that any time-reflection symmetric holographic geometry obeying the Ryu-Takayanagi formula allows us to construct a kinematic space. Its Crofton form is eq.~(\ref{gencrofton}). This form defines a natural measure on the space of geodesics that traverse the geometry.

\subsection{Relation to differential entropy}
\Label{diffent-relation}

A careful reader will notice that our derivation of $\omega(u,v)$ is tantamount to setting
\begin{equation}
\omega(u,v) = d \kappa\,, \Label{omegafromkappa}
\end{equation}
where $\kappa$ is the form\footnote{This is the tautological one-form, which serves as the symplectic potential for the symplectic form $\omega(u,v)$.} integrated in the differential entropy formula:
\begin{equation}
\kappa = - du\, \frac{\partial S(u,v)}{\partial u} \big|_{v = v(u)}
\end{equation}
This derivation parallels the way we obtained the flat space Crofton form from eq.~(\ref{flatcrofton}), in which case we had $\kappa = p \, d\theta$. Eq.~(\ref{omegafromkappa}) makes evident that adding a total derivative term to $\kappa$ has no effect on $\omega(u,v)$. Such total derivative terms have been employed in \cite{holeography, robproof} for proving the differential entropy formula and in \cite{lampros} for accounting for the cutoff dependence of entanglement entropy. This complication drops out of the elegant language of integral geometry.

Eq.~(\ref{ourclaim}) equates the length of a curve with the volume of a region in kinematic space. This region consists of geodesics that intersect the curve. The function $v(u)$ specifies the boundary of this region.\footnote{The same is true for $p(\theta)$ in Sec.~\ref{flatplane}.} Of course, the boundary of the set of geodesics that intersect $\gamma$ consists of geodesics which barely touch $\gamma$. When $\gamma$ is differentiable, this means the tangent geodesics.


\begin{figure}[t]
\centering
\includegraphics[width=.4\textwidth]{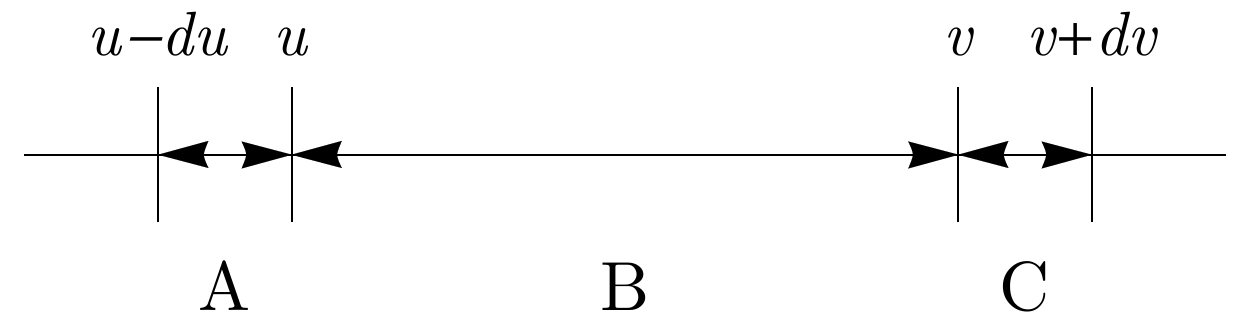}
\hspace{0.05\textwidth}
\includegraphics[width=.4\textwidth]{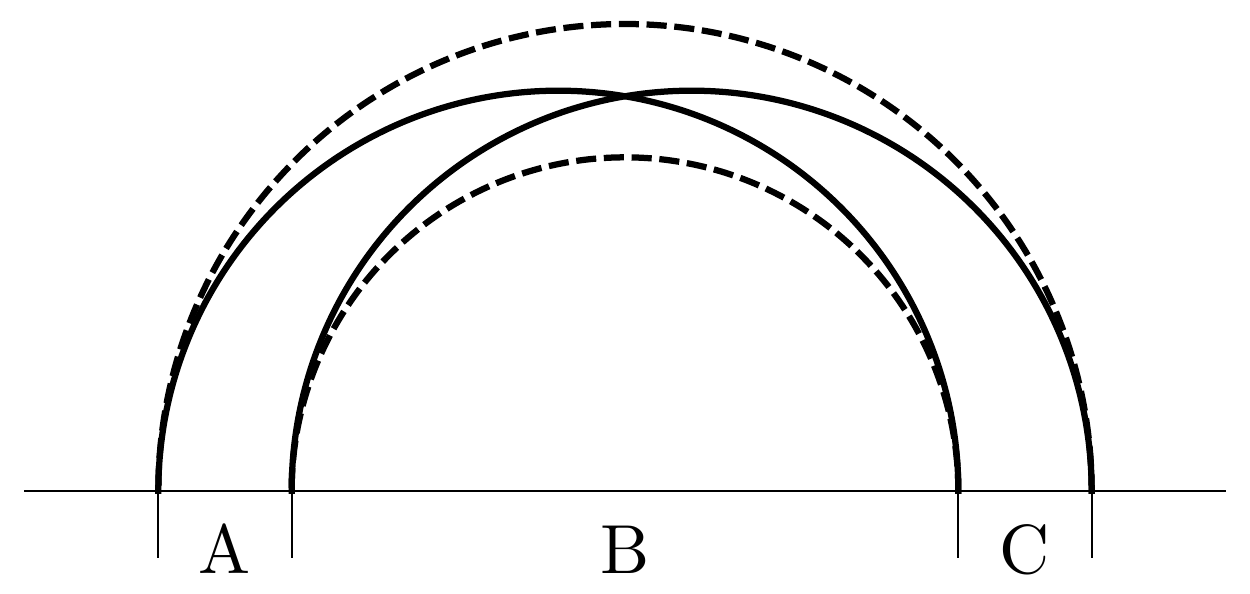}
\caption{The Crofton form (\ref{gencrofton}) can be interpreted as the conditional mutual information $I(A, C | B)$ of the three intervals shown above.  In the bulk it is computed by the four geodesics shown, with the dotted ones counted with negative length.}
\Label{dudv}
\end{figure}

\subsection{Conditional mutual information}
\Label{cmi}

If we follow the Ryu-Takayanagi proposal and identify $S(u,v)$ with the entanglement entropy of the boundary interval $(u,v)$ then $\omega(u,v)$ has a direct interpretation in quantum information theory. It has a definite sign as a consequence of the strong subadditivity of entanglement entropy \cite{ssaproof, strongsub}. Usually, the strong subadditivity inequality is stated in the form
\begin{equation}
S(AB) + S(BC) - S(B) - S(ABC) \geq 0\,,
\Label{ssa}
\end{equation}
where $A,B,C$ are three independent subsystems in the Hilbert space. If we choose $A,B,C$ to be the intervals (see Fig.~\ref{dudv})
\begin{equation}
A = (u - du, u) \qquad {\rm and} \qquad B = (u,v) \qquad {\rm and} \qquad C = (v, v+dv)\,,
\Label{abccoords}
\end{equation}
strong subadditivity mandates that:
\begin{equation}
S(u-du, v) + S(u,v+dv) - S(u,v) - S(u-du, v+dv) 
\approx \frac{\partial^2 S(u,v)}{\partial u \,\partial v} \,du\, dv \geq 0.
\Label{ssaspecial}
\end{equation}
This shows that strong subadditivity is an essential ingredient for the validity of eq.~(\ref{gencrofton}). In its absence, we could not think of $\omega(u, v)$ as a density of geodesics whose integrals compute lengths of convex curves. We will see later that strong subadditivity guarantees the positivity of lengths and the triangle inequality in the bulk geometry.  A related quantity was considered in \cite{markslast} and in \cite{entanglementdensity}, where it was termed the `entanglement density'.  Read in this way, the integral in eq.~(\ref{ourclaim}) quantifies an aggregate degree to which a collection of boundary intervals determined by $\gamma$ fails to saturate strong subadditivity.

\paragraph{Interpreting $\omega(u,v)$ as conditional mutual information}

Recall the definitions of conditional entropy and mutual information, respectively:
\begin{align}
S(A | B) & = S(AB) - S(B) \Label{defcondinf}\\
I(A,B) & = S(A) -S(A|B) \Label{defmut}
\end{align}
In classical terms, the conditional entropy measures a Bayesian reduction in the uncertainty about a composite system $AB$ given the knowledge of its part $B$. Likewise, the mutual information is a measure of correlations between $A$ and $B$. In quantum information theory, the conditional entropy acquires a direct operational meaning \cite{merge1, merge2}; a related operationalization of eq.~(\ref{diffent}) was given in \cite{protocol}.

We may combine the two concepts to obtain the conditional mutual information:
\begin{equation}
I(A,C | B) \equiv S(A|B) - S(A|BC) = I(A, BC) - I(A, B) = \textrm{LHS of~}(\ref{ssa}) 
\Label{ssaconnect}
\end{equation}
Thus strong subadditivity of entropy is just $I(A,C | B)\geq 0$. From a classical point of view, this means that conditioning on larger subsets can only reduce the uncertainty about a system. The strong subadditivity of entanglement entropy is the highly nontrivial extension of this fact to quantum information theory. An operational interpretation of the conditional mutual information was discussed in \cite{condinf}.

Applied to triples listed in (\ref{abccoords}), the nonvanishing of the conditional mutual information gives a definite sign to $\omega(u,v)$. In other words, it yields an orientation on the space of geodesics. This accords with the fact that our $(u,v)$-space is the space of \emph{oriented} geodesics.

\begin{figure}[t]
\centering
\includegraphics[height=.25\textwidth]{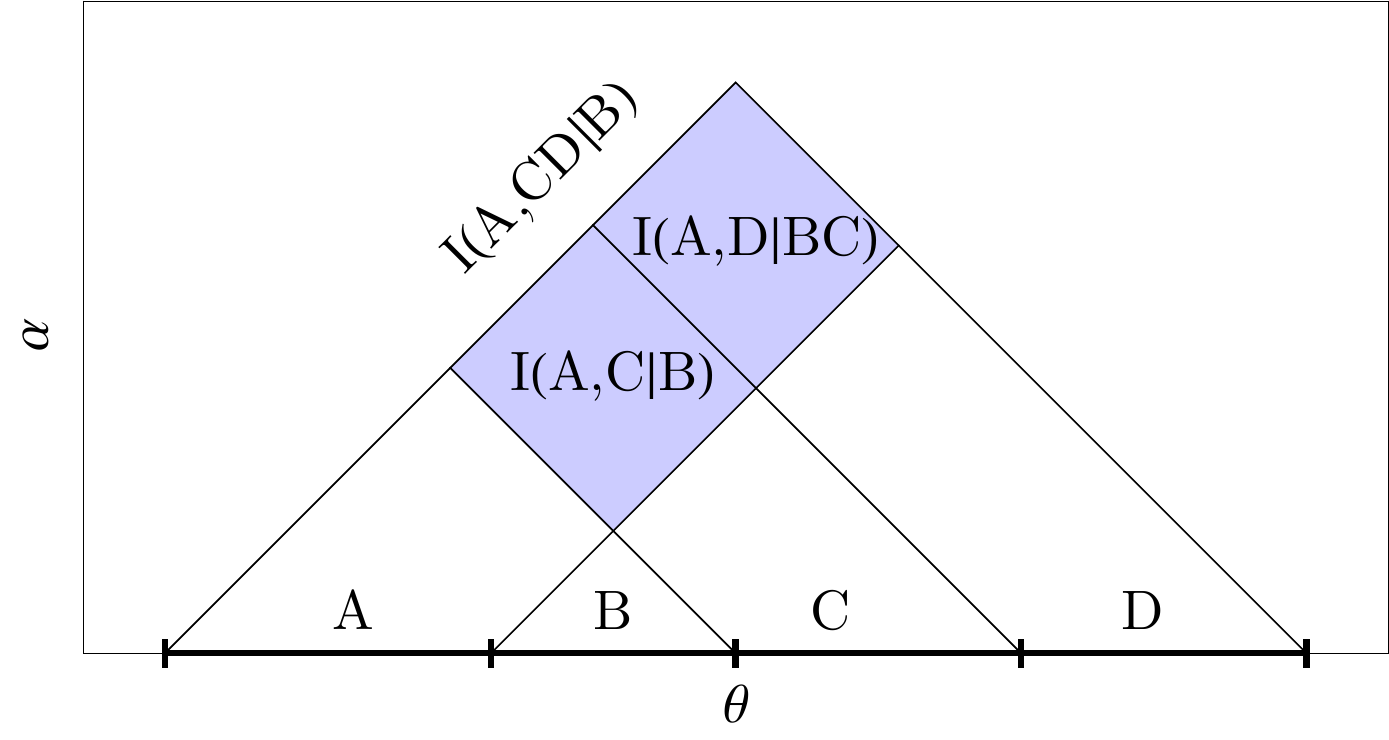}
\hspace{0.02\textwidth}
\includegraphics[height=.25\textwidth]{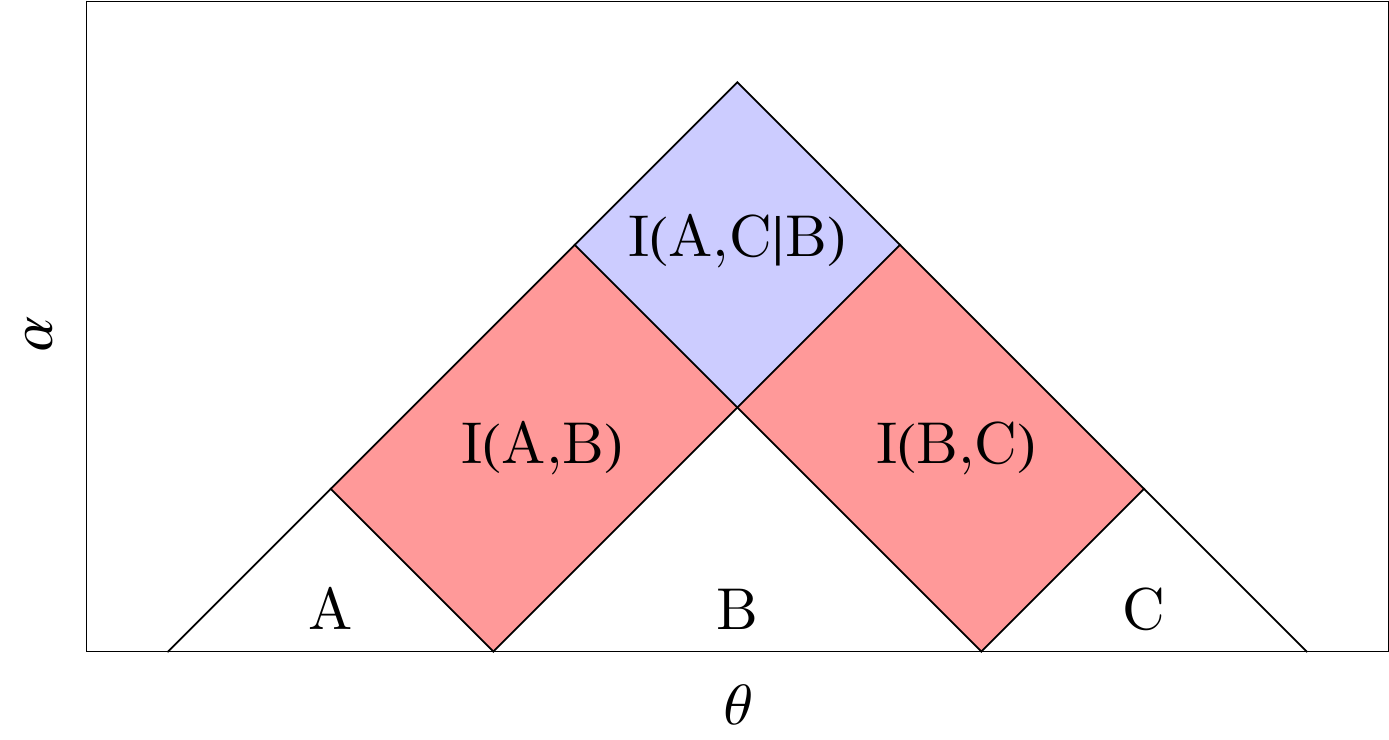}
\caption{Left: Kinematic space makes the chain rule for conditional mutual information (\ref{chainrule}) manifest. Right: The mutual informations and conditional mutual informations are represented by rectangles in kinematic space.}
\Label{allinfo}
\end{figure}

\paragraph{The chain rule for conditional mutual information}

Eq.~(\ref{ssaconnect}) identifies the volume of an infinitesimal rectangle with corners at  $(u, v)$, $(u-du, v)$, $(u-du, v+dv)$ and $(u, v+dv)$ with the conditional mutual information $I(A, C | B)$ of intervals defined in (\ref{abccoords}). Because volumes are additive, we can integrate the volume element (\ref{gencrofton}) to obtain a finite kinematic volume. If our interpretation of $\omega(u,v)$ is to be consistent, a similar additive property must hold for conditional mutual information. This is the chain rule for conditional mutual information:
\begin{equation}
I(A, CD | B) = I (A, C | B) + I(A, D | BC) \Label{chainrule}
\end{equation}
If we supplant the choice of $A, B, C$ from eq.~(\ref{abccoords}) with
\begin{equation}
D = (v+dv, v + \delta v) \qquad {\rm with}~\delta v > dv
\end{equation}
then eq.~(\ref{chainrule}) is simply the addition of areas of adjacent rectangles in kinematic space. This is shown in the left panel of Fig.~\ref{allinfo}. We may now apply the chain rule repeatedly to generate finite size rectangles in $K$, whose interpretation is in terms of conditional mutual informations of three finite sized, adjacent intervals.

The final picture of kinematic space is as shown in the right panel of Fig.~\ref{allinfo}. Every rectangle resting on constant $u$ and $v$ axes determines some conditional mutual information. A special class among those is rectangles which reach all the way to the bottom boundary of kinematic space. The bottom corner of such a rectangle has $u = v$, so in the language of eq.~(\ref{abccoords}) it has $B = \emptyset$. Because $I(A, C\, | \,\emptyset) = I(A, C)$, such rectangles compute \emph{mutual informations} of adjacent intervals.

\subsection{Causal structure}
\Label{causalstr}

In the previous subsection we implicitly associated points in kinematic space with intervals on the boundary. For example, eq.~(\ref{abccoords}) identified the point $(u,v) \in K$ with the interval $(u,v)$ on the cutoff surface. But intervals in one dimension have a partial ordering: one interval may or may not contain another. This means that the kinematic space $K$ is a partially ordered set. 

It is handy to encode this partial ordering and the Crofton form (\ref{gencrofton}) in a single object -- a metric tensor of the form:
\begin{equation}
ds^2 = \frac{\partial^2 S(u,v)}{\partial u\, \partial v} \, du\, dv
\Label{lorentzian}
\end{equation}
This metric gives $K$ a causal structure, with $u$ and $v$ playing the role of null coordinates. As we illustrate in Fig.~\ref{causal-examples} and explain in more detail below, this causal structure encodes the containment relation among boundary intervals. At the same time, the volume form of metric~(\ref{lorentzian}) is the Crofton form (\ref{gencrofton}).  

\begin{figure}[t]
\centering
\includegraphics[height=.3\textwidth]{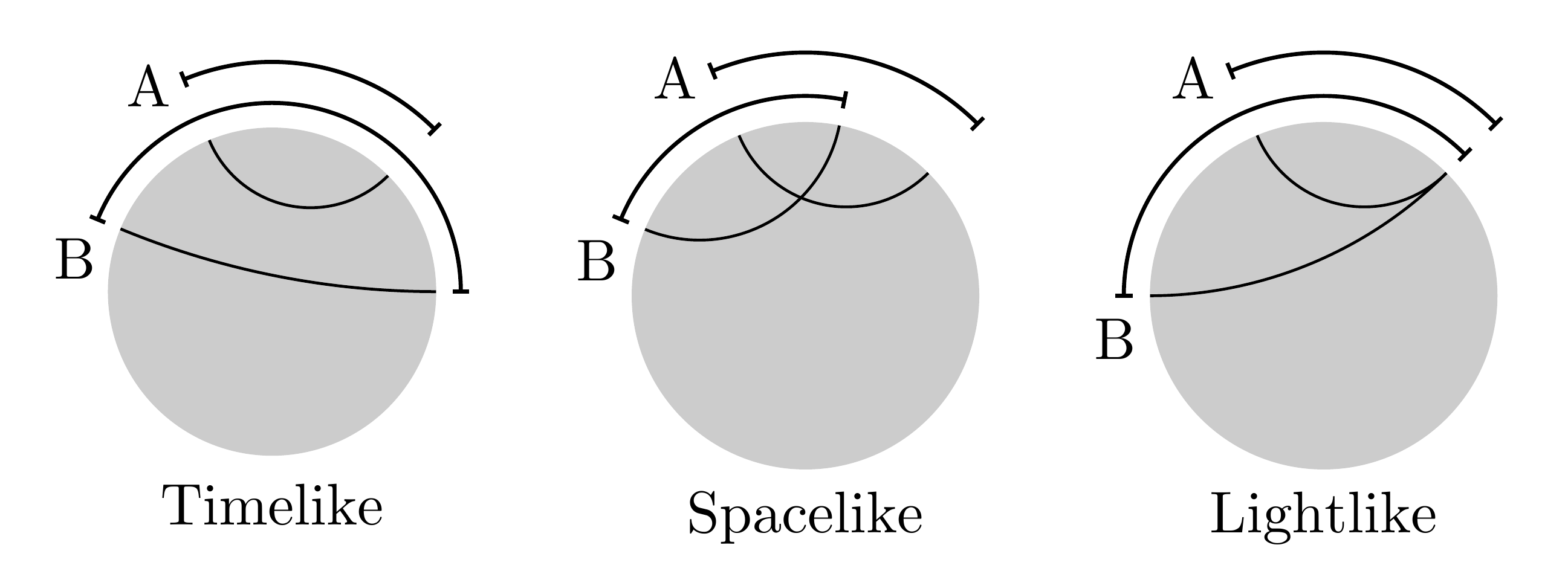}
\caption{Left: The point A is in the causal past of the point B in kinematic space, because the interval B contains the interval A.  Middle: Neither interval contains the other, so A and B are spacelike separated. Right: A and B share an endpoint, so they are lightlike separated (infinitesimally off from timelike and spacelike separation).}
\Label{causal-examples}
\end{figure}

\paragraph{Timelike separation}
Consider two points in $K$ that are timelike separated according to metric~(\ref{lorentzian}). Suppose $(u_2, v_2)$ is in the future of $(u_1, v_1)$. This means that as intervals on the cutoff surface, $(u_1, v_1) \subset (u_2, v_2)$. This is suggestive from an information theoretic perspective: the density matrix of the `past' interval $(u_1, v_1)$ can be obtained from the density matrix of the `future' interval $(u_2, v_2)$ by tracing out the complement. The geodesics that correspond to two timelike separated points in $K$ do not intersect.

The language of timelike separations clarifies why points in $K$ must be oriented geodesics. Changing the orientation of a geodesic means switching its left and right endpoints. Considered as an interval on the cutoff surface, a point in $K$ maps to the complementary interval under orientation reversal. Thus, if one geodesic is in the future of another, the opposite is true for their orientation-reversed counterparts because:
\begin{equation}
(u_1, v_1) \subset (u_2, v_2) \qquad \Leftrightarrow \qquad (u_2, v_2)^c \subset (u_1, v_1)^c\,.
\end{equation}
If points in $K$ did not carry the information about the orientations of geodesics, it would not be possible to have a globally defined notion of past and future. In closing, we note that the asymptotic past $u=v$ consists of zero-size geodesics, i.e. of points on the cutoff surface. The asymptotic future $v=u+2\pi$ comprises almost-360$^\circ$ geodesics: it consists of almost-complete boundary sets, each of which excludes a single boundary point. 

\paragraph{Spacelike separation}
When a pair of points are not timelike separated, they are \emph{generically} spacelike separated. This translates to the statement that considered as boundary intervals, neither one contains the other. In information theoretic terms, it means that unlike the timelike separated case, neither interval contains complete data about the other interval. Consider a pair of spacelike separated points $(u_1, v_1)$ and $(u_2, v_2)$ in $K$ and assume without loss of generality that $u_1 < u_2$. There are two possible cases:
\begin{enumerate}
\item If $v_1 > u_2$, the boundary intervals overlap and the geodesics intersect in the bulk. 
\item If $v_1 < u_2$, the intervals are disjoint and the geodesics do not intersect.
\end{enumerate}

\paragraph{Null separation}
Two null separated geodesics are infinitesimally close to being timelike and spacelike separated. This happens if they have a common left or right endpoint: $u_1 = u_2$ or $v_1 = v_2$. This is why $u$ and $v$ are null coordinates in metric~(\ref{lorentzian}). Two null separated geodesics meet, but their intersection lies on the cutoff surface and not in the bulk. As intervals, one necessarily contains the other.

These relations are obviously respected by Weyl rescalings of the cutoff surface. Thus, a Weyl rescaling in the field theory induces a Weyl rescaling of the kinematic metric~(\ref{lorentzian}).

\paragraph{Conditional mutual information}
In the causal language the rectangles drawn in Fig.~\ref{allinfo} are causal diamonds.
For three adjacent intervals $A, B, C$, the conditional mutual information $I(A, C | B)$ is the area of a causal diamond in kinematic space. The spacelike separated corners of such a rectangle correspond to the intervals $A \cup B$ and $B \cup C$ -- the positive terms in eq.~(\ref{ssa}). The timelike separated corners of the same rectangle correspond to the negative terms -- the intervals $B$ and $A \cup B \cup C$.

\section{Basic geometric objects in kinematic space}
\Label{vackin}

In the previous section we developed a machinery to measure the length of a curve using the lengths of geodesics that intersect it.  In this section we explain how this machinery relates to the familiar geometric notions: points, distances and angles.

\subsection{Points from point-curves}
\Label{point-curves}

Eq.~(\ref{eqschematic}) associates a closed convex curve $\gamma$ in the bulk to a region in kinematic space, which consists of the geodesics that intersect $\gamma$. Consider shrinking $\gamma$ to a single point. The relevant region in kinematic space $K$ undergoes a corresponding shrinkage. When $\gamma$ becomes a point, the volume of the related region in $K$ vanishes: the region becomes a codimension-1 set, i.e. a curve (see Fig.~\ref{point-curve}). We call such a curve a point-curve.


\begin{figure}[t]
\centering
\includegraphics[height=.3\textwidth]{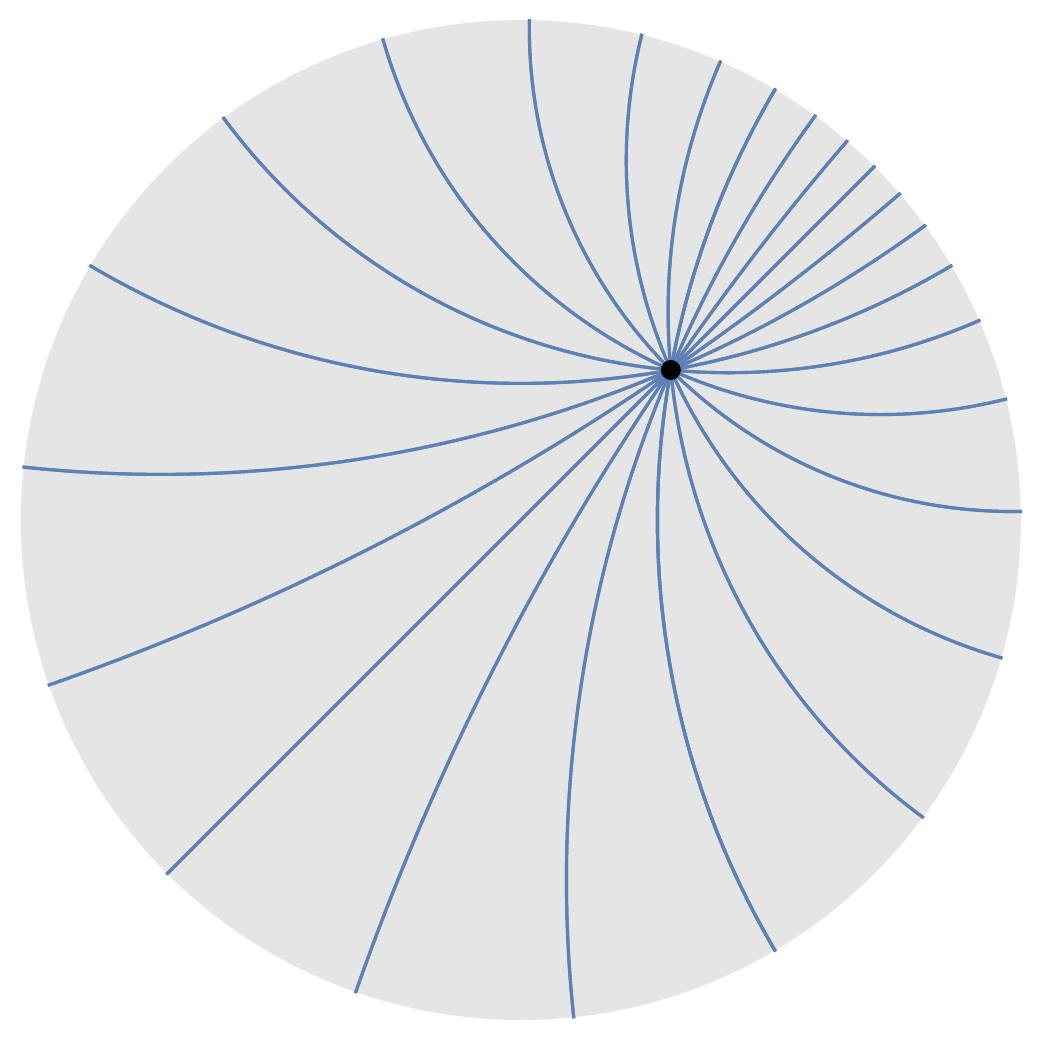}
\includegraphics[height=.3\textwidth]{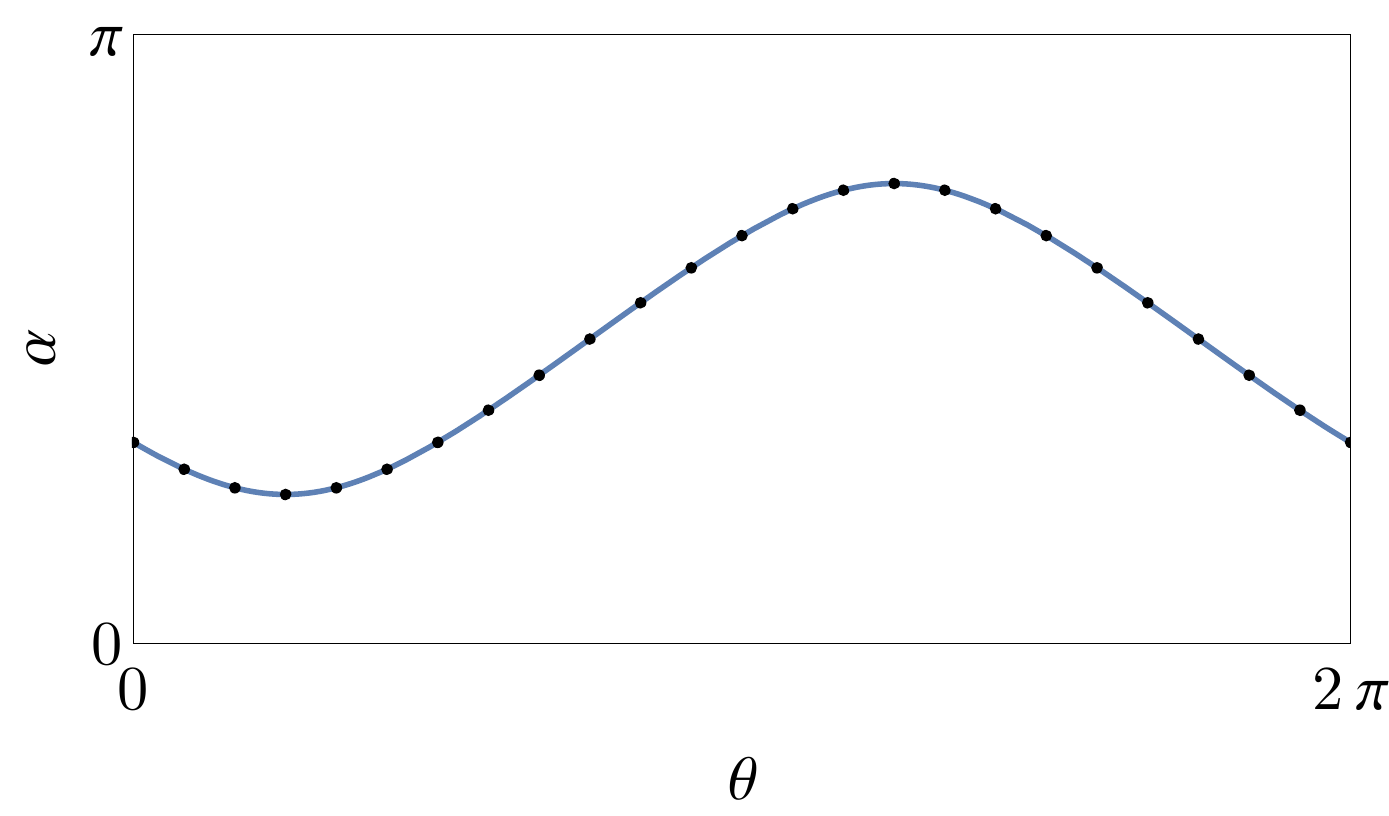}
\caption{The set of geodesics intersecting a single point (left) forms a codimension-1 locus in kinematic space, which we call a point-curve (right). }
\Label{point-curve}
\end{figure}

It is easy to understand why the codimension of a point-curve is 1. A point-curve collects those geodesics, which intersect the $\gamma$ that consists of a single point. 
Such geodesics are selected by a single algebraic equation.

Knowing the point-curves in kinematic space is tantamount to knowing everything about a two-dimensional geometry. As we explain below, this knowledge automatically gives us a bulk distance function and its attendant geometric objects. From a holographic perspective, it is therefore interesting to understand how to derive point-curves from boundary distances alone -- that is, from the knowledge of entanglement entropies in the boundary field theory. 

It turns out that solving this problem requires imposing additional assumptions. In Appendix~\ref{point-curve-derivation} we adopt the most conservative physical assumption: that the bulk geometry be Riemannian. This is a nontrivial assumption, because in general not every distance function on a manifold descends from a Riemannian metric -- a symmetric bilinear form on the tangent space. The solution is eq.~(\ref{pc-master-eqn}) -- a differential equation, which may be used iteratively to derive point-curves of points lying deeper in the bulk given the point-curves of points on a cutoff surface.

\subsection{Distances between points}
\Label{distance-between-points}

Given two point-curves $p_A$ and $p_B$, how do we compute $\ell(A,B)$, the distance between the two bulk points? To answer this question, we refer to our main result, eq.~(\ref{ourclaim}). We must identify which geodesics intersect the geodesic segment stretched between  $A$ and $B$. Said differently, we must identify the geodesics that pass between $A$ and $B$.

The curve $p_A$ divides the kinematic space into two regions. In the causal language of Sec.~\ref{causalstr}, they are the past and the future of $p_A$. The future of $p_A$, which we denote $\tilde{p}_A$, comprises geodesics characterized in the following way: if we deform a geodesic $(u,v) \in \tilde{p}_A$ to its corresponds interval $(u,v)$ on the cutoff surface, we necessarily sweep over the bulk point $A$. More figuratively but less precisely, $\tilde{p}_A$ consists of geodesics, which pass around the point $A$ rather than leaving it outside. This characterization is well-defined, because points in the kinematic space represent oriented geodesics.

Let us assume that all boundary points are connected by single geodesics.
The geodesics that contribute to $\ell(A,B)$ in eq.~(\ref{ourclaim}) are those, which pass around $A$ but leave $B$ outside (or vice versa). As points in $K$, they are in the future of $p_A$ but in the past of $p_B$ (or vice versa), i.e. they live in the region between the curves $p_A$ and $p_B$. 
A parsimonious notation for this region is the symmetric difference:
\begin{equation}
\tilde{p}_A \, \triangle \, \tilde{p}_B = 
\left(\tilde{p}_A\cup \tilde{p}_B\right) \setminus \left(\tilde{p}_A\cap \tilde{p}_B\right).
\end{equation}
Because in this case all the intersection numbers are unity, eq.~(\ref{ourclaim}) gives the distance function as:
\begin{equation}
\frac{\ell(A,B)}{4G}= \frac{1}{4} \int_{\tilde{p}_{A}\,\triangle\,\tilde{p}_{B}}\omega
\Label{metric-definition}
\end{equation}
This result is illustrated in Fig.~\ref{metricdef-fig}. 

\begin{figure}[t]
\centering
\includegraphics[height=.3\textwidth]{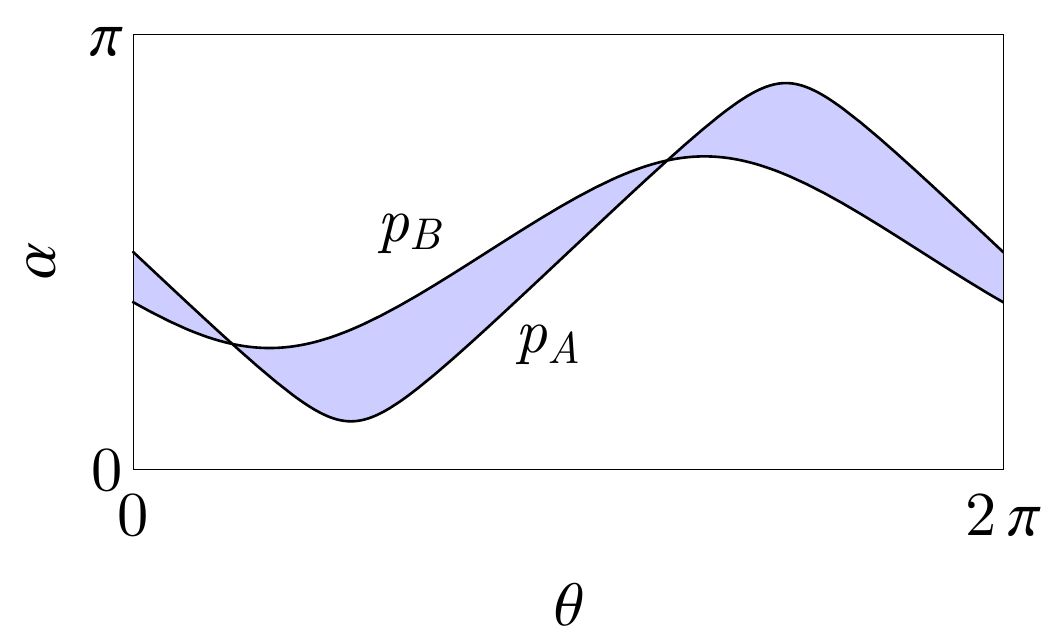}
\hspace{0.1\textwidth}
\includegraphics[height=.3\textwidth]{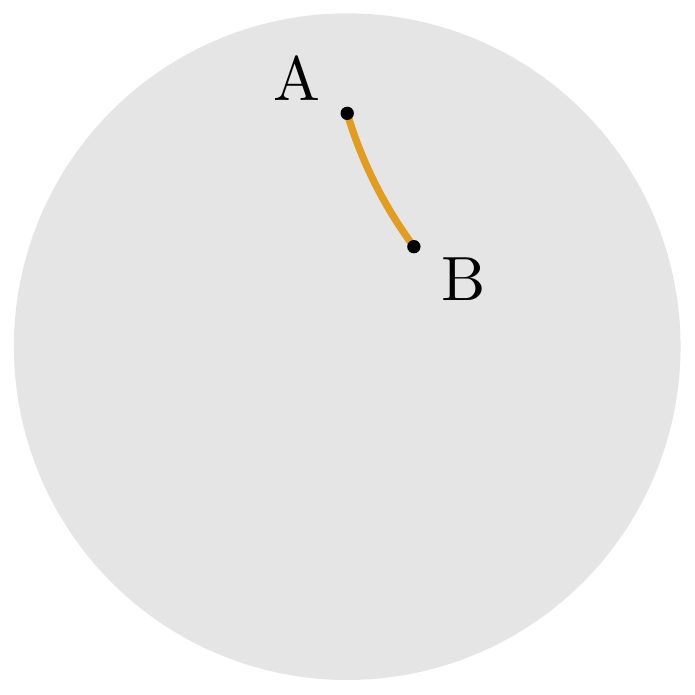}
\caption{The distance between two points is given by the integral of the Crofton form over the region between their two corresponding point-curves.}
\Label{metricdef-fig}
\end{figure}

\paragraph{Boundary points}
\Label{boundarypts}

A point on the cutoff surface can be thought of as a zero-size geodesic whose endpoints are equal. In coordinates, a boundary point $A$ at $\theta = \theta_A$ corresponds in the kinematic space to the point $(u,v) = (\theta_A, \theta_A)$. Such points live on the asymptotic past boundary of $K$, because as intervals they contain no other intervals. In this way, the boundaries of the bulk space and of kinematic space are identified.

The point-curve $p_A$ consists of geodesics, which meet at $A$. When $A$ lives on the cutoff surface, their characterization is simple: they are geodesics which begin or end at $A$. Thus, $p_A$ comprises points in $K$ which have one null coordinate, $u$ or $v$, in common with $A$:
\begin{equation}
p_A = \{ (\theta_A, v) \} \cup \{ (u, \theta_A) \} \Label{boundpointcurve}
\end{equation}
We recognize $p_A$ as the light cone of $A$.

With eq.~(\ref{boundpointcurve}) at hand, we can use (\ref{metric-definition}) to compute the distance between two boundary points. The region $\tilde{p}_A \, \triangle \, \tilde{p}_B$ consists of two causal diamonds; see Fig.~\ref{boundary-pts-fig}. Their spacelike separated corners are $(u,v) = (\theta_A, \theta_B)$ and $(\theta_B, \theta_A + 2\pi)$, which correspond to both orientations of the geodesic connecting $A$ and $B$. Their timelike separated corners are $(\theta_A, \theta_A)$ and $(\theta_B, \theta_B + 2\pi)$ (and vice versa), which means that points in $\tilde{p}_A \,\triangle\, \tilde{p}_B$ considered as intervals include either $A$ or $B$ but not both. Explicitly computing the integral, we have:
\begin{align}
\frac{\ell(A,B)}{4G} & = 
\frac{1}{4} \int_{\theta_A}^{\theta_B} du \int_{\theta_B}^{\theta_A+2\pi} dv \, \frac{\partial^2S(u,v)}{\partial u \, \partial v} 
\, + \,
 \frac{1}{4} \int_{\theta_A}^{\theta_B} dv \int_{\theta_B}^{\theta_A+2\pi} du \, \frac{\partial^2S(u,v)}{\partial u \, \partial v} \Label{kgammageod} \\
& = 
\frac{1}{2}\, \big( S(\theta_B, \theta_A+2\pi) + S(\theta_A, \theta_B) - S(\theta_B, \theta_B) - S(\theta_A, \theta_A+2\pi) \big) = S(\theta_A, \theta_B) \nonumber
\end{align}

\begin{figure}[t]
\centering
\includegraphics[height=.25\textwidth]{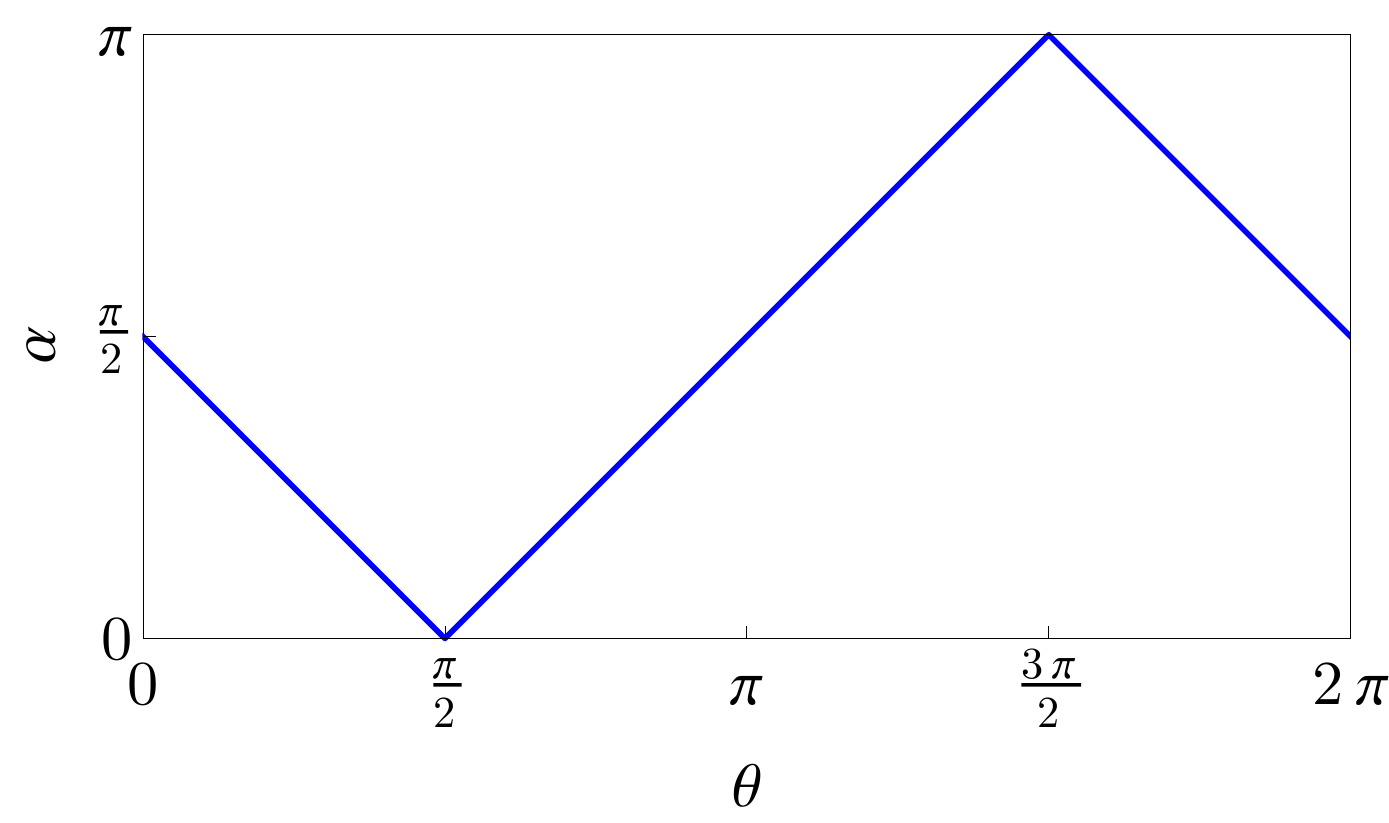}
\hspace{0.05\textwidth}
\includegraphics[height=.25\textwidth]{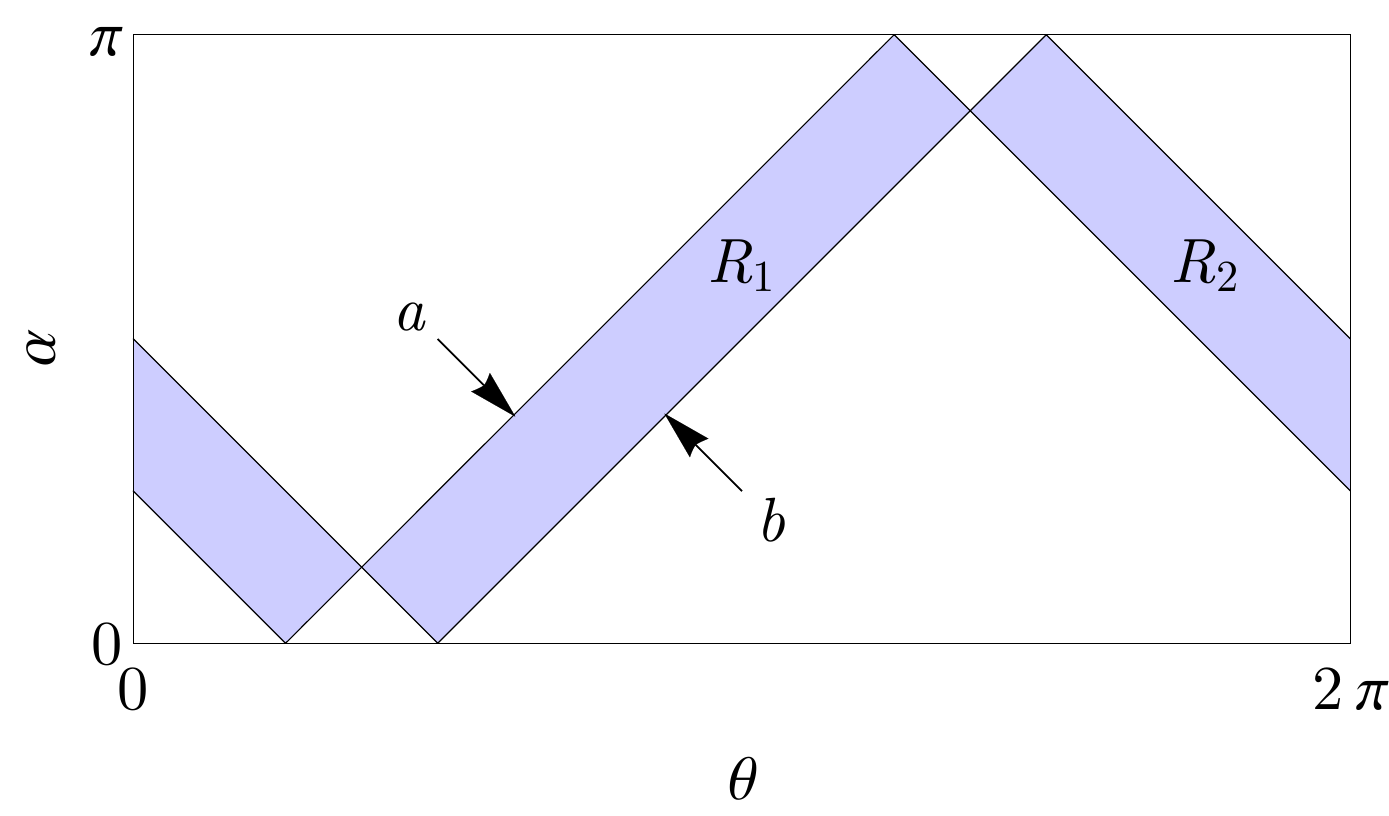}
\caption{Left: the point-curve of a boundary point is a causal cone in kinematic space.  Right: The region $\tilde{p}_A \,\triangle\, \tilde{p}_B$ consists of two causal diamonds.}
\Label{boundary-pts-fig}
\end{figure}

Of course, carrying this computation out explicitly was unnecessary. In Secs.~\ref{cmi} and \ref{causalstr} we recognized the area of a causal diamond with timelike separated corners at $(\theta_A, \theta_A)$ and $(u, v)$ as the mutual information of intervals $(u, \theta_A)$ and $(\theta_A, v)$. Both rectangular regions in eq.~(\ref{kgammageod}) fit this template with $v = u + 2\pi$ and therefore both individually compute mutual informations of complementary intervals $(\theta_A, \theta_B)$ and $(\theta_B, \theta_A)$. Eq.~(\ref{kgammageod}) states simply:
\begin{equation}
\frac{\ell(A, B)}{4G} = \frac{1}{4} \cdot 2 \cdot I\big( (\theta_A, \theta_B), (\theta_B, \theta_A + 2\pi) \big) = S\left(\theta_A, \theta_B\right) \,.
\end{equation}

\paragraph{Triangle inequality} 
It is easy to see that eq.~(\ref{metric-definition}) is a consistent distance function: 
\begin{align}
\ell(A,A) = 0 \qquad & \Leftarrow \qquad \tilde{p}_A \, \triangle \,\tilde{p}_A = \emptyset \\
\ell(A,B_{\neq A}) > 0 \qquad & \Leftarrow \qquad \tilde{p}_A \,\triangle\, \tilde{p}_B \neq \emptyset ~~{\rm and}~\omega~\textrm{is nondegenerate} 
\end{align}
The triangle inequality requires slightly more attention. It is a consequence of the set-theoretic inequality:
\begin{equation}
\tilde{p}_{A}\,\triangle\,\tilde{p}_{C} \subseteq
\left(\tilde{p}_{A}\,\triangle\,\tilde{p}_{B}\right)
\cup \left(\tilde{p}_{B}\,\triangle\,\tilde{p}_{C}\right).
\end{equation}
This inequality is easily understood geometrically: if a geodesic intersects the line segment connecting $A$ and $C$, then it must necessarily intersect either the line segment between $A$ and $B$ or the line segment between $B$ and $C$.

\subsection{Curves} 
\Label{discusscurves}

\begin{figure}[t]
\centering
\includegraphics[height=.35\textwidth]{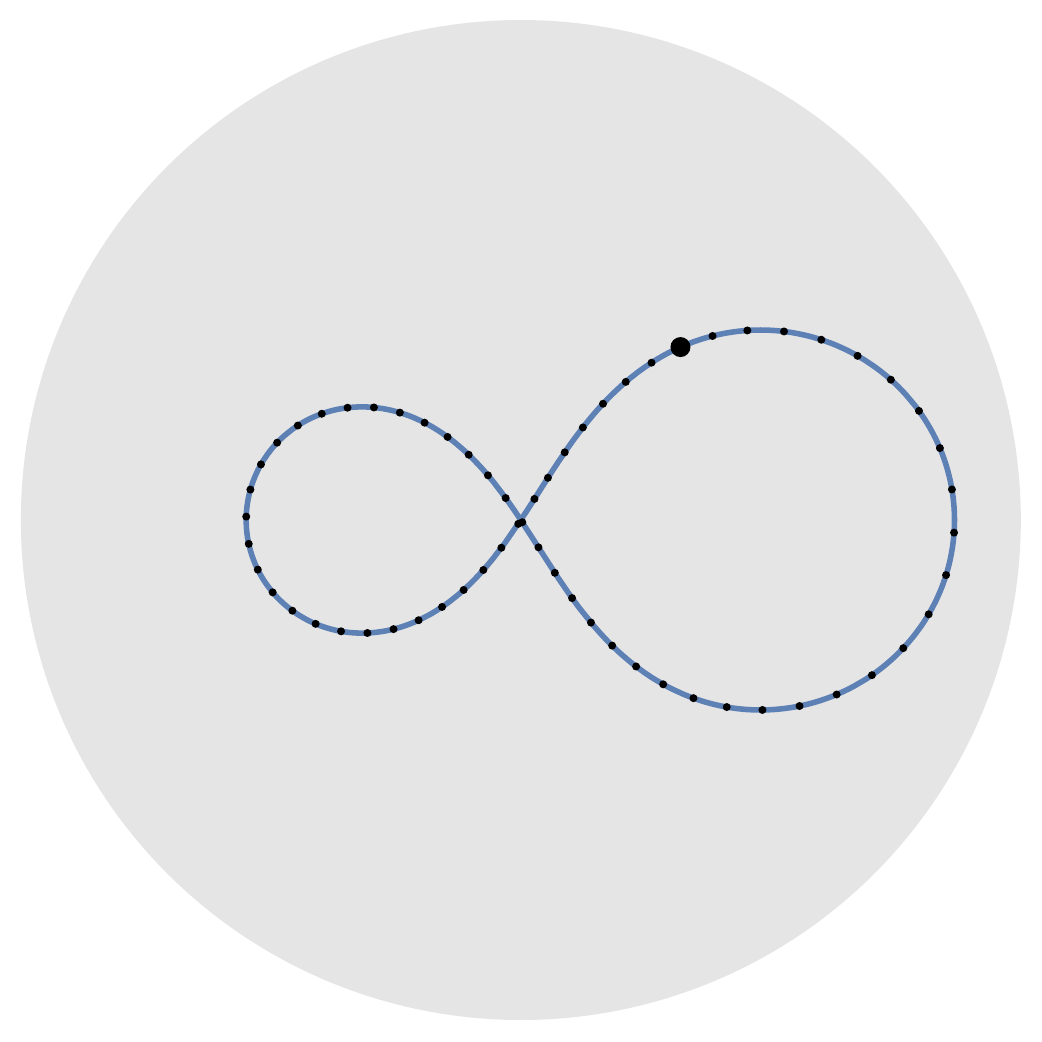}
\hspace{0.02\textwidth}
\includegraphics[height=.35\textwidth]{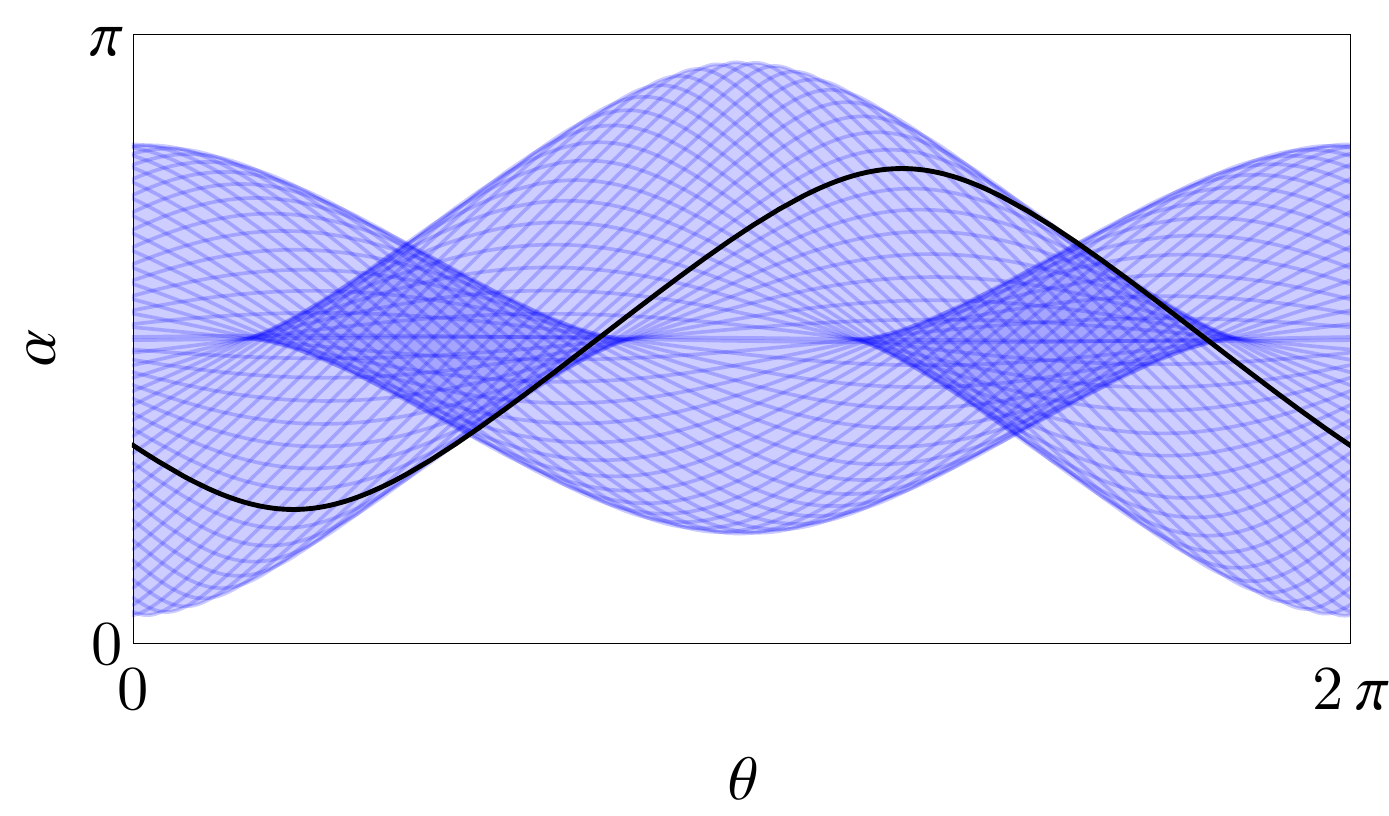}
\caption{Left: A curve $\gamma$ in hyperbolic space. Right: Its corresponding region in kinematic space, with a sample of point-curves shown for points on $\gamma$.  The point-curve for one point, highlighted on the left, is shown in black on the right. The kinematic region can be found by taking the weighted union of the point-curves for each point on $\gamma$.  Two point-curves intersect each point in the light-shaded region, which indicates that the corresponding geodesics intersect $\gamma$ twice. Four point-curves intersect each point in the dark-shaded region, so those points correspond to geodesics intersecting $\gamma$ in four places.}
\Label{curves-to-regions}
\end{figure}

A curve $\gamma$ defines a region $K_\gamma \subset K$, which consists of geodesics that intersect $\gamma$. In general $K_\gamma$ must be defined with multiplicity set by the intersection number of the geodesic and $\gamma$. We can split this characterization up into contributions from individual points that comprise $\gamma$. Thus, $K_\gamma$ is simply the union (counted with multiplicity) of the point-curves of all points in $\gamma$. 

When $\gamma$ is closed and convex, every geodesic that intersects $\gamma$ does so at two points. The only exception are the tangent geodesics, but these form a set of measure zero and can be ignored. So we have a simple relation:
\begin{equation}
K_\gamma = \bigcup_{A \in \gamma} p_A
\Label{bulkcurvedef}
\end{equation}
In this case the entire region counts with multiplicity 2. An interactive demonstration provided in the supplemental material illustrates concretely how to construct the kinematic region for any bulk curve.

When a region in kinematic space cannot be built out of a (weighted) union of point-curves, it fails to define a positive length curve in the bulk. In general, such a region can be understood as a weighted combination of curves, where the weights are allowed to be positive or negative. Such weighted curves may be combined to form other combinations, as is explored in Exercises~\ref{lastqn} and \ref{addandsubtract}.

\subsection{Angles}
\Label{angles-sec}

\begin{figure}[t]
\centering
\includegraphics[height=.25\textwidth]{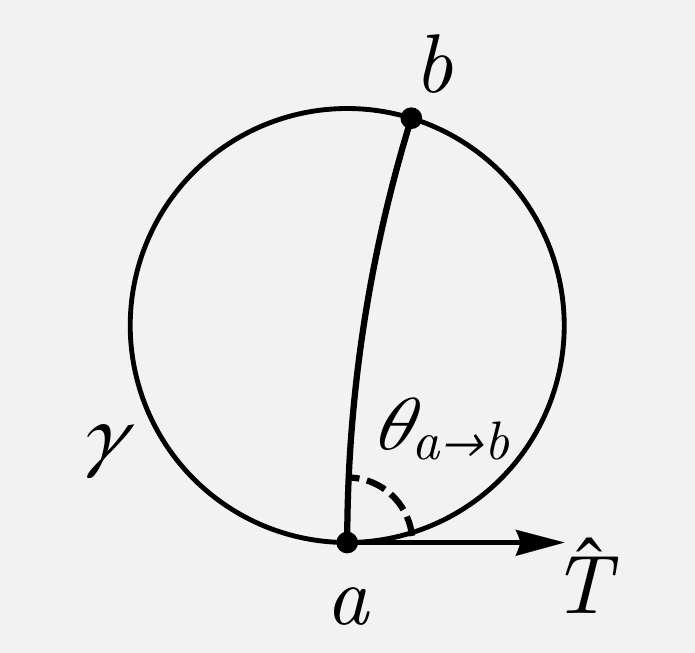}
\hspace{0.15\textwidth}
\includegraphics[height=.25\textwidth]{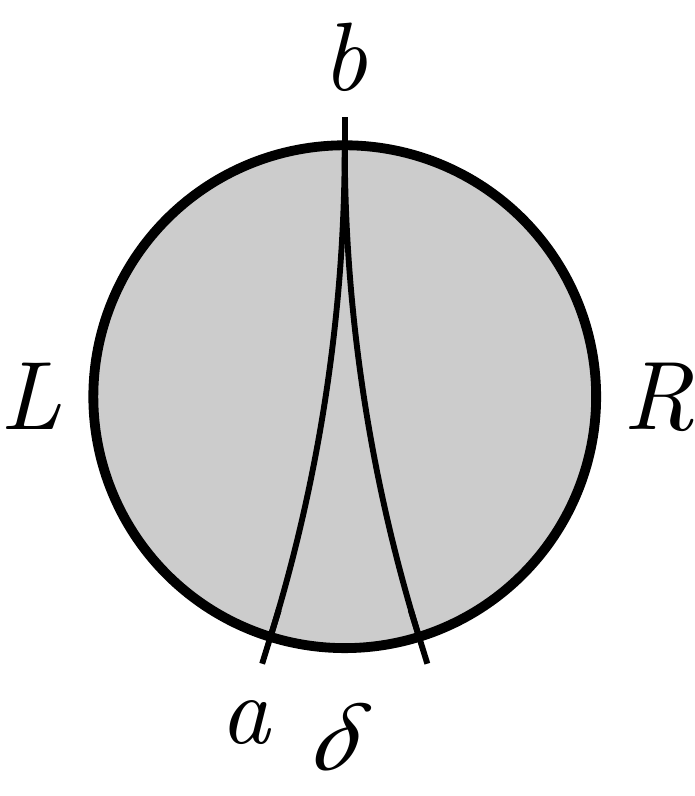}
\caption{Left: The angle between a geodesic and the boundary is determined by the variation of the entropy under movement of one endpoint. Right: The angle can also be determined as a ratio of mutual informations; see eq.~(\ref{anglemutinf}).}
\Label{angles}
\end{figure}

The angle between two bulk curves intersecting at a point has a simple
expression in terms of entanglement entropies. The Riemannian distance
function between two points is defined as
\begin{equation}
\ell\left(a,b\right)=\min_{X\left(s\right)_{a\rightarrow b}}\int ds\sqrt{g_{\mu\nu}\left(X\right)X_{,s}^{\mu}X_{,s}^{\nu}}\,,
\end{equation}
where $X\left(s\right)$ is some path from $a$ to $b$. Using the
Hamilton-Jacobi argument that the variation of a minimized action
with respect to a coordinate at its endpoint is given by the conjugate
momentum at that endpoint, we obtain
\begin{equation}
\frac{\partial}{\partial a^{\mu}}\ell\left(a,b\right)=g_{\mu\nu}\left(a\right)\hat{G}_{a\rightarrow b}^{\nu}
\end{equation}
where $G_{a\rightarrow b}^{\nu}=\hat{X}_{,s}^{\nu}\big|_{a}$ is the
unit tangent vector to the geodesic at $a$. If we now take the derivative
in the direction $\hat{T}$, we find:
\begin{equation}
\hat{T}\cdot\nabla_{a}\ell\left(a,b\right)=\hat{T}\cdot\hat{G}_{a\rightarrow b}\,.
\Label{dotproduct}
\end{equation}
Treating $a$ (resp. $b$) as a coordinate on some bulk curve $\gamma$, let $\hat{T}$ be the tangent
vector to $\gamma$ at $a$.  Then, the right hand side computes the cosine of $\theta_{a\rightarrow b}$, the angle between $\gamma$ and the $a\to b$ geodesic (see Fig.~\ref{angles}). Then eq.~(\ref{dotproduct}) becomes:
\begin{equation}
\cos\left(\theta_{a\rightarrow b}\right)=
\frac{\partial \ell}{\partial a}\big|_{\left(a,b\right)}
\Big/
\frac{\partial \ell}{\partial a}\big|_{\left(a,a\right)} \Label{hamilton-jacobi-angles}\,.
\end{equation}
The factor $\frac{\partial \ell}{\partial a}\big|_{\left(a,a\right)}$
on the right sets the normalization of $\hat{T}$. When $\gamma$ is a cutoff surface in some radial slicing of a holographic geometry, we may use the Ryu-Takayanagi proposal $S(a,b) = \ell(a,b)/4G$ to rewrite this expression in a more physical way. Let us divide the boundary into three
regions $L$, $R$ and $\delta$ (see the right panel of Fig. \ref{angles}). Then the angle
between the geodesic and the cutoff surface can be rewritten as:
\begin{equation}
\cos\left(\theta_{a\rightarrow b}\right)=
\lim_{\delta\rightarrow0}\,
\frac{S(\delta|R)-S(\delta|L)}{2S(\delta)}
=\lim_{\delta\rightarrow0}\,
\frac{I\left(L,\delta\right)-I\left(R,\delta\right)}{I\left(L,\delta\right)+I\left(R,\delta\right)}\,,
\Label{anglemutinf}
\end{equation}
Of course, in the limit that $\gamma$ tends to the asymptotic boundary, this quantity vanishes.

Using result~(\ref{hamilton-jacobi-angles}), we can compute
the geodesic curvature of the cutoff surface. Let us approximate the boundary
as a series of points $a_{i}$ connected by geodesic segments. At
the point $a_{i}$ there is a deflection angle $\delta\theta_{i}$
between the two geodesics intersecting it. This angle $\delta\theta_{i}$
is just the sum of the two angles $\delta\theta_{i}^{L},\delta\theta_{i}^{R}$
at which these two geodesics intersect the boundary (see Fig.~\ref{boundary-curvature}).
These two angles are easily computed using eq.~(\ref{hamilton-jacobi-angles}):
\begin{align}
\cos\delta\theta_{i}^{R} = &\, \frac{\partial S\left(a,b\right)}{\partial a}\big|_{\left(a_{i},a_{i+1}\right)}\Big/
\frac{\partial S\left(a,b\right)}{\partial a}\big|_{\left(a_{i},a_{i}\right)}\nonumber\\
\cos\delta\theta_{i}^{L} = &\, -\frac{\partial S\left(a,b\right)}{\partial b}\big|_{\left(a_{i-1},a_{i}\right)}\Big/
\frac{\partial S\left(a,b\right)}{\partial a}\big|_{\left(a_{i},a_{i}\right)}
\end{align}
Expanding in $\delta\theta$ and $\delta a$ and using the fact $\frac{\partial^{2}S}{\partial a\partial b}\big|_{\left(a,a\right)}=0$,
we find 
\begin{align}
\delta\theta_{i}^{R} = &\, \delta a\sqrt{-\frac{\partial^{3}S\left(a,b\right)}{\partial a\partial b^{2}}\Big/\frac{\partial S\left(a,b\right)}{\partial a}}\big|_{\left(a_{i},a_{i}\right)}\nonumber\\
\delta\theta_{i}^{L} = &\, \delta a\sqrt{\frac{\partial^{3}S\left(a,b\right)}{\partial a^{2}\partial b}\Big/\frac{\partial S\left(a,b\right)}{\partial a}}\big|_{\left(a_{i},a_{i}\right)}.
\end{align}
Finally, we recall that the geodesic curvature is generally defined
such that the contribution of a sharp angle $\phi$ to the integral
$\int k_{g}ds$ is just $\phi$. Taking the limit $\delta a\rightarrow0$,
we find the geodesic curvature at $a$:
\begin{equation}
k_{g}\left(a\right) = 
\frac{1}{4G}\frac{\delta\theta^{L}+\delta\theta^{R}}
{-\delta a\,\frac{\partial S\left(a,b\right)}{\partial a}\big|_{\left(a,a\right)}}
= \frac{1}{4G}\left(\sqrt{\frac{\partial^{3}S}{\partial a\partial b^{2}}}+\sqrt{-\frac{\partial^{3}S}{\partial a^{2}\partial b}}\right)\left(-\frac{\partial S}{\partial a}\right)^{-3/2}\big|_{b=a}
\end{equation}
This quantity can be related to the stress tensor of the dual CFT, as explained in \cite{CFTstress}.

\begin{figure}[t]
\centering
\includegraphics[height=.1\textwidth]{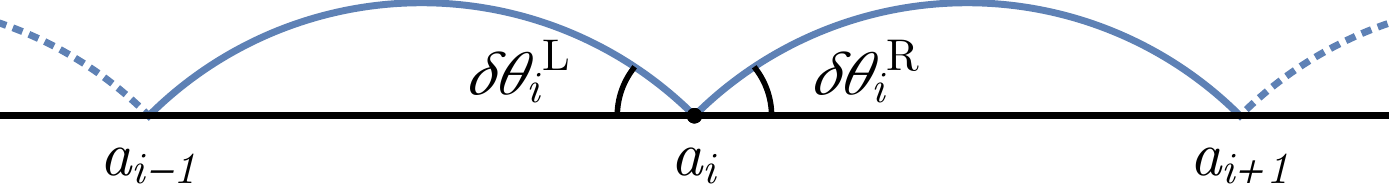}
\caption{The curvature of the cutoff surface can be determined by computing the angle between infinitesimal geodesics and the boundary.}
\Label{boundary-curvature}
\end{figure}

\section{Canonical example: pure AdS$_3$}
\Label{exads3}

We illustrate the material of Secs.~\ref{croftongeneral} and \ref{vackin} with the simplest example: the kinematic space of a static slice of pure AdS$_3$, i.e. the hyperbolic plane.
This example affords an independent derivation of the kinematic space, which starts from:

\subsection{The hyperboloid model}
The hyperbolic plane $\mathbb{H}^2$ with curvature scale $L/{4G}$ can be constructed as the locus
\begin{equation}
- t^2 + x^2 + y^2 = - \left(\frac{L}{4G}\right)^2 \,\,\,\,\Label{h2embed}
\end{equation}
in an auxiliary space with metric:
\begin{equation}
ds^2 = -dt^2 + dx^2 + dy^2\,. \,\,\,\Label{ambient}
\end{equation}
In this construction, the $SO(2,1)$ invariance of the hyperbolic plane is simply the group of rotations and boosts in the ambient $\mathbb{R}^{2,1}$, which leave the locus~(\ref{h2embed}) invariant. A useful fact \cite{adsnotes} is that geodesics on the hyperbolic plane are intersections of (\ref{h2embed}) with hyperplanes through the origin in $\mathbb{R}^{2,1}$. The $SO(2,1)$ symmetry acts on the geodesics by rotating and boosting the normal of a hyperplane in $\mathbb{R}^{2,1}$. The construction is illustrated in Fig.~\ref{hyperboloid}.

\begin{figure}[t]
\centering
\includegraphics[height=.3\textwidth]{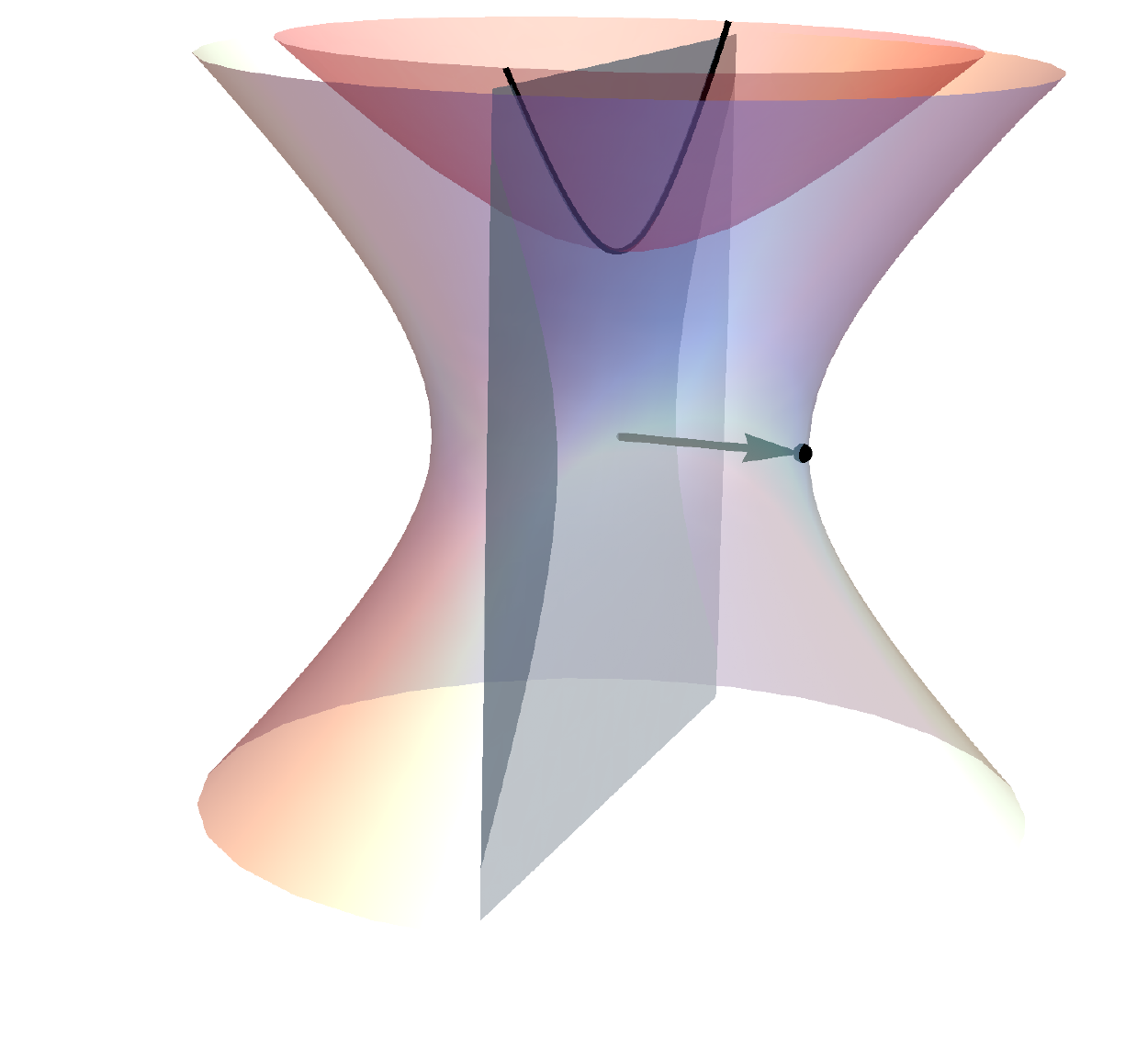}
\caption{The hyperbolic space $\mathbb{H}^2$ and the de Sitter space $\text{dS}_2$ are shown embedded in $\mathbb{R}^{2,1}$.  A geodesic in $\mathbb{H}^2$ lies on a unique plane in $\mathbb{R}^{2,1}$ and its normal vector at the origin identifies a corresponding point in $\text{dS}_2$.  This establishes $\text{dS}_2$ as the natural geometry for the kinematic space of $\mathbb{H}^2$.}
\Label{hyperboloid}
\end{figure}

Another locus invariant under the $SO(2,1)$ symmetry is:
\begin{equation}
- t^2 + x^2 + y^2 = \frac{L}{4G} = \frac{c}{6}. \,\,\,\,\Label{ds2embed}
\end{equation}
The right hand side is chosen with a view to normalization~(\ref{lorentzian}).
This is the two-dimensional de Sitter space, dS$_2$. The normals of hyperplanes in $\mathbb{R}^{2,1}$ (projected from the origin) intersect this locus at two points. Thus, there is a canonical identification between geodesics on $\mathbb{H}^2$ and pairs of antipodal points in dS$_2$ and this identification is invariant under symmetry transformations. Consequently, dS$_2$ is the kinematic space of the hyperbolic plane. For a given geodesic, the two points in dS$_2$ correspond to the two orientations of the geodesic.

To confirm this, evaluate the metric~(\ref{lorentzian}) substituting the entanglement entropy of an interval $(u,v)$ in the vacuum of a 1+1-dimensional CFT:
\begin{equation}
S(u,v) = \frac{c}{3} \log\frac{\sin (v-u)/2}{\mu}\,.
\end{equation}
We obtain
\begin{equation}
ds^2 = \frac{c}{6}\, \frac{du\, dv}{2 \sin^2(v-u)/2}\,, \Label{ds2units}
\end{equation}
which is the two-dimensional de Sitter space from eq.~(\ref{ds2embed}). 
The volume form on this space is $\omega(u,v)$, which may be compared with eq.~(\ref{uvcroftonform}) from Sec.~\ref{introhyp}. In that section we considered a unitless hyperbolic plane, thereby implicitly setting $L / 4G = c / 6 \equiv 1$. Thus, eqs.~(\ref{uvcroftonform}) and (\ref{ds2units}) agree.

\subsection{Point-curves}

In Sec.~\ref{point-curves} we defined point-curves in kinematic space: collections of geodesics that meet at a bulk point. Point-curves of the hyperbolic plane are particularly neat. The reason for their elegant form can be traced to the equivalence of the causal wedge and the entanglement wedge in AdS$_3$.

Consider the center of the hyperbolic plane in coordinates~(\ref{h2metric}). The point-curve of the center consists of all diameters of the hyperbolic plane, that is of points $v = u + \pi$. Referring to metric~(\ref{ds2units}), we see that this locus is the `waist' of the contracting-and-expanding de Sitter space. In other words, it is a spacelike geodesic in dS$_2$.

We may now apply a boost in the ambient space~(\ref{ambient}) to map the center to any other point on the hyperbolic plane. Such a boost is an isometry of eq.~(\ref{ds2embed}), so it maps spacelike geodesics to spacelike geodesics. We conclude that every point-curve of the hyperbolic plane is a spacelike geodesic in the kinematic dS$_2$.

Two of us previously reported this finding in \cite{lampros}. An alternative derivation of this fact and its higher-dimensional generalizations, which uses the moving frames formalism, is given in Sec.~2.3.1 of \cite{solanes}. In coordinates $\alpha = (v-u)/2$ and $\theta = (v+u)/2$, we may explicitly obtain the point-curves by applying the transformation~(\ref{boost1action}-\ref{boost2action}) to the point-curve of the center, $\alpha = \pi/2$. The result is:
\begin{equation}
\alpha(\theta) = \cos^{-1} \big( \tanh \rho_0 \cos(\theta-\theta_0) \big),
\end{equation}
where $\rho_0$ and $\theta_0$ determine the magnitude and direction of the boost. These parameters act as coordinates (\ref{h2metric}) on the hyperbolic plane.

\section{Discussion}

The Ryu-Takayanagi proposal highlights an important yet mysterious connection between quantum entanglement and geometry. The strikingly simple encoding of CFT entanglement entropies as bulk geodesics hints at the existence of a novel, promising framework for describing geometry---and maybe even gravitational dynamics---from the perspective of information theory. The Ryu-Takayanagi formula should be read as the first entry in the ``dictionary'' of such a new formulation of the theory of gravity. There exists, however, a ``language barrier,''  which hinders progress in further extending the translation: traditional general relativity is formulated in the \emph{local} language of differential geometry whereas the fundamental information theoretic quantities are represented holographically by non-local geometric data. The present paper makes a first step toward bridging this gap.

\paragraph{Kinematic space as a dual geometry.}
Our goal was to look at geometry with a fresh eye attuned to holography and information theory. We developed a mathematical formalism for understanding the bulk only in terms of boundary data by exploiting results and ideas from the field of \emph{integral geometry}. Of pivotal importance is \emph{kinematic space}: the space of all boundary anchored geodesics. When the bulk geometry has no conjugate points, the data contained in the kinematic space is sufficient to reconstruct the bulk.

This data consists of the Crofton form (\ref{gencrofton})---an exact 2-form which defines a measure on the space of geodesics---and a partial ordering, which reflects the containment relation among intervals subtended by the bulk geodesics (Sec.~\ref{causalstr}). These structures endow the kinematic space with a preferred Lorentzian metric~(\ref{lorentzian}); its causal structure geometrizes the containment relation. Bulk lengths and geodesic distances obtain a surprisingly simple equivalent description: they correspond to volumes of regions in the auxiliary metric geometry of kinematic space.

\paragraph{Kinematic space as a CFT information geometry.} 
The prime utility of kinematic space, however, lies in the way it marries geometry with information theory. A geometer thinks of kinematic space as the space of geodesics; an information theorist will read it as the space of boundary intervals. From the information theoretic perspective the causal structure of kinematic space means the following: interval $B$ is in the causal past of interval $A$ if the reduced density matrix on $B$ can be obtained from the density matrix on $A$ by partial tracing.

Geometric considerations---requiring that bulk lengths map to volumes in kinematic space---fix the Crofton density (\ref{gencrofton}). An information theorist recognizes this volume form---the `area' of an infinitesimal causal diamond in kinematic space---as the conditional mutual information of three neighboring intervals (Sec.~\ref{cmi}). 
This is a highly nontrivial identification. For example, computing the area of a larger causal diamond by integrating the Crofton form translates into assembling the conditional mutual information of three large intervals from smaller components using the chain rule for conditional mutual information (\ref{chainrule}). Similarly, so long as the strong subadditivity inequality $I(A,C|B)\geq 0$ is not saturated, the Crofton density remains nondegenerate. In light of this interpretation of the Crofton form, the length of every bulk curve comprises a collection of conditional mutual informations: it is an aggregate measure of how far the infrared degrees of freedom are away from saturating the strong subadditivity of entanglement. Enlarging the kinematic region -- that is, collecting more conditional mutual informations -- produces longer curves (see Exercises~\ref{lastqn} and \ref{addandsubtract} for an illustration). These statements can be viewed as a generalization of the Ryu-Takayanagi relation.

\paragraph{Relation to MERA}
The features of the kinematic space reviewed in the previous paragraphs are reminiscent of the Multi-scale Entanglement Renormalization Ansatz (MERA) network used to approximate ground states of critical systems \cite{mera, merathermal}. A key virtue of MERA is that it geometrizes the strong subadditivity inequality. The network contains isometric tensors, which isolate UV degrees of freedom; the (appropriately weighted) number of isometries becomes the conditional mutual information of some three adjacent spatial regions. In effect, the ``density of isometries'' plays the same role as the Crofton form (\ref{gencrofton}) in the kinematic space, albeit in a discrete setting. Other aspects of the kinematic space also find faithful counterparts on the MERA side. For example, the causal structure we defined in Sec.~\ref{causalstr} mimics the causal cones discussed in the context of MERA e.g.~in \cite{mera2, minupdates}. We will discuss these similarities in detail in an upcoming paper \cite{upcoming}. To whet the reader's appetite, we state here its main point: that MERA should best be thought of as a discretization of the kinematic space rather than of the spatial slice of a dual holographic spacetime as suggested in \cite{briansessay, brianspaper}. Note that associating MERA to the metric space (\ref{lorentzian}) has a broader scope of applicability, because the latter may be defined entirely in the information theoretic language, without presuming the existence of a holographic dual.

\paragraph{Point-curves and sub-AdS locality}
We would like to emphasize another aspect of our dual construction, one whose full potential has not yet unfolded in the present work and which deserves further exploration. It is the way in which the kinematic space encodes bulk locality. A key intermediate step in retrieving the bulk geometry from kinematic space is the identification of point-curves, the kinematic avatars of bulk points. We do this in Appendix~\ref{point-curve-derivation} by demanding that the kinematic formula for the bulk distance function (\ref{metric-definition}) come from a Riemannian metric. This
requirement, which uniquely determines the point-curves, is sensitive to the \emph{local} structure of the bulk geometry. In effect, the point-curves mediate the translation of bulk locality into the language of the boundary theory. This is a hint that integral geometry carries lessons about the emergence of sub-AdS locality. For example, it would be interesting to use integral geometry in order to understand the construction of local bulk operators in the boundary theory as in \cite{hkll2}.

\paragraph{A privileged background for CFT?} 
In the special case of pure AdS$_3$ we may identify the kinematic space with the asymptotic boundary. The conformal factor from eq.~(\ref{lorentzian}) arises from approaching the boundary in the de Sitter slicing of AdS$_3$. The map between geodesics on a spatial slice of AdS$_3$ and points in the kinematic space can then be understood in terms of a null sheet projected from the bulk geodesic, which arrives on the asymptotic boundary at a single point. We stress that this way of thinking about the kinematic space does not extend to a more general context. It does, however, raise the question: what happens if we put the CFT on a background metric given by the kinematic metric (\ref{lorentzian})? Perhaps this choice of metric, which is selected based on the state of the CFT, makes some properties of the state manifest?

\paragraph{Relevance to dS/CFT} 
The kinematic space of the hyperbolic space is the de Sitter geometry.\footnote{In higher dimensions, we generalize the space of geodesics to the space of codimension-1 totally geodesic surfaces \cite{solanes}.} To our knowledge, this fact has not been exploited in studies of the dS/CFT correspondence \cite{dscft} (but see \cite{janandvijay}). It may be worthwhile to reexamine the dS/CFT program from the perspective of integral geometry. This perspective offers new insights: for example, the elliptic de Sitter space discussed in \cite{elliptic} is simply the space of unoriented geodesics. One question of particular interest is the following: What is the meaning of bulk equations of motion in de Sitter gravity when the background de Sitter spacetime is interpreted as the kinematic space of an auxiliary hyperbolic geometry?

\acknowledgments

We are particularly grateful to Michael Freedman for bringing integral geometry to our attention. 
We thank Ahmed Almheiri, Vijay Balasubramanian, Alex Belin, Jan de Boer, Xi Dong, Patrick Hayden, Micha{\l} Heller, Byungwoo Kang, Esperanza Lopez, Raghu Mahajan, Benjamin Mosk, Rob Myers, Hirosi Ooguri, John Pardon, John Preskill, Xiao-liang Qi, Grant Salton, Joan Sim{\'o}n, Steve Shenker, Lee Smolin, Leonard Susskind, Brian Swingle, and Tadashi Takayanagi for useful discussions. SM would especially like to thank Eva Silverstein for her advice and support. 

BC and JS are grateful for the support from the National Science Foundation Grant No. PHYS-1066293 during the ``Emergent Spacetime in String Theory'' workshop at the Aspen Center for Physics, where this work was initiated; and for hospitality and support of Caltech and Perimeter Institute for Theoretical Physics (supported by the Government of Canada through Industry Canada and by the Province of Ontario through the Ministry of Economic Development \& Innovation), where this work was partially completed. BC also thanks the organizers of the workshops ``Holographic Renormalization Group and Entanglement'' at Paris Diderot University and ``Entangle This: Space, Time and Matter'' at ITP Madrid. SM was supported in part by an award from the Department of Energy (DOE) Office of Science Graduate Fellowship Program.

\appendix



\section{Review of differential entropy}
\Label{diffent-review}

The differential entropy formula was written down in \cite{holeography} and developed further in \cite{robplusstudents, xi, entwinement, robproof, wien, lampros}. Here we present a proof, which is similar to \cite{wien}.

Let $X$ be a smooth manifold with boundary $\partial X$. Consider
a general reparameterization-invariant action functional of the form 
\begin{equation}
S\left[\gamma\right]=\int_{0}^{1}dt\,\mathcal{L}\left(\gamma,\dot{\gamma}\right)\Label{length-functional}
\end{equation}
defined for a curve $\gamma:\left[0,1\right]\rightarrow X$. In particular,
we require that $\mathcal{L}$ not depend on any higher derivatives
of $\gamma$, and reparameterization-invariance requires that $\mathcal{L}$
be homogeneous of degree 1 in $\dot{\gamma}$. The example of most
immediate interest is the Riemannian length functional, 
\begin{equation}
\ell[\gamma] = \int_{0}^{1}dt\sqrt{g_{\mu\nu}\left(\gamma\right)\dot{\gamma}^{\mu}\dot{\gamma}^{\nu}},\Label{Riemannian-length}
\end{equation}
which via the Ryu-Takayanagi prescription is related to entanglement entropies of boundary intervals.  We could also consider alternate entropy functionals for higher-derivative gravity \cite{donghigherdgravity}.

Reparameterization-invariance implies that the Hamiltonian corresponding
to $\mathcal{L}$ vanishes, so we may write 
\begin{equation}
S\left[\gamma\right] = \int_{0}^{1}dt\,\dot{\gamma}^{\mu}
\frac{\partial\mathcal{L}}{\partial\dot{\gamma}^{\mu}}
= \int_{\gamma}p_{\mu}dx^{\mu},
\end{equation}
where $p_{\mu}=\frac{\partial\mathcal{L}}{\partial\dot{\gamma}^{\mu}}$
is the momentum conjugate to $\gamma^{\mu}$. This is easily verified
for the Riemannian length functional (\ref{Riemannian-length}).

For two points $x,y\in X$, define $S\left(x,y\right)$ to be the
action of the extremal curve connecting $x$ and $y$. In the case
of the Riemannian length functional (\ref{Riemannian-length}), this
is just the usual geodesic distance.
\footnote{In fact, with the restriction $\mathcal{L}>0$, other actions of the
form \ref{length-functional} define what are known in the mathematical
literature as \emph{Finsler manifolds} \cite{finsler}. A Finsler manifold is a generalization
of a Riemannian manifold in which the norm on the tangent space is
no longer required to be the square root of a quadratic form. The
function $\mathcal{L}\left(\gamma,\dot{\gamma}\right)=\left|\dot{\gamma}\right|$
is known as a \emph{Finsler function}. Given a Finsler function, the
length of a curve $\gamma$ is defined to be $S\left[\gamma\right]$,
and the geodesic distance between two points $x,y$ is defined to
be $S\left(x,y\right)$.} 
We will often refer to $S\left(x,y\right)$ as the distance between $x$
and $y$, even though in general it could have another meaning.

Our goal is to compute $S\left[\gamma\right]$ for an arbirary curve
$\gamma$ using only the knowledge of the distance
$S_\text{bdy}(x,y) = S\left(x,y\right)\big|_{\partial X}$
between points $x,y$ on the \emph{boundary} of $X$.  In the context
of holographic duality, this corresponds to computing the bulk quantity
$S\left[\gamma\right]$ using only boundary data.

Let us assume that for any point $y$ lying on $\gamma$, there
is some point $x\in\partial X$ such that the extremal curve from
$x$ to $y$ is tangent to $\gamma$ at $y$.
\footnote{This will not be the case, for instance, when there are \emph{trapped}
geodesics tangent to $\gamma$.  If a geodesic tangent to $\gamma$ is nonminimal, 
additional boundary data about lengths of nonminimal geodesics \cite{entwinement} is required to determine the length of $\gamma$.} 
Denote by $G_{\gamma}$ this set of geodesics $(x,y)$ tangent to $\gamma$,
labeled by their endpoints on $\partial X$ and $\gamma$.
Without loss of generality, we can assume that $\gamma$ is closed;
if not, consider the curve following $\gamma$ forwards then backwards.
Then $G_{\gamma}$ is also a closed curve in $\partial X\times X$.

Consider the exterior derivative of $S$, 
\begin{equation}
dS=\frac{\partial S}{\partial x^{\mu}}dx^{\mu}+\frac{\partial S}{\partial y^{\mu}}dy^{\mu},\Label{total-derivative}
\end{equation}
and integrate both sides over $G_{\gamma}$. Since $G_{\gamma}$ is
closed, the first term integrates to zero:
\begin{equation}
\int_{G_{\gamma}}dS=0.
\end{equation}
Using the standard Hamilton-Jacobi argument, we find that the variation
of $S$ with respect to its endpoint coordinate is given by the conjugate momentum at that
endpoint:
\begin{equation}
\frac{\partial S(x,y)}{\partial y^{\mu}}=p_{\mu}\big|_{y}. \Label{hamilton-jacobi}
\end{equation}
The tangency of the $x\to y$ geodesic to $\gamma$ guarantees that this momentum is the same as we would have obtained varying $S[\gamma]$.
\footnote{In fact, we only require that $p_{\mu}dy^{\mu}$ take the correct
value. For the pseudo-Riemannian length functional, we have $p_{\mu}=g_{\mu\nu}\hat{T}^{\nu}$,
where $\hat{T}^{\nu}$ is the unit tangent vector to the geodesic
at its endpoint. Hence, we are free to add a null vector to $T^{\nu}$
so long as it has zero dot product with $T^{\nu}$. See \cite{robproof}
for a more detailed discussion of this ``null vector alignment''
condition.} 
The integral of the second term in (\ref{total-derivative})
then becomes:
\begin{equation}
\int_{G_{\gamma}}\frac{\partial S}{\partial y^{\mu}}dy^{\mu} = 
\int_{\gamma}p_{\mu}dy^{\mu} = S\left[\gamma\right].
\end{equation}
Lastly, define the set $\tilde{G}_{\gamma}\subset\partial X\times\partial X$,
which is obtained from $G_{\gamma}$ by extending each extremal curve
to the boundary. The same Hamilton-Jacobi argument (\ref{hamilton-jacobi}) tells us that $\frac{\partial S}{\partial x^{\mu}}\big|_{G_{\gamma}}=\frac{\partial S}{\partial x^{\mu}}\big|_{\tilde{G}_{\gamma}}$.
Since $S\big|_{\tilde{G}_{\gamma}}$ is a subset of the boundary data
we have access to, and we only take its derivative in directions along the boundary,
 we can now compute $S\left[\gamma\right]$ in terms of boundary data:
\begin{equation}
S\left[\gamma\right]=-\int_{\tilde{G}_{\gamma}}\frac{\partial S_\text{bdy}}{\partial x^{\mu}}dx^{\mu}. \Label{general-diffent}
\end{equation}
This is the differential entropy formula. Referring back to eq.~(\ref{total-derivative}), we recognize that it arises from applying integration by parts to an ordinary Riemannian length integral.

In two dimensions, we have the familiar differential entropy formula:
\begin{equation}
S[\gamma] = - \int_{\partial X} du\, \frac{\partial S_\text{bdy}(u,v)}{\partial u} \big|_{v = v(u)}.
\Label{diffentapp}
\end{equation}
When $S[\gamma] = \ell[\gamma]$ is the Riemannian length functional, applying the Hamilton-Jacobi equation again as in eq.~(\ref{hamilton-jacobi-angles}) yields a simple but surprising geometric identity:
\begin{equation}
\ell[\gamma] = \int_{\partial X} ds \cos \theta_\text{bdy}(s)
\Label{diffent-geometric}
\end{equation}
Here $ds$ is the proper length along $\partial X$ and $\theta_\text{bdy}$ is the angle at which the geodesic tangent to $\gamma$ hits the boundary. When $\partial X$ is the \emph{asymptotic} boundary, the cosine vanishes but the range of integration over $s$ is infinite. Defining the right hand side of (\ref{diffent-geometric}) via the limit as $\partial X$ is sent to the asymptotic boundary recovers the correct, finite length of $\gamma$.


\section{Point-curves and bulk reconstruction}
\Label{point-curve-derivation}

In Sec.~\ref{vackin} we showed how to use the Crofton formula on
kinematic space to describe the holographic bulk geometry using entanglement
entropies. The fundamental objects in this description are the point-curves,
which identify a bulk point in a gauge-invariant way by the collection
of geodesics that pass through it (see Fig.~\ref{point-curve}). When the bulk geometry is known, it is trivial to find the point-curves. The goal of this appendix is to determine the
point-curves using only boundary data. We restrict ourselves
to the case where the Ryu-Takayanagi formula computes lengths of geodesics
and receives no quantum corrections; higher curvature or higher-derivative
gravity theories will require more analysis. When the geometry contains nonminimal geodesics, our method will require a generalization along the lines of \cite{entwinement, lampros}.

Refs.~\cite{bilson1, bilson2, bilson3} gave another algorithm for recovering non-trivial components of the metric in symmetric setups from analytic properties of field theory quantities, including entanglement entropies. The results of the present appendix are complementary to that development: they highlight the information-theoretic structure underlying the organization of the bulk and can accommodate less symmetric inputs. 
Other recent work on bulk reconstruction includes \cite{berkeley, chi, rhodual, errorcorrect}.

We start by discussing
boundary rigidity theorems, which motivate imposing a
Riemannian condition. We then show how imposing this condition leads to
a method for determining the point-curves from boundary data.  To test the method, we apply it in the case of a rotation-symmetric bulk and confirm
that it yields the same results as does integrating the geodesic equations.

\subsection{Boundary rigidity}

The ability to reconstruct Riemannian manifolds from boundary-to-boundary
distances, a question first posed by Michel \cite{michel}, has been
explored significanly in the mathematical literature. Pestov and Uhlmann
\cite{pestovuhlmann} proved that the metric structure of a two-dimensional, compact, simply
connected Riemannian manifold with convex boundary and with no conjugate
points is uniquely determined (up to diffeomorphism) by the lengths
of boundary-to-boundary geodesics. This is known as \emph{boundary
distance rigidity}. Some of these restrictions can be dropped with
access to nonminimal geodesics (i.e. entwinement \cite{entwinement}).
Our work will not make use of these theorems,
but they provide useful information on when bulk reconstruction is
possible \cite{porratirabadan}. 

Given the rigidity theorem, we expect that point-curves should be
determined from boundary data in many cases by demanding that the distance function on the manifold descend from a \emph{Riemannian
metric}. Indeed, this
condition severely constrains the point-curves. When varying the endpoints of the geodesics on the boundary sweeps the whole tangent space of every point, this procedure fixes the point-curves completely.

\subsection{Riemannian condition}
\Label{riemannian-condition}

\begin{figure}[t]
\centering
\includegraphics[height=.25\textwidth]{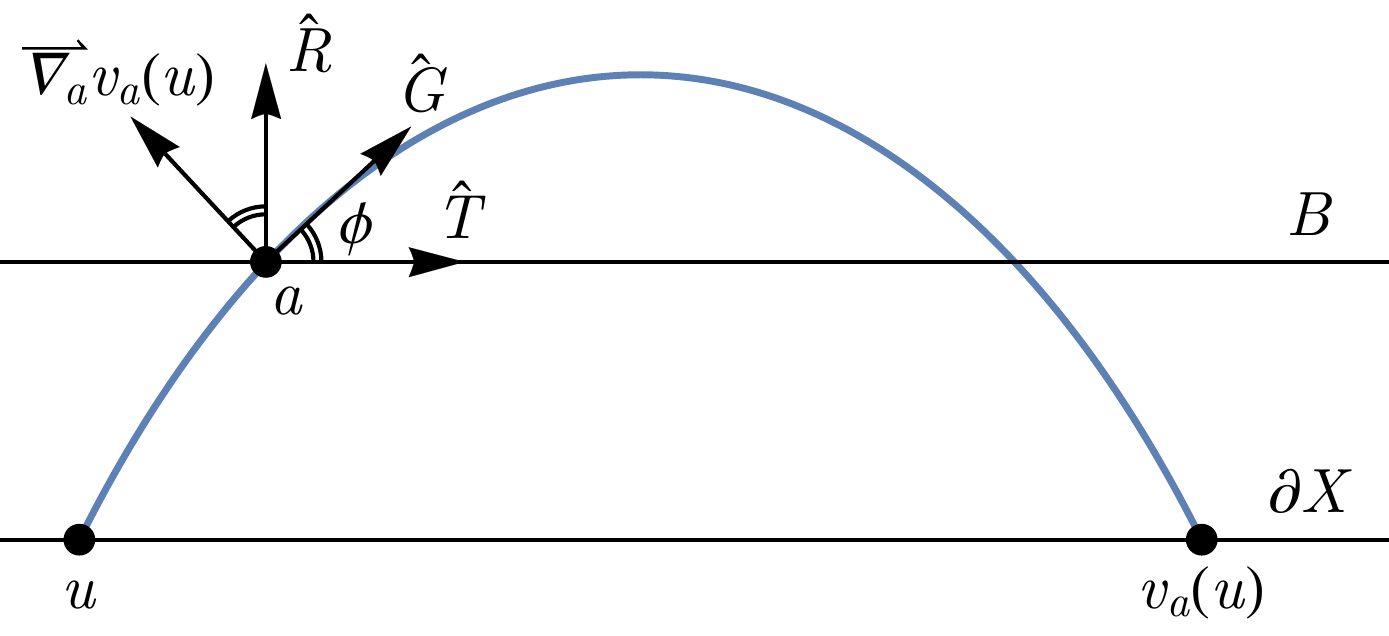} 
\caption{An illustration of the bulk reconstruction method of Sec.~\ref{riemannian-condition}.}
\Label{riemannian-condition-fig}
\end{figure}

We take an iterative approach and assume that we already know the point-curves of all points, which live on some given
cutoff surface. Using this initial data we compute the point-curves lying on a new cutoff surface, which lies deeper in the bulk.

Consider a two-dimensional Riemannian manifold $X$. The point-curve of every point
lying on the cutoff surface $\partial X$ is already known to be a light cone in kinematic space (see Sec. \ref{distance-between-points}).
Now suppose we have already constructed the point-curves for some
cutoff surface $B$ in the bulk. This is the initial data that seeds our construction.
Note that we already know the distances between points on $\partial X$ and $B$ from eq.~(\ref{metric-definition}).

Choose some point $a\in B$ on the current cutoff surface. This
point has an associated point-curve $v_{a}\left(u\right)$. In order
to find the point-curve for a deeper point $a+\delta a$, we would
like to compute the quantity $\nabla_{a}v_{a}\left(u\right)$. We
already know the value of $\hat{T}\cdot\nabla_{a}v_{a}\left(u\right)$,
where $\hat{T}$ is the tangent vector to the surface $B$ at $a$,
so what we need to compute is the component $\hat{R}\cdot\nabla_{a}v_{a}\left(u\right)$,
where $\hat{R}$ is the normal vector to $B$, pointing inward from
$a$.

To start, consider the vector $\nabla_{a}\ell\left(u,a\right)$ for
a given $u$. As shown in Sec.~\ref{angles-sec}, this is just the
unit tangent vector $\hat{G}$ pointing along the geodesic from $u$
to $a$ (see Fig.~\ref{riemannian-condition-fig}). It is perpendicular to $\nabla_{a}v_{a}\left(u\right)$
since moving $a$ along the geodesic from $u$ to $v_{a}\left(u\right)$
does not change the endpoint $v$. This gives the result 
\begin{equation}
\frac{\hat{R}\cdot\nabla_{a}v_{a}\left(u\right)}{\left|\hat{T}\cdot\nabla_{a}v_{a}\left(u\right)\right|}=\frac{\cos\phi}{|\sin\phi |}\,,
\Label{pc-master-eqn}
\end{equation}
where $\cos\phi=\hat{G}\cdot\hat{T}=\frac{\partial\ell\left(u,a\right)}{\partial a^{\mu}}\cdot\hat{T}^{\mu}$
(see Fig.~\ref{riemannian-condition-fig}). Since $\cos\phi$ and $\hat{T}\cdot\nabla_{a}v_{a}\left(u\right)$
can both be computed already, this gives us access to the full vector
$\nabla_{a}v_{a}\left(u\right)$. In principle this equation is enough to construct the full bulk geometry.

\subsection{Rotationally symmetric bulk}

We would like to apply eq.~(\ref{pc-master-eqn}) to a rotationally symmetric
manifold. We restrict our attention to the case without nonminimal geodesics. Define the metric to be 
\begin{equation}
ds^{2}=d\rho^{2}+\left(\frac{C\left(\rho\right)}{2\pi}\right)^{2}d\theta^{2}
\Label{rotsym-metric}
\end{equation}
and consider a cutoff surface at $\rho=\rho_{\rm UV}$. Here $C\left(\rho\right)$
denotes the circumference of a centered circle with radius $\rho$. 

In the presence of rotational symmetry it is convenient to parameterize
the space of geodesics using the coordinates $\left(\alpha,\theta\right)$,
the opening angle and the angular coordinate of the center (see Fig.~\ref{alphatheta}).
We write the length of a geodesic as $\ell\left(\alpha\right)$ and
instead of expressing a point-curve as a function $v\left(u\right)$
we use $\alpha\left(\theta\right)$. Writing eq.~(\ref{pc-master-eqn})
in these coordinates, we have:
\begin{equation}
\frac{\hat{R}\cdot\nabla_{a}\alpha_{a}\left(\theta\right)}{\left|\hat{T}\cdot\nabla_{a}\alpha_{a}\left(\theta\right)\right|}=\frac{\cos\phi_{a}\left(\theta\right)}
{\sin\phi_{a}\left(\theta\right)}.
\end{equation}
Now symmetry requires that $\alpha_{\left(\rho,\theta_{0}\right)}\left(\theta\right)=\alpha_{\left(\rho,0\right)}\left(\theta-\theta_{0}\right)$,
so we only need to find a one-parameter family of point-curves, the
function $\alpha_{\rho}\left(\theta\right)$, to construct the geometry.
In addition, the angle formula simplifies to 
\begin{equation}
\cos\phi_{\rho}\left(\theta\right)=\frac{\pi\ell^{\prime}(\alpha_{\rho}\left(\theta\right))}{C\left(\rho\right)},
\end{equation}
which can be seen either from eq.~(\ref{hamilton-jacobi-angles}) or
from (\ref{diffent-geometric}). If we choose the minimum of $\alpha_{\rho}\left(\theta\right)$
to be at $\theta=0$, we can replace $C\left(\rho\right)=\pi\frac{d\ell}{d\alpha}\big|_{\alpha_{\rho}\left(0\right)}$.
This yields the point-curve equation:
\begin{equation}
\frac{\partial_{\rho}\alpha_{\rho}\left(\theta\right)}{|\partial_{\theta}\alpha_{\rho}\left(\theta\right)|}=\frac{2\pi}{\pi\ell^{\prime}(\alpha_{\rho}\left(0\right))}\left(\left(\frac{\pi\ell^{\prime}(\alpha_{\rho}\left(0\right))}{\pi\ell^{\prime}(\alpha_{\rho}\left(\theta\right))}\right)^{2}-1\right)^{-1/2}
\Label{rotsym-pc-eqn}
\end{equation}
This is a first-order equation in $\rho$. It can be solved subject to the boundary condition $\alpha_{\rho_{\rm UV}}\left(\theta\right)=\cos^{-1}\left(\cos\theta\right) = |\theta|$,
the light-cone curves of Sec.~\ref{distance-between-points}. 

Eq.~(\ref{rotsym-pc-eqn}) takes a simpler form if we change variables
from $\alpha$, an explicitly UV quantity, to the IR quantity $\pi\ell^{\prime}\left(\alpha\right)$,
the signed circumference of the circle which is the outer envelope of all $\alpha$-sized geodesics. Defining $c_{\rho}\left(\theta\right)=\pi\ell^{\prime}\left(\alpha_{\rho}\left(\theta\right)\right)$,
we have
\begin{equation}
\frac{\partial_{\rho}c_{\rho}\left(\theta\right)}{\left|\partial_{\theta}c_{\rho}\left(\theta\right)\right|}=\frac{2\pi}{c_{\rho}\left(0\right)}\left(\left(\frac{c_{\rho}\left(0\right)}{c_{\rho}\left(\theta\right)}\right)^{2}-1\right)^{-1/2}.\label{eq:rotsym-ir-pc-eqn}
\end{equation}
Note that all explicit dependence on boundary data $\ell\left(\alpha\right)$
has vanished. This data has been shifted to the boundary conditions,
which now take the form $c_{\rho_{\rm UV}}
\left(\theta\right)=\pi\ell^{\prime}\left(|\theta|)\right)$.

The geodesic equation for $\theta\left(\rho\right)$ in
geometry~(\ref{rotsym-metric}) can be written as:
\begin{equation}
\frac{d\theta\left(\rho\right)}{d\rho}=\pm\frac{2\pi}{C\left(\rho\right)}\left(\left(\frac{C\left(\rho\right)}{C\left(\rho_{0}\right)}\right)^{2}-1\right)^{-1/2}\,.
\end{equation}
Here $\rho_{0}$ is the deepest point of the geodesic. The equivalence of the right side with that of (\ref{eq:rotsym-ir-pc-eqn}) is immediate with the identifications $C\left(\rho\right)=c_{\rho}\left(0\right)$ and $C\left(\rho_{0}\right)=c_{\rho}\left(\theta\right)$. The left sides both equal $\frac{C(\rho)}{2\pi}\cot\phi$ (see Fig.~\ref{riemannian-condition-fig}, where $\nabla_a c_\rho(\theta)$ takes the place of $\nabla_a v_a(u)$).  Thus, our procedure has the effect of iteratively solving the geodesic equations from the boundary to find the point-curves.

\section{Curves in kinematic space}
\Label{geometric-concepts}

This appendix explains how to translate between bulk curves and regions in kinematic space. We begin with the simplest case of closed convex curves and then move on to nonconvex and open curves. A general region in kinematic space corresponds to a weighted combination of curves, possibly with negative coefficients.

\begin{figure}[t]
\centering
\raisebox{0.02\textwidth}{\includegraphics[width=.27\textwidth]{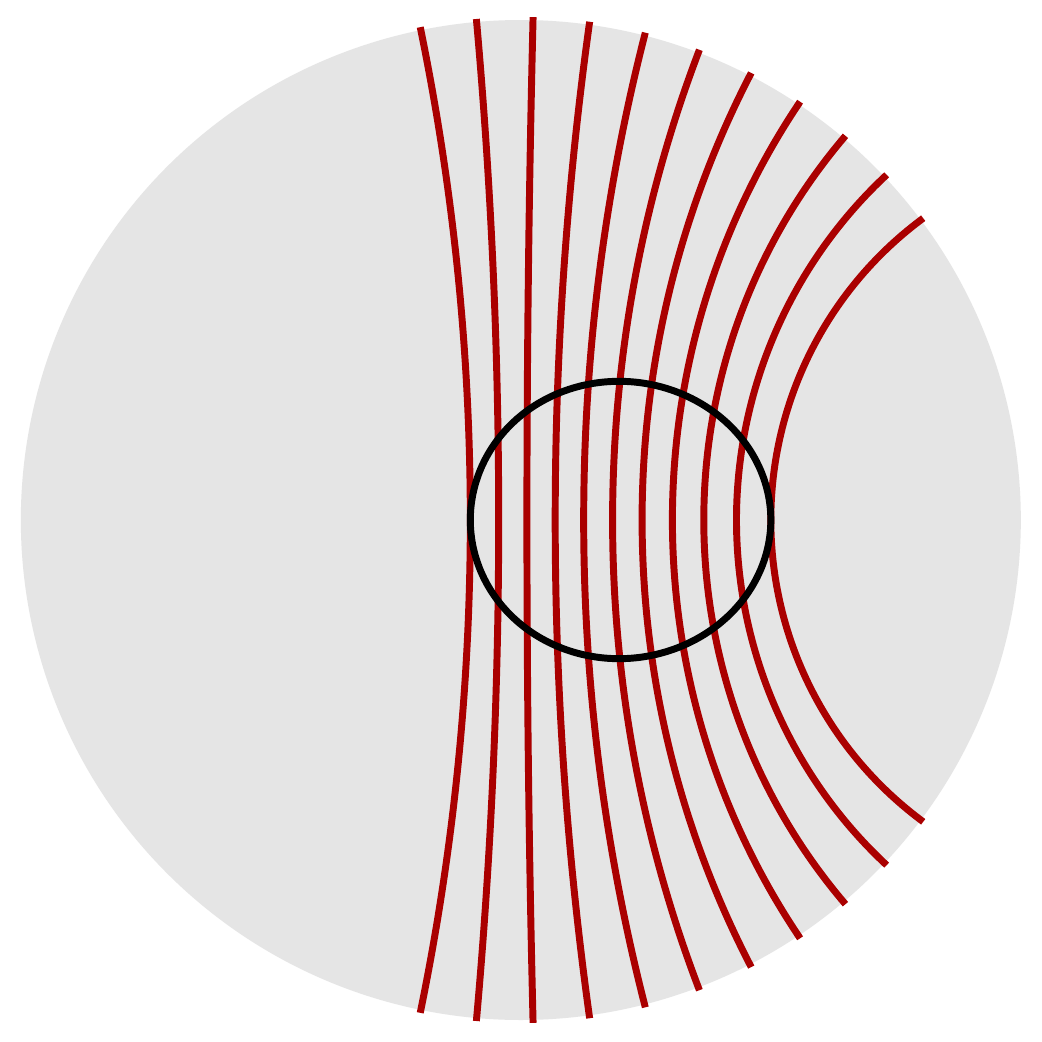}}
\hspace{0.02\textwidth}
\raisebox{0.02\textwidth}{\includegraphics[width=.27\textwidth]{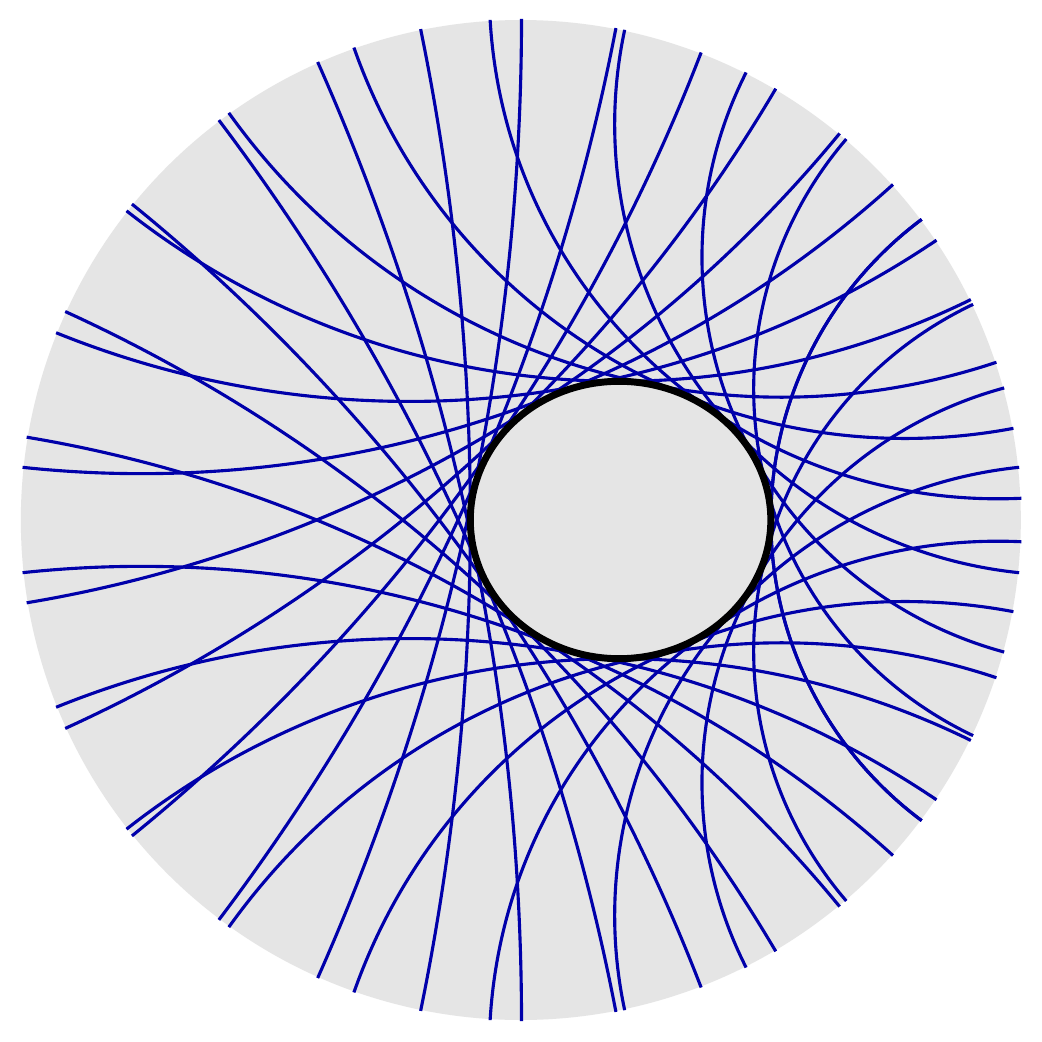}}
\hspace{0.01\textwidth}
\includegraphics[width=.40\textwidth]{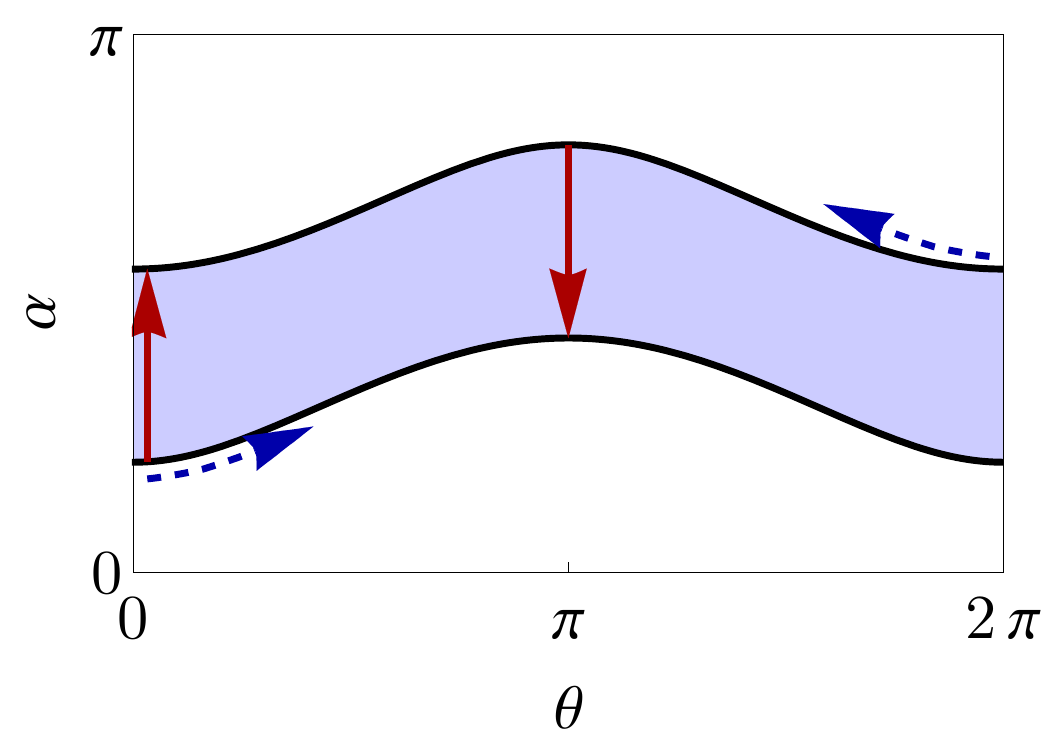}
\caption{A closed convex bulk curve $\gamma$ with a (timelike) family of intersecting geodesics with a common center (left) and a (spacelike) family of geodesics tangent to it (center). The full set of geodesics that intersect $\gamma$ forms the shaded region in kinematic space (right). The timelike family from the left panel is shown with solid red arrows while the spacelike family of tangent geodesics, which forms the boundary of the requisite kinematic region, is shown with blue dotted arrows. Stokes' theorem relates the length of the curve to the differential entropy over this boundary of the region, with the orientation given by the blue dotted arrows. All arrows are shown together with their $\mathbb{Z}_2$-related partners.}
\label{convexbulk}
\end{figure}

\subsection{Closed convex curves}
\Label{smoothconvexcurves}
We will make use of the $(\alpha,\theta)$ parametrization of kinematic space (see Fig.~\ref{alphatheta}). Consider a closed convex bulk curve $\gamma$. What is the region in kinematic space whose volume computes the length of this curve? Eq.~(\ref{ourclaim}) gives the answer: this region consists of geodesics that intersect $\gamma$. It is bounded by the family of geodesics that are tangent to $\gamma$.

Let us elaborate. Any geodesic segment that connects 2 points along a convex curve is fully enclosed by the curve. As a result, a geodesic intersects a closed convex curve at most at two points. The geodesics intersecting it exactly once belong in the family of tangent geodesics, a lower dimension set. For each angle $\theta$ there are 2 tangent geodesics: $\alpha_{\text{lower}}(\theta)$ and $\alpha_{\text{upper}}(\theta)$. All geodesics centered at $\theta$ with opening angle smaller than $\alpha_{\text{lower}}(\theta)$ or larger than $\alpha_{\text{upper}}(\theta)$ lie completely outside the curve of interest while the ones with opening angle in the interval $(\alpha_{\text{lower}}(\theta), \alpha_{\text{upper}}(\theta))$ intersect it twice (Fig.~\ref{convexbulk}). Therefore, the kinematic volume dual to a curve is bounded by the family of tangent geodesics.

The upper and lower boundaries of this kinematic region, however, are related by the $\mathbb{Z}_2$ symmetry of kinematic space: 
\begin{equation}
\theta \rightarrow \theta+\pi 
\qquad {\rm and} \qquad
\alpha \rightarrow \pi-\alpha\,. \label{z2kinem}
\end{equation}
The symmetry arises from the reversal of a geodesic's orientation, which corresponds to taking the complement of each boundary interval. Namely, the same geodesic can either be treated as centered at $\theta$ with opening angle $\alpha$ or -- with opposite orientation -- as centered at $\theta + \pi$ with opening angle $\pi - \alpha$. This symmetry relates the upper and lower boundaries of the kinematic region corresponding to $\gamma$:
\myeq{\alpha_{\text{upper}}(\theta)= \pi-\alpha_{\text{lower}}(\theta+\pi).}
As a consequence, the said region is invariant under the $\mathbb{Z}_2$ transformation.

\subsection{Concave and open curves}
\label{concaveappendix}

\begin{figure}[t]
\centering
\includegraphics[height=.35\textwidth]{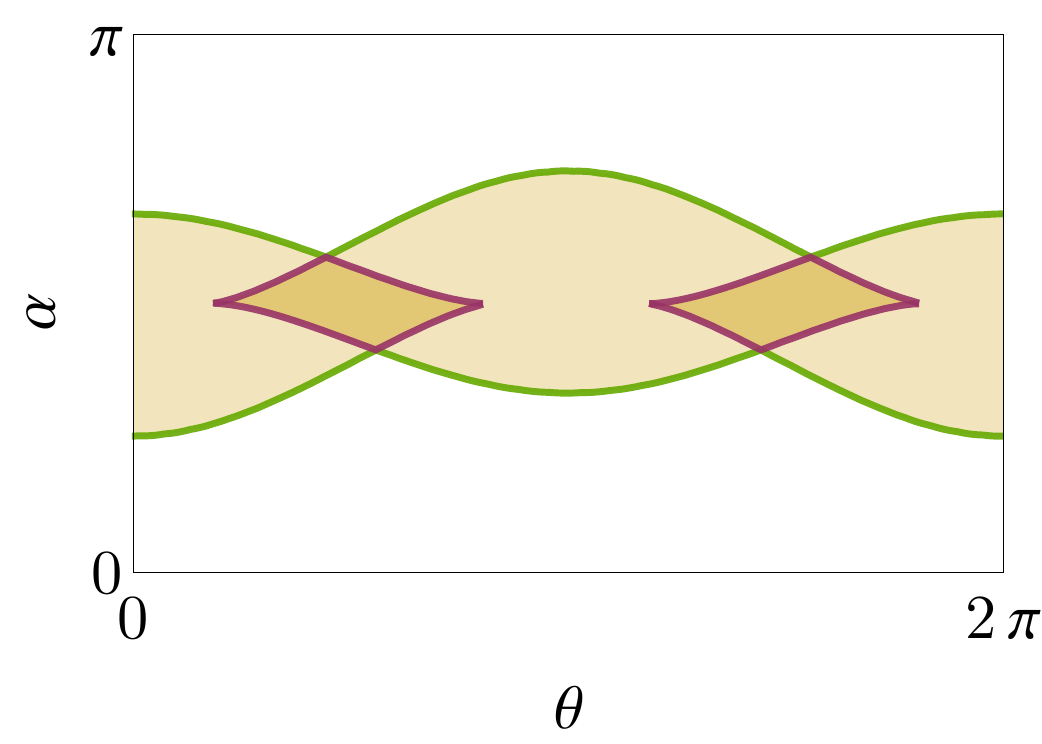}
\hspace{.05\textwidth}
\includegraphics[height=.4\textwidth]{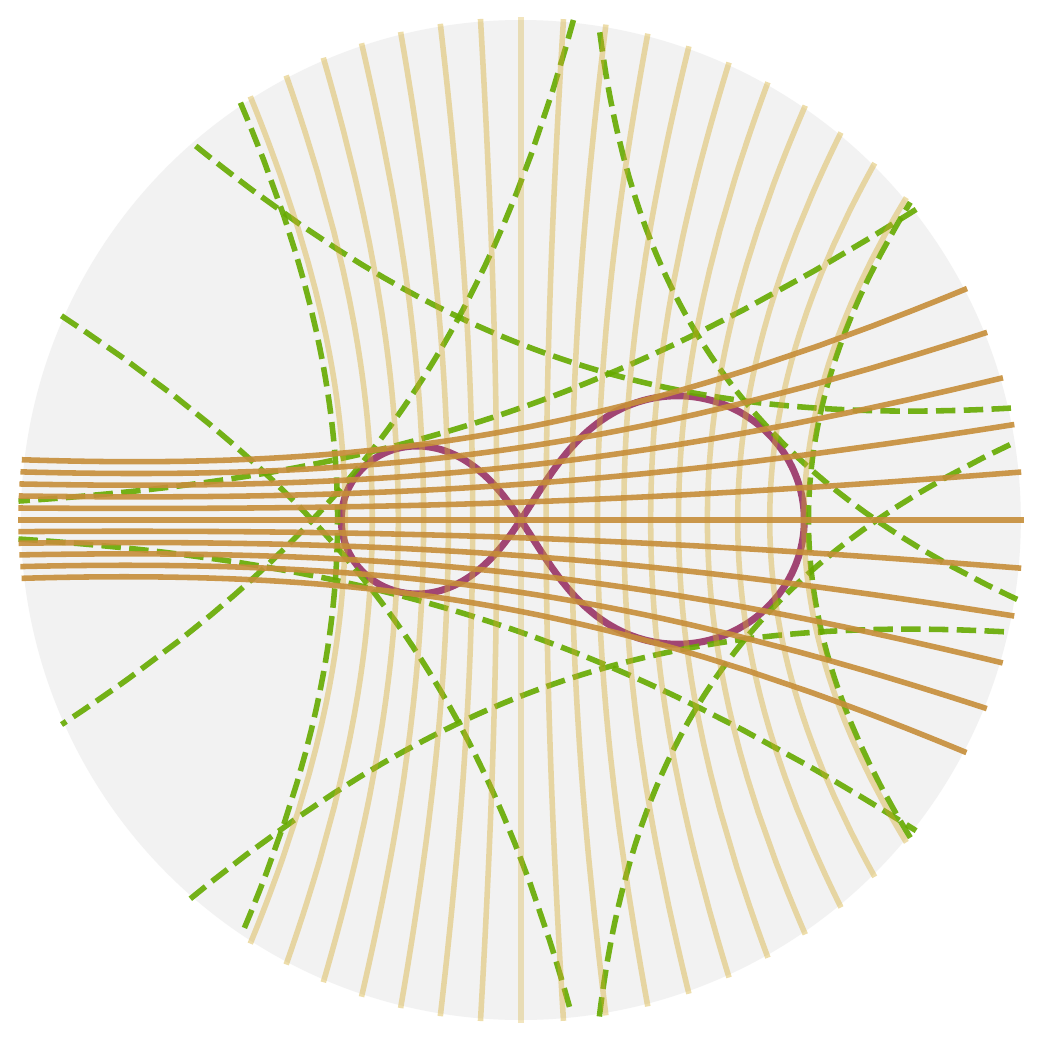} 
\caption{Left: The kinematic region for the curve shown on the right.  The boundary of the region, shown in green, is the set of geodesics tangent to the curve's convex hull.  Some tangent geodesics, shown in purple, also intersect the curve.  The light and dark shaded regions correspond to geodesics that intersect the curve 2 and 4 times, respectively.  Right: A nonconvex curve in hyperbolic space.  Geodesics shown in dashed green are tangent to its convex hull. Light and dark geodesics that intersect the curve live in light and dark shaded regions on the left.}
\label{concave}
\end{figure}

What is different in the dual description of a concave curve? There are two modifications: 1) the number of intersection points of a geodesic with the curve can be larger than two, and consequently, 2) some tangent geodesics are not on the boundary of the dual kinematic region, because they intersect the curve at other points.

An example of a closed bulk curve along with its dual kinematic region is shown in Fig.~\ref{concave}. There are two types of regions in the kinematic strip: light-colored regions that correspond to geodesics with 2 intersection points with the bulk curve and dark-colored regions that correspond to geodesics with 4 intersection points. These are in direct correspondence to the coloring of the bulk geodesics that intersect the curve shown in the left panel of Fig.~\ref{concave}.

If the kinematic region dual to a concave curve is not bounded by the tangent geodesics, then what determines its boundary (thick green line in Fig.~\ref{concave})? The answer is the geodesics tangent to the \emph{convex hull} of the concave curve. All geodesics that intersect the convex hull of a curve also intersect the curve itself, and vice versa.  The only difference is the multiplicity of intersections. All geodesics will intersect the convex hull at most twice as discussed above, but higher multiplicities will occur for subset of these geodesics in the case of a concave curve. Thus, if we count all kinematic regions with multiplicity 2, we obtain the dual of a curve's convex hull.

\begin{figure}[t]
\centering
\includegraphics[height=.35\textwidth]{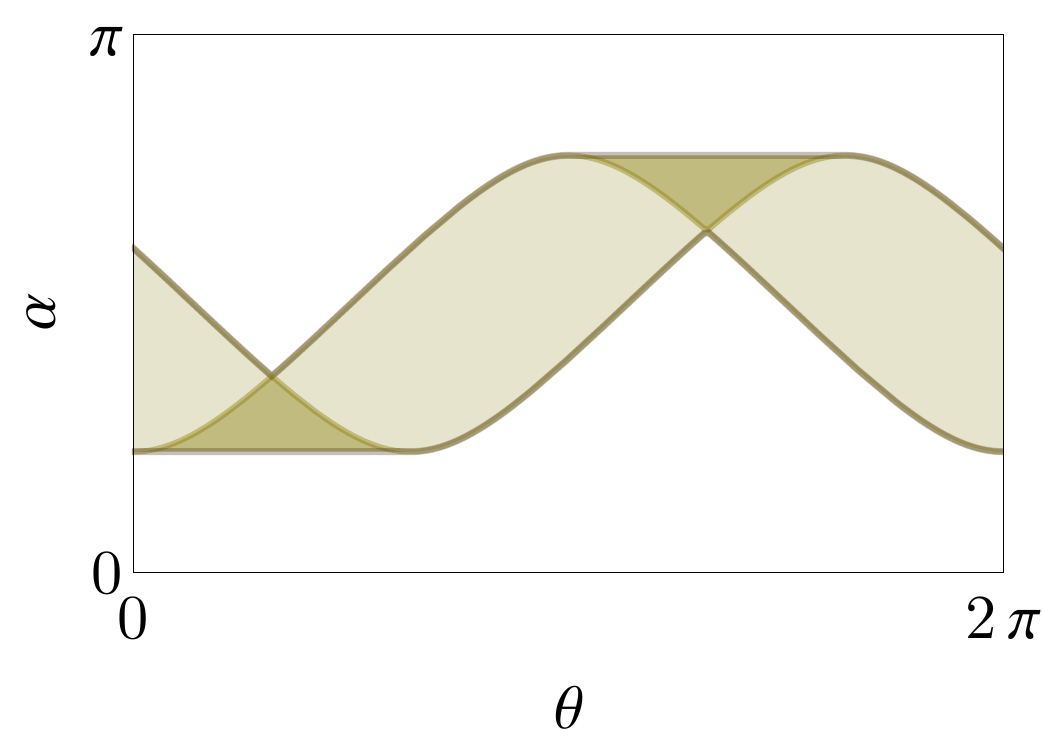}
\hspace{.07\textwidth}
\includegraphics[height=.4\textwidth]{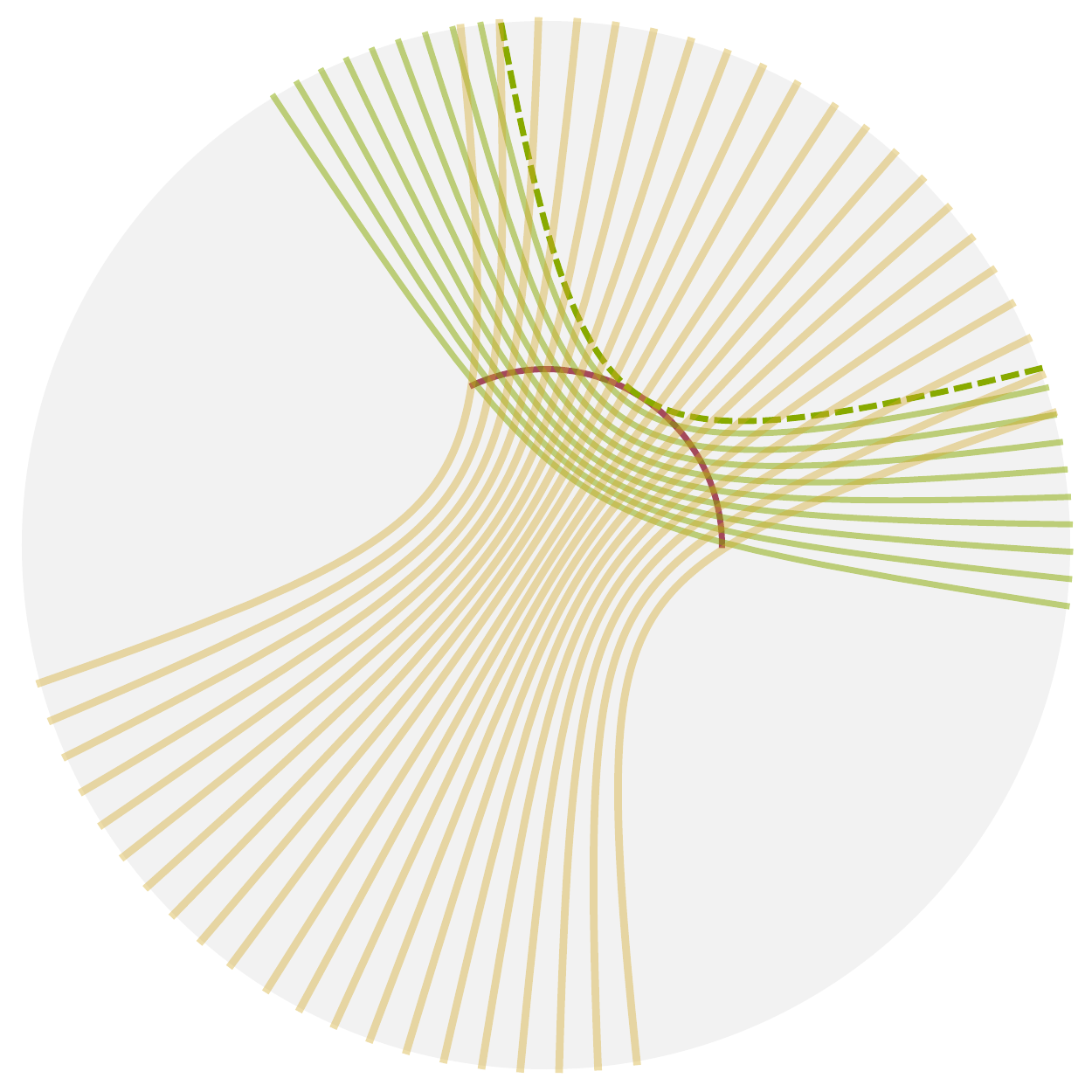} 
\caption{Left: Kinematic region for the open bulk curve shown in the right. Light-shaded regions represent geodesics intersecting the curve once, while darker-shaded regions contain geodesics with double intersection multiplicity. The point-curves of the endpoints appear as the boundary of the light-shaded regions, while the dark region corresponds to the added length from nonconvexity of the curve. Right: The corresponding bulk open curve (solid dark brown) along with two sets of intersecting geodesics, shaded light and dark as in the corresponding kinematic region on the left.  Only the geodesics with odd intersection number pass between the two endpoints.}
\label{opencurve}
\end{figure}

The discussion above immediately extends to open curves (see Fig.~\ref{opencurve}).  This can be seen by viewing an open curve as a closed curve, which turns around and retracts its path back to the starting point.  Geodesics intersecting an open curve an odd number of times pass between the two endpoints and must intersect the geodesic connecting the endpoints.  Hence, as in Fig.~\ref{metricdef-fig}, the odd-multiplicity regions are bounded by the point-curves for the open curve's endpoints.

\subsection{Signed lengths of bulk curves}

So far we have discussed how bulk curves are represented as regions in kinematic space.  We now discuss the inverse ``boundary-to-bulk'' problem of finding the bulk curve corresponding to an arbitrary kinematic region (see \cite{robproof, wien} for a detailed discussion in the language of differential entropy).  We will see here that a general kinematic region corresponds not to a simple bulk curve of positive length, but a collection of curves with \emph{signed} weight.

\begin{figure}[t]
\centering
\includegraphics[height=.35\textwidth]{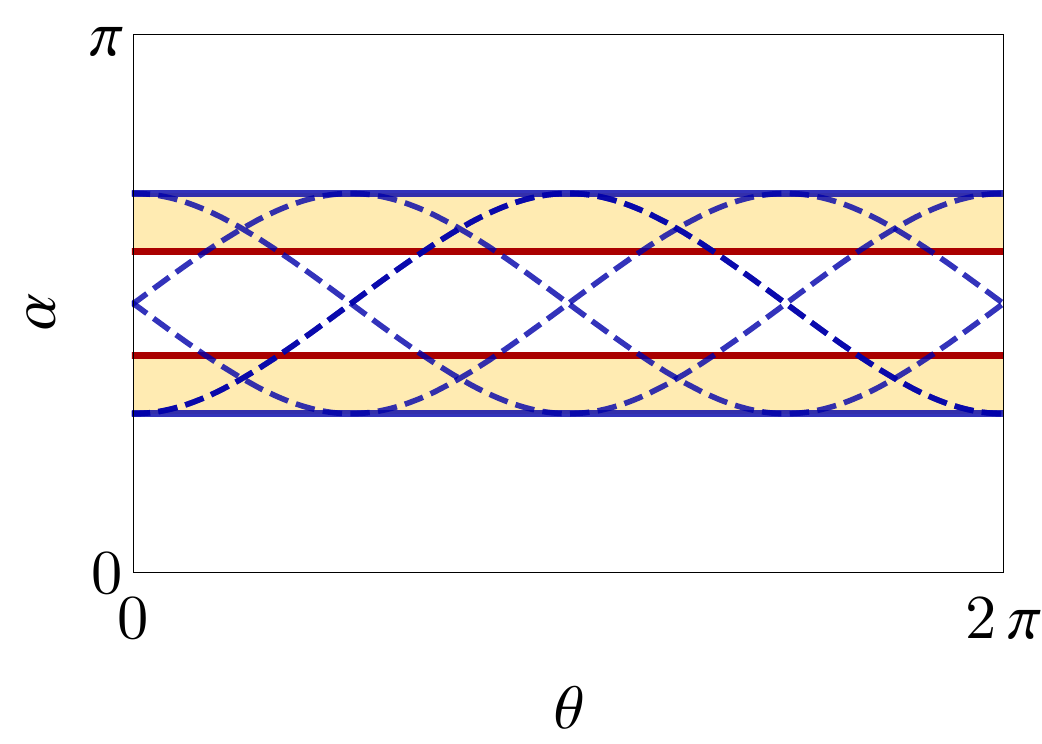}
\hspace{0.12\textwidth}
\includegraphics[height=.35\textwidth,angle=90]{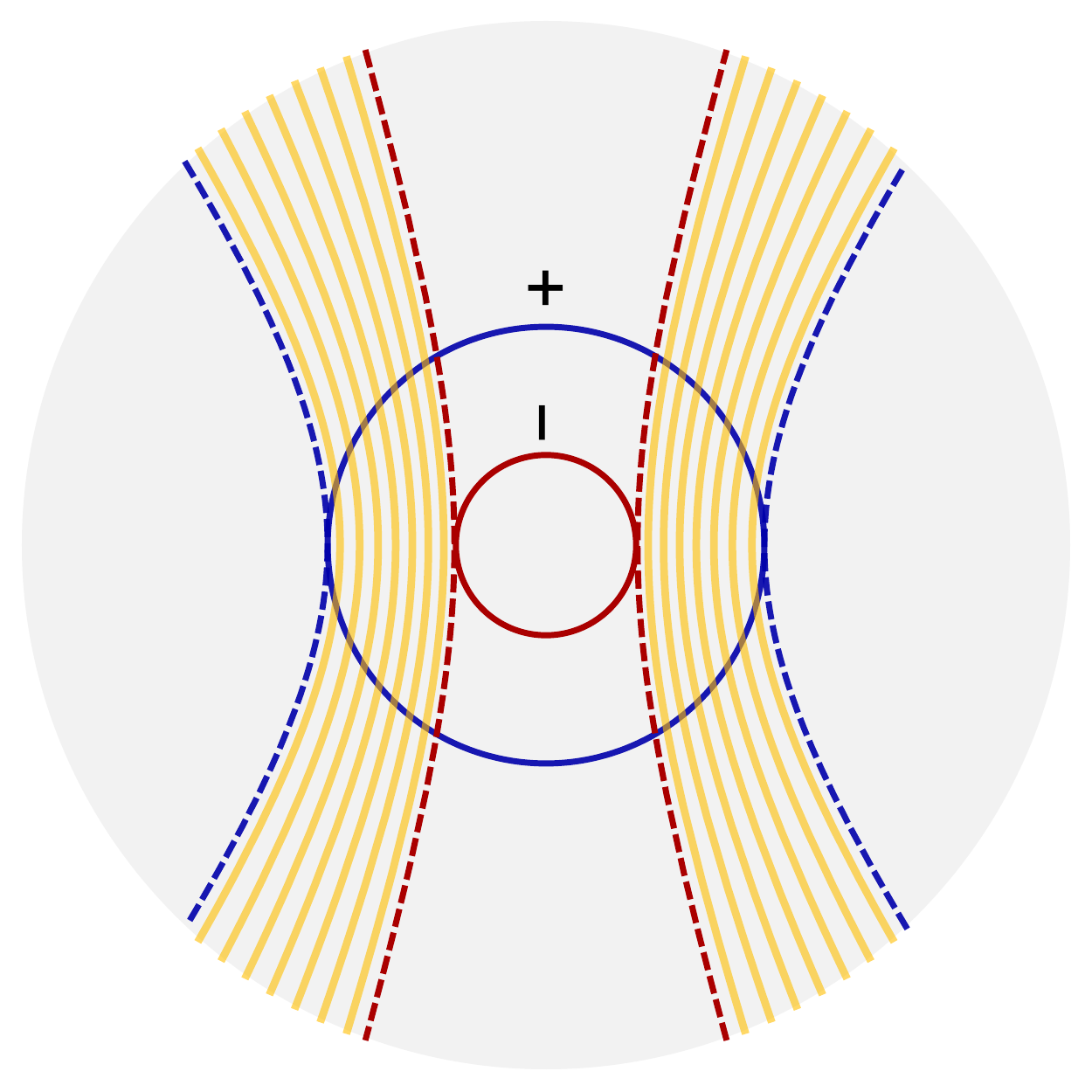} 
\caption{Left: The kinematic region corresponding to a collection of bulk curves with signed weight. The yellow colored regions are the set of geodesics with positive intersection number. We can read off that the bulk image of this region contains curves with nonpositive coefficients, because the tangent point-curves (blue dashed lines) are not entirely contained in the yellow region. Right: the corresponding signed curve consists of a large circle with positive weight and a small circle with negative weight. 
We can think of the kinematic region on the left as the difference of the kinematic regions of the constituent circles.}
\label{neglength}
\end{figure}

A necessary condition for a kinematic region to correspond to a simple bulk curve is that it be invariant under the $\mathbb{Z}_2$ symmetry (\ref{z2kinem}).  In other words, the intersection number between a geodesic and a curve is independent of the orientation of that geodesic.  However, this condition is not sufficient; see Fig.~\ref{neglength} for a simple example.  If the full strip between the outer blue lines had been included, it would correspond to a closed convex bulk curve. The removal of the inner strip corresponds to the subtraction of an additional curve, the inner circle in Fig.~\ref{neglength}. Fig.~\ref{dudv} instantiates the same conclusion: 
a $u,v$ rectangle in kinematic space corresponds to four geodesics, weighted so as to compute the conditional mutual information of three neighboring intervals.

We would like to state sufficient conditions for a kinematic region to correspond to a positive-length bulk curve.  The answer comes from eq.~(\ref{bulkcurvedef}): the kinematic region must be a weighted union of point-curves.  Based on Sec.~\ref{smoothconvexcurves}, this condition can be restated as follows: a given kinematic region is dual to a positive length bulk curve if and only if all the point-curves tangent to the region's boundaries are fully included in the region.  This clarifies why the kinematic strip considered in Fig.~\ref{neglength} does not define a single bulk curve with constant positive weight: all point-curves tangent to the lower (blue) boundary of the region (dashed blue curves) are only partially contained in the region whereas all point-curves tangent to the red boundary lie completely outside it.

\section{Exercises}
\Label{exercises}

\begin{problem}[Flat-space length formula]
Derive eq.~(\ref{flatcrofton}) for closed, convex, smooth curves $\gamma$.
\Label{deriveflatcrofton}
\end{problem}

\begin{solution}
To start, we find a coordinate representation of $\gamma$ in terms of $p(\theta)$. When a straight line (\ref{straight}) is tangent to $\gamma$ at a point with coordinates $(x,y)$, then $dx/d\theta = dy/d\theta = 0$. Therefore, differentiating eq.~(\ref{straight}) gives:
\begin{equation}
- x \sin\theta + y \cos\theta - dp/d\theta = 0\,. \Label{deriv}
\end{equation}
Isolating $x$ and $y$ from eqs.~(\ref{straight}) and (\ref{deriv}), we obtain a parametric form of the curve $\gamma$:
\begin{equation}
(x, y) = \big(p\cos\theta - (dp/d\theta)\sin\theta,\, p \sin\theta + (dp/d\theta) \cos\theta \big)
\Label{defgamma}
\end{equation}
Differentiating (\ref{defgamma}) with respect to $\theta$, we obtain the length element along the curve $\gamma$:
\begin{equation}
ds = \sqrt{dx^2 + dy^2} = \big(p + d^2p/d\theta^2\big)\, d\theta
\Label{dsdtheta}
\end{equation}
Because $\gamma$ is assumed closed, smooth and convex, the polar angle of its normal vector, $\theta$, sweeps the values from 0 to $2\pi$. Therefore the circumference of $\gamma$ can be written as:
\begin{equation}
\textrm{circumference of $\gamma$} = \oint ds = \int_0^{2\pi} d\theta\, p(\theta). \Label{flatcroftonex}
\end{equation}
The term involving $d^2p/d\theta^2$ drops out, because it is a total derivative.
\end{solution}

\begin{problem}[Flat-space area formula]
For the same $\gamma$, prove that:
\begin{equation}
\textrm{area enclosed by $\gamma$} = \frac{1}{2} \int_0^{2\pi} d\theta\, \big(p(\theta)^2 - p'(\theta)^2\big). \Label{flatareaex}
\end{equation}
\end{problem}

\begin{solution}
Without loss of generality, assume that the origin is enclosed by $\gamma$. Then the area enclosed by $\gamma$ consists of small triangles of height $p(\theta)$ and width $ds$ given in eq.~(\ref{dsdtheta}) and becomes:
\begin{equation}
\textrm{area enclosed by $\gamma$} = \frac{1}{2} \int_0^{2\pi} d\theta\, \big(p(\theta) + p''(\theta)\big)\, p(\theta) = \frac{1}{2} \int_0^{2\pi} d\theta\, \big(p(\theta)^2 - p'(\theta)^2\big)
\end{equation}
The last step follows from integration by parts.
\end{solution}


\begin{problem}[Flat-space Crofton formula]
For the same $\gamma$, prove eq.~(\ref{flatfinalcrofton}).
\Label{derivefinalflat}
\end{problem}

\begin{solution}
We begin with a simple rewriting of eq.~(\ref{flatcrofton}):
\begin{align}
\textrm{circumference of $\gamma$} 
= & \,\int_0^{2\pi} d\theta\, p(\theta) = \int_0^{2\pi} d\theta\, \int_0^{p(\theta)} dp 
\nonumber \\
= & \,\,\,\frac{1}{2} \int_0^{2\pi} d\theta\, \int_0^{p(\theta)} dp \, +\, 
\frac{1}{2} \int_0^{2\pi} d\theta\, \int_{-p(\theta+\pi)}^0 dp 
\nonumber \\
= & \,\,\,\frac{1}{2} \int_0^{2\pi} d\theta\, \int_{-p(\theta+\pi)}^{p(\theta)} dp
\end{align}
Now recognize that at fixed $\theta$, the limits of integration over $p$ define the two $\theta$-sloped lines tangent to $\gamma$. Integrating over these values of $p$ amounts to integrating over all lines that intersect $\gamma$. Because we assume that $\gamma$ is closed and convex, such lines intersect $\gamma$ at 2 points: $n_\gamma(\theta, p) = 2$. All other lines are either tangent to $\gamma$ (a set of measure zero) or do not intersect $\gamma$ so $n_\gamma(\theta, p) = 0$. This proves the result.
\end{solution}


\begin{problem}[Kinematic measure for hyperbolic plane]
Guess the kinematic measure for the hyperbolic plane starting from the flat space kinematic measure (\ref{flatcroftonform}).
\Label{makeguess}
\end{problem}

\begin{solution}
Rotations in $\theta$ are symmetries of the hyperbolic plane, so we may start with an ansatz:
\begin{equation}
\omega(\theta, \alpha) = df(\alpha)\wedge d\theta\,.
\end{equation}
This ought to reduce to the flat space measure (\ref{flatcroftonform}) in a neighborhood of the origin $\rho \ll 1$. In this neighborhood, the geodesic (\ref{geodesic}) becomes:
\begin{equation}
\rho\, \cos\tilde\theta \cos\theta + \rho\, \sin\tilde\theta \sin\theta = \cos\alpha
\qquad \to \qquad x \cos\theta + y \cos\theta = \cos\alpha\,.
\Label{badlimit}
\end{equation}
Na{\"\i}vely this suggests $f(\alpha) = \cos\alpha$, but one has to be careful. The limit (\ref{badlimit}) only works for $\alpha \approx \pi/2$, because other geodesics never get near the origin. Noting that $f(\alpha)$ must be odd about $\alpha = \pi/2$ so that $\omega(\theta, \alpha)$ becomes even, this implies that:
\begin{equation}
f(\alpha) = \cos\alpha \left( 1 + \mathcal{O}\big( (\alpha - \pi/2)^2 \big) \right)
\end{equation}
At the same time, $f(\alpha)$ should blow up as $\alpha \to 0$ and $\pi$, because geodesics become infinitely dense near the boundary. This does not uniquely fix $f(\alpha)$, but the simplest guess is $f(\alpha) = \cot\alpha$, which leads to eq.~(\ref{atcroftonform}).\end{solution}


\begin{problem}[Invariance of kinematic measure for hyperbolic space]
Prove that measure (\ref{atcroftonform}) is invariant under the isometries of the hyperbolic plane.
\Label{proveguess}
\end{problem}

\begin{solution}
Measure (\ref{atcroftonform}) is clearly invariant under rotations; the nontrivial part is the invariance under boosts. To verify it, we first have to find how $\theta$ and $\alpha$ in eq.~(\ref{geodesic}) change under the action of a boost. 

In the hyperboloid model of the hyperbolic plane,
\begin{align}
t^2 -x^2 - y^2 = 1 \qquad & {\rm with} \\
t = \cosh\rho \qquad & {\rm and} \qquad x = \sinh\rho \cos\tilde\theta 
\qquad {\rm and} \qquad y = \sinh\rho \sin\tilde\theta\,, \nonumber
\end{align}
the boost along the $\tilde\theta = 0$ axis is a Lorentz transformation in the $t-x$ directions. Denoting the rapidity of the boost with $\rho_0$, its effect on the coordinates $\rho$ and $\tilde\theta$ in metric (\ref{h2metric}) is:
\begin{eqnarray}
t: \quad & \phantom{\cos\theta} \cosh\rho' 
& = \, \cosh\rho_0 \cosh\rho - \sinh\rho_0 \sinh\rho  \cos\tilde\theta \nonumber \\
x: \quad & \sinh\rho' \cos\tilde\theta' 
& = \, -\sinh\rho_0 \cosh\rho  + \cosh\rho_0 \sinh\rho  \cos\tilde\theta 
\Label{coordchange} \\
y: \quad & \sinh\rho' \sin\tilde\theta' 
& = \,\sinh\rho \sin\tilde\theta \nonumber 
\end{eqnarray}
Consider a geodesic characterized by $\theta'$ and $\alpha'$ in coordinates $\rho'$ and $\tilde\theta'$: 
\begin{equation}
\tanh\rho' \cos(\tilde\theta' - \theta') = \cos\alpha'.
\end{equation}
After changing coordinates according to (\ref{coordchange}), the geodesic may again be brought to the canonical form (\ref{geodesic}) involving $\rho$ and $\tilde\theta$, but with new parameters given by:
\begin{align}
\alpha & = \cos^{-1} \frac{\cosh\rho_0 \cos\alpha' + \sinh\rho_0 \cos\theta'}{\sqrt{\sin^2\theta' + (\cosh\rho_0 \cos\theta' +\sinh\rho_0 \cos\alpha')^2}} 
\Label{boost1action}
\\
\theta & = \tan^{-1} \frac{\sin\theta'}{\cosh\rho_0 \cos\theta' +\sinh\rho_0 \cos\alpha'}
\Label{boost2action}
\end{align}
It remains to verify that:
\begin{equation}
\omega(\theta, \alpha) = - \frac{d\alpha \wedge d\theta}{\sin^2\alpha} =
- \frac{d\alpha' \wedge d\theta' }{\sin^2\alpha}\, 
\left| \begin{array}{cc} 
\frac{\partial \theta}{\partial \theta'} & \frac{\partial \alpha}{\partial \theta'} \\
\frac{\partial \theta}{\partial \alpha'} & \frac{\partial \alpha}{\partial \alpha'}
\end{array} \right|
\stackrel{?}{=} 
- \frac{d\alpha' \wedge d\theta'}{\sin^2\alpha'} = \omega(\theta', \alpha').
\end{equation}
The resulting equation is a straightforward trigonometric identity.
\end{solution}


\begin{problem}[Weighted combinations of curves] 
\Label{lastqn}
Fig.~\ref{problemfig} shows a concave bulk curve and its corresponding kinematic region. What (weighted) bulk curves correspond to the regions A and B in the right panel?
\end{problem}

\begin{figure}[h!]
\centering
\includegraphics[width=.3\textwidth]{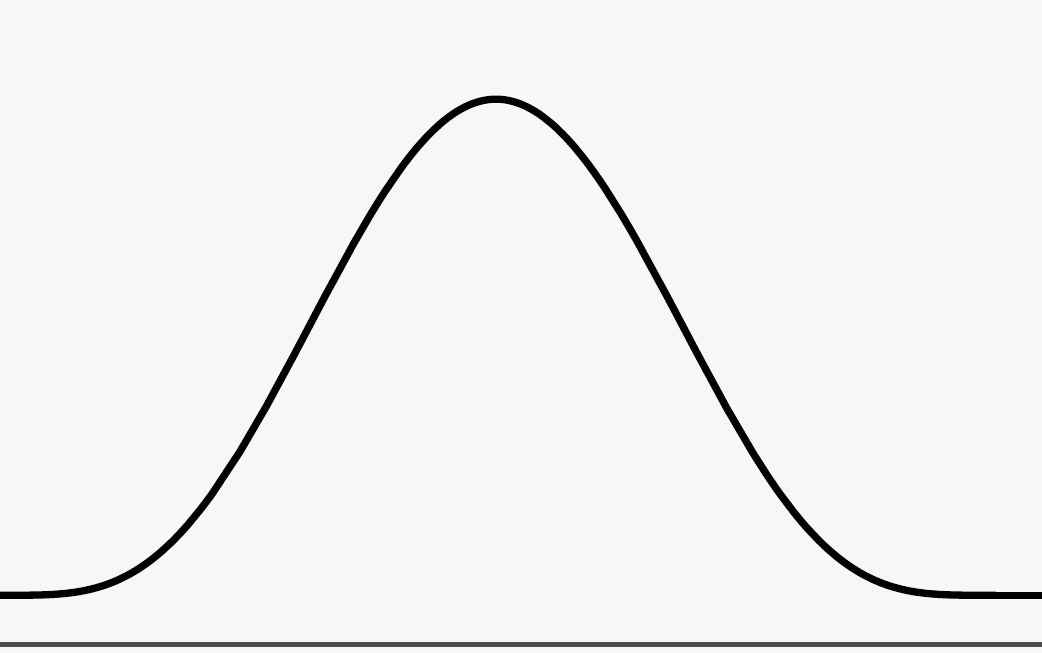}
\hspace{0.02\textwidth}
\includegraphics[width=.3\textwidth]{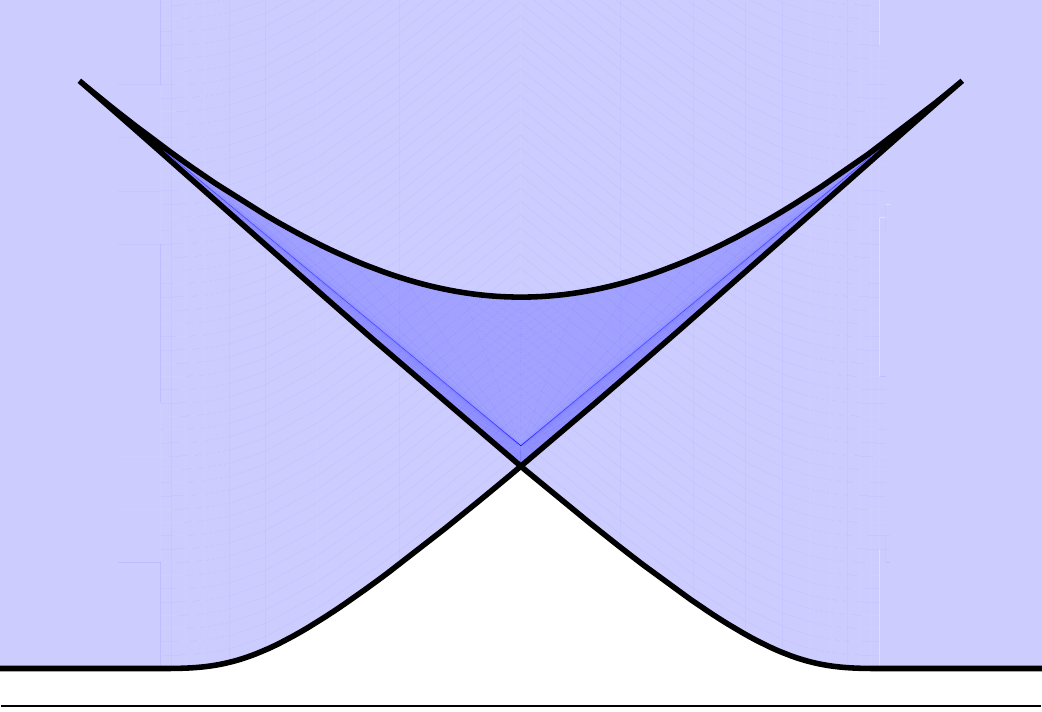}
\hspace{0.02\textwidth}
\includegraphics[width=.3\textwidth]{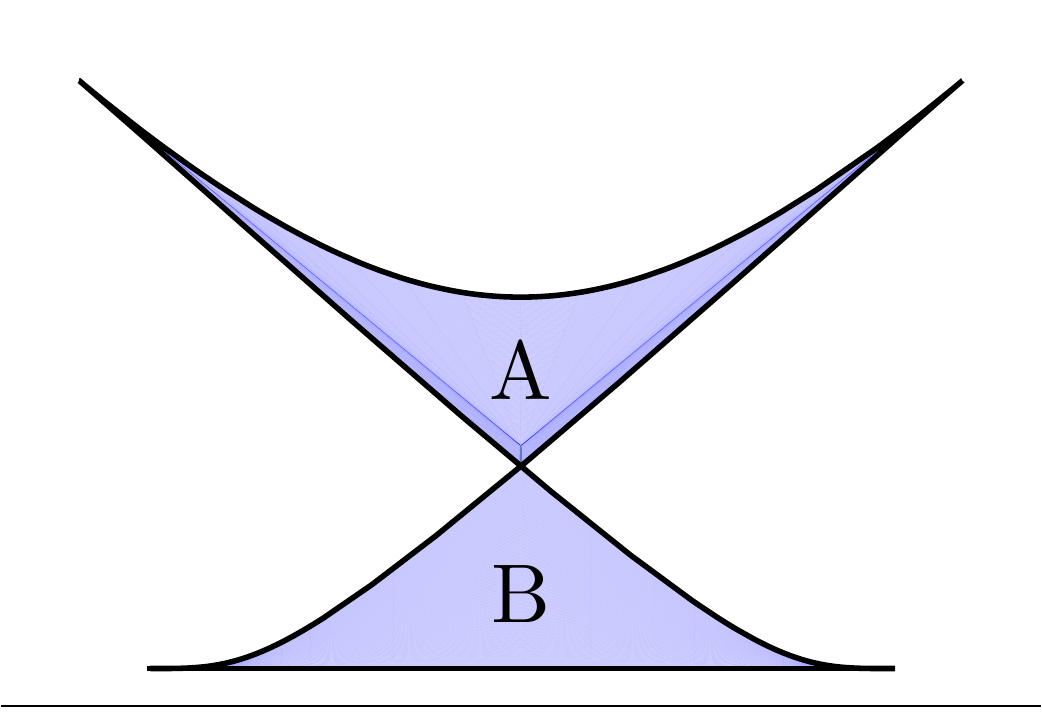}
\caption{Exercise~\ref{lastqn}: A concave bulk curve and its corresponding kinematic region (with multiplicities) distinguishes two swallowtail regions A and B in kinematic space.}
\Label{problemfig}
\end{figure}

\begin{solution}
The answer is shown in Fig.~\ref{solutionfig}. To understand it, start with the kinematic region in the middle panel of Fig.~\ref{problemfig}. If we subtract region A from it, we end up counting every intersecting geodesic with multiplicity 1, which produces the convex cover of the bulk curve (see Appendix~\ref{concaveappendix}). Thus, region A corresponds to the difference between the given curve and its convex cover, which is shown on the left of Fig.~\ref{solutionfig}. 

If we then add region B, we will obtain a kinematic region with a `horizontal' boundary. This corresponds to a bulk curve at a constant radial coordinate. The difference between that and the convex cover of the original curve is shown in the right panel of Fig.~\ref{solutionfig}. 
\end{solution}

\begin{figure}[t]
\centering
\includegraphics[width=.42\textwidth]{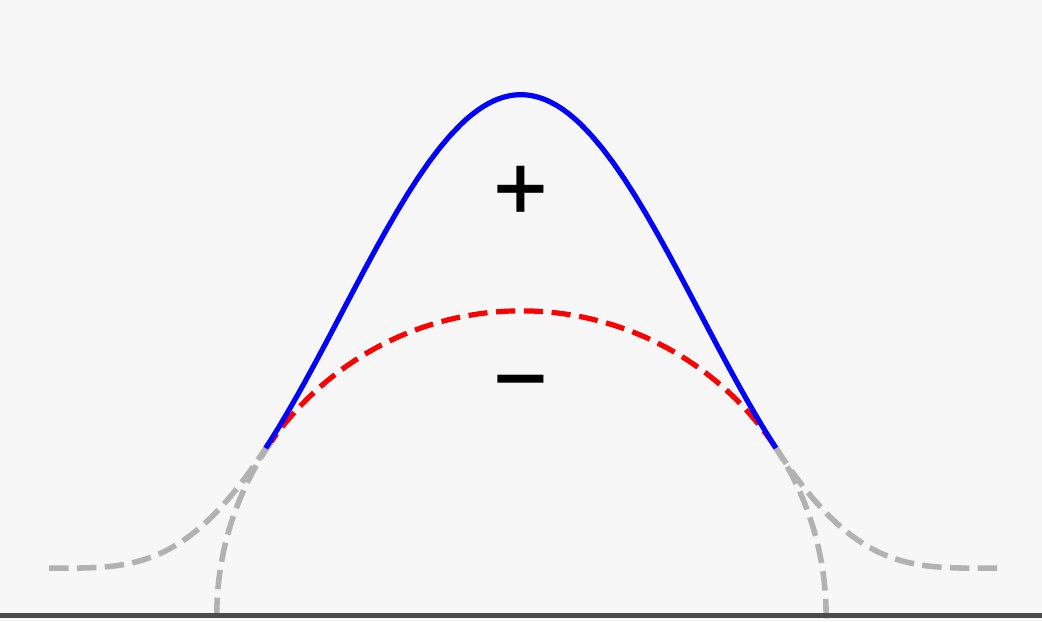}
\hspace{0.1\textwidth}
\includegraphics[width=.42\textwidth]{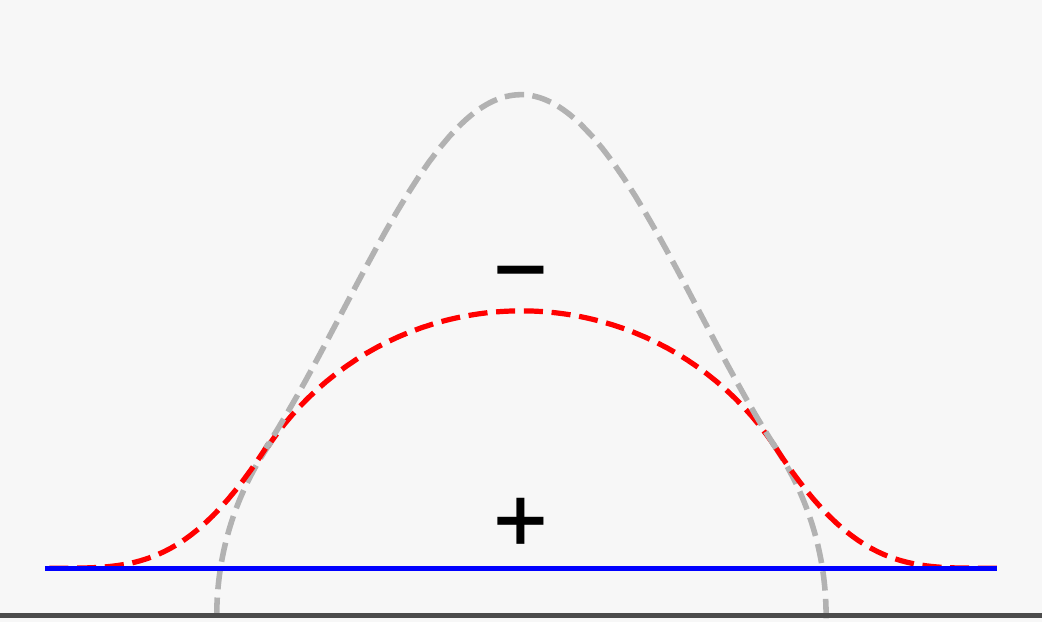}
\caption{Answers to Exercise~\ref{lastqn}; see text. Both answers are differences of two curves. The positive-weight segments are shown in continuous blue while the negative-weight segments are displayed in dashed red. The circular shape is the geodesic, which closes the convex cover of the initial bulk curve.}
\Label{solutionfig}
\end{figure}


\begin{problem}[An explicit calculation] 
\Label{addandsubtract}
Compute the volume of a triangle with vertices at $(\theta, \alpha) = (0, \alpha_0)$ and $(\pm \alpha_1 \mp \alpha_0, \alpha_1)$ in kinematic space. Interpret the result in terms of lengths of bulk curves. Assume $\alpha_0 > \alpha_1$.
\end{problem}

\begin{solution}
In $u,v$-coordinates (viz. eq.~\ref{defuv}) the integral becomes:
\begin{align}
\int_{-\alpha_0}^{\alpha_0-2\alpha_1} du & \int_{2\alpha_1+u}^{\alpha_0} dv \,
\frac{\partial^2S(u,v)}{\partial u \, \partial v} 
\Label{explcomp} \\
 = & \, \textcolor{red}{-S(-\alpha_0, \alpha_0)} 
 \textcolor{blue}{\, + \,  
2 \cdot \frac{1}{2}\, S(\alpha_0-2\alpha_1, \alpha_0) -
\int_{-\alpha_0}^{\alpha_0-2\alpha_1} du\, 
\frac{\partial S(u,v)}{\partial u} \Big|_{v = 2\alpha_1+u}}
\nonumber
\end{align}
The three terms on the right hand side compute lengths of the bulk curves shown in Fig.~\ref{finalfig}. The first term is minus the geodesic from $u = -\alpha_0$ to $v = \alpha_0$. The other two terms compute the length of the outer envelope of all geodesics of opening angle $\alpha_1$, which `fit in between' $u = -\alpha_0$ and $v = \alpha_0$. Specifically, the last term is the differential entropy of the open bulk curve that joins the deepest points of these geodesics.

Thus, the volume of our kinematic triangle again computes the difference of the lengths of two bulk curves. Both the positive-weight and the negative-weight contributions are infinite, but the difference is finite. Geometrically, the triangle comprises those geodesics, which intersect the blue curve in Fig.~\ref{finalfig} but do not reach the $\alpha_0$-sized red geodesic. All intersection numbers are 1.
\end{solution}

\begin{figure}[t!]
\centering
\includegraphics[width=.56\textwidth]{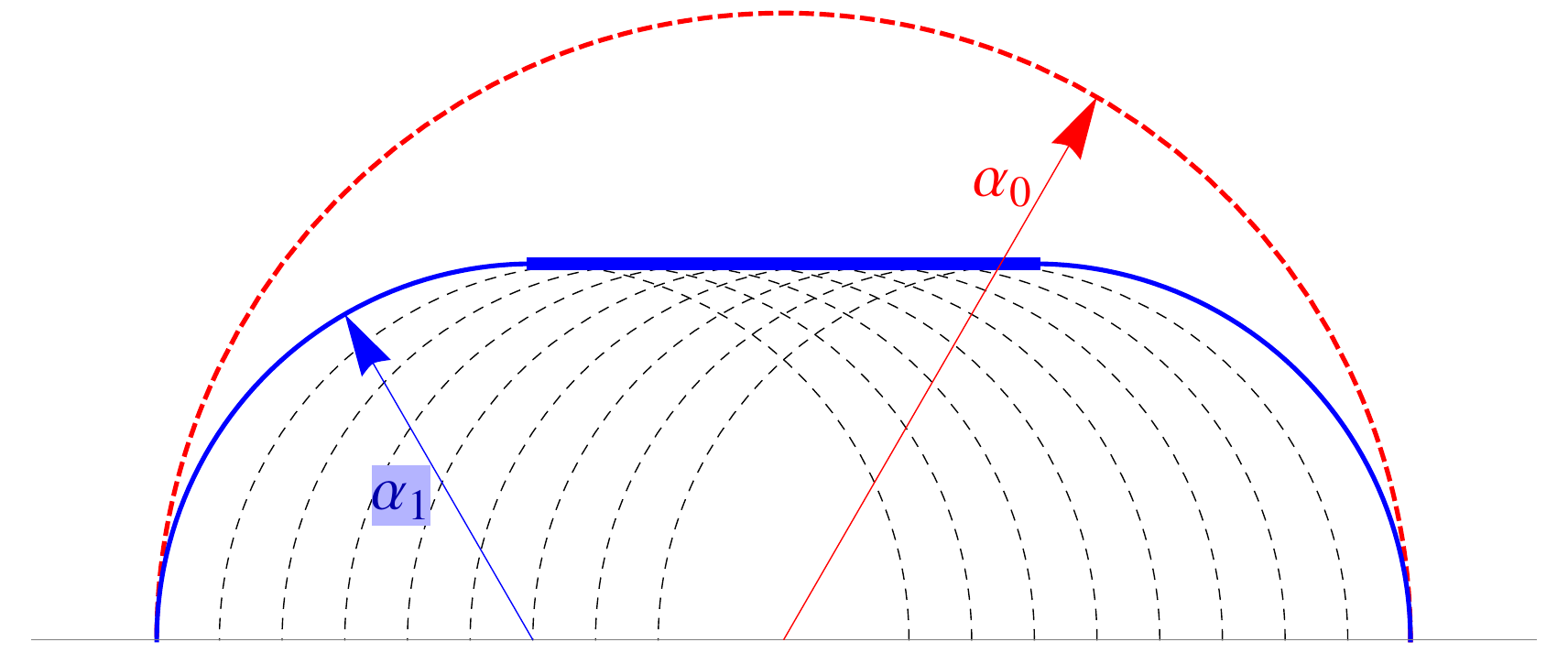}
\caption{Answer to Exercise~\ref{addandsubtract}: the combination of bulk curves, whose signed length is computed by the color-coded eq.~(\ref{explcomp}). The positive contribution (the outer envelope of $\alpha_1$-sized geodesics) is shown in continuous blue while the negative contribution (an $\alpha_0$-sized geodesic) is displayed in dashed red. The last term in eq.~(\ref{explcomp}) is the differential entropy formula of the bulk curve that joins the deepest points of the $\alpha_1$-sized geodesics; it is marked thicker.
}
\Label{finalfig}
\end{figure}


\end{document}